\documentclass[letterpaper]{JHEP3}
\usepackage{graphicx}
\usepackage{amsmath}
\usepackage{amsfonts}
\usepackage{amssymb}
\usepackage{dsfont}
\usepackage{dcolumn}
\usepackage{bm}
\def \cha{\widetilde{\chi}^{\pm}_1}

\def \na{\widetilde{\chi}^{0}_1}
\def \nb{\widetilde{\chi}^{0}_2}
\def \nc{\widetilde{\chi}^{0}_3}

\def \g{\widetilde{g}}

\def \ta{\widetilde{t}_1}

\def \sta{\widetilde{\tau}_1}
\def \stb{\widetilde{\tau}_2}

\def \slr{\widetilde{l}_R}

\def \snl{\widetilde{\nu}_{\tau}}
\def \snm{\widetilde{\nu}_{\mu}}

\def \hc{H^{\pm}}

\def\beq{\begin{equation}}
\def\be{\begin{equation}}
\def\beqn{\begin{eqnarray}}
\def\ee{\end{equation}}
\def\eeq{\end{equation}}
\def\eeqn{\end{eqnarray}}

\def \non{nonuniversalities~}
\title
{Sparticles at the LHC}
\author{Daniel Feldman, Zuowei Liu, and Pran Nath\\
Department of Physics, Northeastern University, Boston, MA 02115, USA\\
E-mails:\email{ feldman.da@neu.edu,   liu.zu@neu.edu, nath@neu.edu
}\\}
\abstract{ Sparticle mass hierarchies will play an important role in
the type of signatures that will be visible at the Large Hadron
Collider. We analyze these hierarchies for the four lightest
sparticles for a general class of supergravity unified models
including nonuniversalities in the soft breaking sector. It is shown
that out of nearly $10^4$ possibilities of sparticle mass
hierarchies, only a small number  survives the rigorous constraints
of radiative electroweak symmetry breaking, relic density and other
experimental constraints. The signature space of these mass patterns
at the Large Hadron Collider is investigated using a large set of
final states including   multi-leptonic states, hadronically
decaying $\tau$s,  tagged $b$ jets and other hadronic jets. In all,
we analyze more than 40 such lepton plus jet and missing energy
signatures along with several kinematical signatures such as missing
transverse momentum,  effective mass, and invariant mass
distributions of final state observables. It is shown that a
composite analysis can produce significant discrimination among
sparticle mass patterns allowing for a possible identification of
the source of soft breaking. While the analysis given is for
supergravity models, the techniques based on mass pattern analysis
are applicable to wide class of models including string and brane
models. }
\preprint{Journal-Ref: JHEP04(2008)054\\Published: April 14, 2008}
\begin{document}
\section{Introduction}
~~~~~Supersymmtery (SUSY) remains a leading candidate to describe new
physics beyond that of the Standard Model (SM).
Recently, an approach for identifying supersymmetric particles (sparticles)
was proposed involving sparticle mass hierarchies, or sparticle mass
patterns.  Such patterns could yield distinct identifiable signatures
at the Fermilab's Tevatron and at the CERN Large Hadron Collider (LHC)\cite{Feldman:2007zn,Feldman:2007fq}.
At the same time, the hierarchical mass patterns are
model dependent and the determinations of such patterns could be  helpful
in extrapolating the data back to the theoretical model.
This new approach has been investigated  within the framework of gravity mediated breaking of supersymmetry\cite{msugra,barbi,hlw}
and specifically within the minimal supergravity grand unified model, the
mSUGRA model \cite{msugra} (for a review see \cite{review})
with sparticle mass ranges that lie within reach of the present colliders
(for a review of recent  search strategies see \cite{Yamamoto:2007it}).
The analysis of \cite{Feldman:2007zn,Feldman:2007fq}
was  a rather brief introduction to the technique.
Here we carry out a more in depth analysis within models with both
universal and nonuniversal soft
supersymmetry breaking \cite{Kaplunovsky:1993rd,Brignole:1993dj,Nath:1997qm}.
Thus, in the minimal supersymmetric extension of the Standard Model (MSSM)
there are 32 supersymmetric particles. We list them here to set notation.
There are 4 Higgs boson states,  of which three  $(h,H,A)$ are neutral,
the first two being CP even and the third CP odd, and one
charged Higgs $H^{\pm}$. In the gaugino-Higgsino sector there are two charged mass eigenstates
(charginos) $\tilde\chi^{\pm}_{i=1,2}$, four charge neutral states (neutralinos) $\tilde\chi^0_{i=1,4}$,
and the gluino $\tilde g$.
In the sfermion sector, before diagonalization, there are
9  scalar leptons (sleptons) which are superpartners of the leptons with left and right chirality and
are denoted as:
$\{\tilde e_{L,R},\tilde \mu_{L,R}, \tilde\tau_{L,R} ,
  \tilde\nu_{e_L},\tilde \nu_{{\mu}_L},\tilde \nu_{{\tau}_L} \}$.
Finally there are 12 squarks which are the superpartners of the
quarks and are represented by:
$\{\tilde u_{L,R},\tilde c_{L,R}, \tilde t_{L,R},
  \tilde d_{L,R},\tilde s_{L,R}, \tilde b_{L,R} \}$.
Mass diagonal slepton and squark states will in general be mixtures of
$L,R$ states.

If the 32 masses are treated as essentially all independent, aside
from sum rules (for a pedagogical analysis on sum rules in the
context of unification and RG analysis see \cite{Martin:1993ft}) on
the Higgs, sfermions, chargino and neutralino masses, then without
imposition of any phenomenological constraints, the number of
hierarchical patterns for the sparticles could be as many as
$O(10^{28})$ or larger. This represents a mini landscape in a loose way  reminiscent
of  the string landscape (which, however, is much larger with as
many as
 $O(10^{1000})$ possibilities).
[Here we refer to the landscape of mass hierarchies and not to the
landscape of vacua as is  the case when one talks of a string
landscape. For the string case the landscape consists of a countably
discrete set, while for the case considered here, since the
parameters can vary continuously, the landscape of vacua is indeed
much larger.  However, our focus will be the landscape of mass
hierarchies.] Now, the number of
possibilities can be reduced by very significant amounts in
supergravity models with the imposition of  the constraints of
radiative electroweak symmetry breaking (REWSB)\footnote{EWSB can be
realized non-radiatively  for certain choices of parameters in the
presence of \non in the Higgs sector. Since in this analysis
boundary conditions have been imposed at the GUT scale and RGEs have
been used to obtain the low energy physics, we will
 retain this terminology in the subsequent descriptions of EWSB.},  and other
phenomenological constraints.
This was precisely what was accomplished in the analysis of \cite{Feldman:2007zn,Feldman:2007fq}.
The analysis of Ref.\cite{Feldman:2007zn,Feldman:2007fq} focused on the mass hierarchies for
the first four lightest sparticles, and found the
residual number of  hierarchies to be 22 in mSUGRA.
Here, the possible signatures from some of the patterns
were also discussed along with the prospects for direct detection of dark matter within various
mass hierarchies.

%%%%%%%%%%%%%%%%%%%%%%%%%%%%%%%%%%%%%%%%%%%%%%%%%%%%%%%%%%%%%%%%%%%%%%

The phenomenology of supergravity (SUGRA)  models has been discussed since
their inception and there exists  a considerable amount of  literature
regarding the implications of SUGRA
(for early works see
\cite{earlypheno, early}, for more recent works see
\cite{modern,Trotta,Allanach0607}, for
works with \non  see \cite{NU}, and for works with hierarchical breaking and with
$U(1)$ gauge extensions
see \cite{Kors:2004hz,Feldman:2006wd,Barger:2004bz}).
While many analyses of the mSUGRA parameter space have been
limited to the  case of vanishing trilinear couplings,  several recent works
\cite{Ellis:2004tc,Stark:2005mp,Djouadi:2006be,
Ellis:2006ix,Trotta,Allanach0607,uc,Bringmann:2007nk}
 have appeared relaxing this assumption, and new portions of
the parameter space have been found consistent with all known
experimental constraints on the model.

%%%%%%%%%%%%%%%%%%%%%%%%%%%%%%%%%%%%%%%%%%%%%%%%%%%%%%%%%%%%%%%%%%%%%%
 In this paper we give  a more exhaustive  analysis of sparticle mass hierarchies
 for SUGRA models including nonuniversalities and also carry out a more
 detailed  analysis  of the signatures arising from these patterns. We further focus on ways in which
 patterns can be discriminated from each other using the relevant distinctive  features of the signature
 space.  It is found that for some model points  one encounters the phenomenon where two distinct points in the parameter
 space of soft  breaking may yield the same signatures within a 2$\sigma$ error bar.
 We also discuss in this paper how such  signature degeneracies  can sometimes
 be lifted by an increased integrated luminosity.  Finally,  we  discuss the
 issue of   how
 well the soft parameters $m_0$ and $m_{1/2}$  (where $m_0$, $m_{1/2}$ are the mass
 parameters in mSUGRA models defined in Sec.(2))
 may  eventually be determined at
 the LHC which allows one to obtain an estimate on the resolution of these parameters
 using  optimal LHC luminosities.

 The outline of the rest of the paper is as follows:
In Sec.(\ref{A})
we give a discussion of the sparticle landscape. Specifically, the
landscape for the  4 lightest sparticles (in addition to the lightest
Higgs boson) for the mSUGRA case is discussed in Sec.(\ref{A1}) and
the landscape for the  4 lightest sparticles for the nonuniversal
SUGRA case is  discussed in Sec.(\ref{A2}). This includes cases with
\non in the Higgs sector, \non in the third generation sector, and
\non in the gaugino sector. In
Sec.(\ref{A4}), we
also discuss the possible number of mass patterns that can arise for the
above case, as well as for the case when all 32 sparticle masses are
taken into account.
An analysis of the patterns and their origin in the space of
soft breaking parameters is given in Sec.(\ref{B}).
An analysis of the benchmarks  for  the landscape of 4 sparticle
patterns is given in Sec.(\ref{B3}).
Sec.(\ref{D}) is devoted to the
discussion of the sparticle signatures at the LHC.  In Sec.(\ref{Dsub11}) we  give
a discussion of the various  SUSY tools  that are utilized in this analysis. We
discuss technical details of the analysis of the LHC signatures we
have investigated in Sec.(\ref{D1}). We then move on to discuss how
one can distinguish sparticle mass patterns arising in mSUGRA in Sec.(\ref{D2}),
and sparticle  mass patterns in SUGRA with \non in Sec.(\ref{D3}).
The trileptonic signal as a tool to distinguish patterns is discussed in Sec.(\ref{D6}).
We utilize both event counting signatures and  kinematical signatures
 in our analysis, the latter being discussed in Sec.(\ref{D5}).
In Sec.(\ref{D7}), a method for distinguishing patterns utilizing a
large set of signatures is also given.
We discuss the signature space degeneracy among  different models and how to lift it
in Sec.(\ref{E1}), and then we generalize our analysis to investigate  the resolving power of the LHC
with regards to its ability to probe the soft  parameter space in Sec.(\ref{E2}).
Conclusions are  given in Sec.(\ref{F}). Some of our longer tables
have been relegated to the Appendix. \\
%%%%%%%%%%%%%%%%%%%%%%%%%%%%%%%%%%%%%%%%%%%%%%%%%%%%%%%%%%%%%%%%%%%%%%
%%% --------------- The Sparticle Landscape ---------------
%%%%%%%%%%%%%%%%%%%%%%%%%%%%%%%%%%%%%%%%%%%%%%%%%%%%%%%%%%%%%%%%%%%%%%
\section{The Sparticle  Landscape \label{A}}
The analysis proceeds by specifying the model input parameters at
the GUT scale, $M_{G}\sim 2\times 10^{16}$ GeV, (no flavor mixing is allowed at the GUT scale)
and using the renormalization group equations (RGEs)
to predict the sparticle masses and
mixing angles  at the electroweak scale. The RGE code used to
obtain the mass spectrum is SuSpect  2.34 \cite{SUSPECT}, which is the default RGE calculator in
MicrOMEGAs version 2.0.7 \cite{MICRO}.  We have also investigated
other RGE programs including ISASUGRA/ISAJET \cite{ISAJET}, SPheno
\cite{SPHENO} and SOFTSUSY \cite{SOFTSUSY}. We have cross checked
our analysis using different codes and find no significant disagreement
in most regions of the parameter space. The largest
sensitivity appears to arise for the case of large $\tan\beta$ and
the analysis is also quite sensitive to the running bottom mass and to the
top pole mass (we take
$m^{\overline{\rm MS}}_b(m_b)= 4.23~ {\rm GeV} $ and
$m_t(\rm pole) = 170.9 ~{\rm GeV}$  in this analysis).
Such sensitivities and their implications for the analysis of relic density calculations  are
well known in the literature\cite{sensitivity} and a detailed comparison for
various codes  can be found
in  Refs.~(\cite{Baer:2005pv}, \cite{Belanger:2005jk}, \cite{Allanach:2004rh}, \cite{Allanach:2003jw}).

Below we give the relevant constraints from collider and astrophysical data
that are applied throughout the analysis unless stated otherwise.
\begin{enumerate}
\item WMAP 3 year data:
The lightest R-Parity odd supersymmetric particle (LSP) is assumed charge neutral.
The constraint on the relic abundance of
dark matter under the assumption the relic abundance of neutralinos is the dominant component
places the bound: $0.0855<\Omega_{\na} h^2<0.1189~~(2\sigma)$ \cite{Spergel:2006hy}.
\item
As is well known sparticle loop exchanges make a contribution to the FCNC process
$b\rightarrow s\gamma$ which
is of the same order as  the Standard Model contributions (for an update of SUSY contributions
see  \cite{susybsgamma}).  The
 experimental limits on  $b\to s\gamma$ impose severe constraints
 on the SUSY parameter space and we use here the  constraints from
the Heavy Flavor Averaging Group (HFAG) \cite{hfag} along with the
BABAR, Belle and CLEO
   experimental results: ${\mathcal Br}(b\rightarrow s\gamma) =(355\pm 24^{+9}_{-10}\pm 3) \times 10^{-6}$.
A new estimate of ${\mathcal Br}(\bar B \to X_s \gamma)$ at
$O(\alpha^2_s)$ gives \cite{Misiak:2006zs} ${\mathcal
Br}(b\rightarrow s\gamma) =(3.15\pm 0.23) \times 10^{-4}$  which
moves the previous SM mean value of $3.6\times 10^{-4}$ a bit lower. In order
to accommodate this recent analysis on the SM mean, as well as the previous analysis, we have taken a wider
$3.5\sigma$ error corridor around the HFAG
value in our numerical analysis. The total ${\mathcal Br}(\bar B \to X_s \gamma)$ including
the sum of SM and SUSY contributions  are constrained by  this
corridor. With a 2$\sigma$ corridor, while some of the allowed points in our analysis will be eliminated,
the main results of our pattern analysis remain unchanged.
\item
The process $B_s\to \mu^+\mu^-$ can become significant for large
$\tan\beta$ since
the decay  has a leading $\tan^6\beta$ \cite{tb}
dependence and thus large $\tan\beta$ could be constrained by the experimental
limit ${\mathcal Br}( B_s \to \mu^{+}\mu^{-})$ $< 1.5 \times10^{-7}$ (90\% CL), $ 2.0 \times
10^{-7}$  (95\% CL) \cite{Abulencia:2005pw}.  This limit has
just recently been updated  \cite{Abazov:2007iy}
and gives ${\mathcal Br}( B_s \to \mu^{+}\mu^{-}) < 1.2 \times10^{-7}$  (95\% CL).
Preliminary analyses \cite{bsmumu07} have reported
the possibility of even more stringent constraints by a factor of 10.
We take a more conservative
approach in this analysis and allow model points subject to the bound
${\mathcal Br}( B_s \to \mu^{+}\mu^{-})  < 9\times 10^{-6}$ (for a review see \cite{Anikeev:2001rk}).

\item Additionally, we also impose a
lower limit  on the lightest  CP even  Higgs boson mass. For the Standard
Model like Higgs boson this limit is
$\approx$ 114.4 {~\rm GeV} \cite{smhiggs}, while a limit of 108.2 {\rm GeV} at 95\% CL
is set on the production of an invisibly decaying Standard Model like Higgs by
OPAL \cite{OPAL2007}. For the MSSM we take the constraint to be $m_h> 100 ~{\rm GeV}$.
A relaxation of the light Higgs mass constraint by 8 - 10 GeV
affects mainly the analysis of SUGRA models where the stop mass can be light.
However, light stops are possible even with the strictest imposition of the LEP bounds on the SM Higgs Boson.
We take the other sparticle mass
constraints to be $m_{\cha}>104.5 ~{\rm GeV}$ \cite{lepcharg} for the lighter
chargino,  $m_{\ta}>101.5 ~{\rm GeV}$  for the lighter stop,  and $m_{\sta}>98.8 ~{\rm  GeV}$ for the
lighter stau.
\end{enumerate}
In addition to the above one may also consider the constraints from the anomalous magnetic moment
of the muon. It  is known that the supersymmetric electroweak corrections to $g_{\mu}-2$ can be as large or
larger than the Standard Model electroweak corrections\cite{yuan}.
The implications of recent  experimental data  has
been discussed in several works (see, e.g.\cite{g2T}). As in \cite{Djouadi:2006be}, here we use a rather
conservative bound $-11.4\times 10^{-10}<g_{\mu}-2<9.4\times 10^{-9}$.
%%%%%%%%%%%%%%%%%%%%%%%%%%%%%%%%%%%%%%%%%%%%%%%%%%%%%%%%%%%%%%%%%%%%%%
\subsection{The mSUGRA landscape  for the 4 lightest sparticles
\label{A1}}
One mSUGRA model is a point in a 4 dimensional parameter space spanned by
$m_0$, $m_{1/2}$, $A_0$,  $\tan\beta$, and the sign of $\mu$,
where $m_0$ is the universal scalar mass, $m_{1/2}$ is the universal gaugino mass,
$A_0$ is the universal trilinear coupling, $\tan\beta$ is the ratio of the two
Higgs VEVs in the MSSM, and $\mu$
is the Higgs mixing parameter that enters via the term $\mu H_1 H_2$ in the superpotential.
Typically scans of the parameter space
are done by taking a vanishing
trilinear coupling, and/or by looking at fixed values of
$\tan\beta$ while varying ($m_0$, $m_{1/2}$). In this work we carry
out a random scan in the 4-D  input parameter space for fixed signs  of $\mu$
with Monte Carlo
simulations using flat priors under the following ranges of the
input parameters
\beqn
0 < m_0 < 4 {~\rm TeV}, ~~~ 0 < m_{1/2} < 2 {~\rm TeV}~~~
|A_0/m_0| < 10,~~~ 1 < \tan\beta < 60. \label{softranges}
\eeqn
\begin{table}[htbp]
    \begin{center}
\begin{tabular}{||l||l||c||c||}
\hline\hline
mSP&     Mass Pattern & $\mu >0$ & $\mu<0$
\\\hline\hline
mSP1    &   $\na$   $<$ $\cha$  $<$ $\nb$   $<$ $\nc$   & Y  & Y    \cr
mSP2    &   $\na$   $<$ $\cha$  $<$ $\nb$   $<$ $A/H$  & Y  & Y  \cr
mSP3    &   $\na$   $<$ $\cha$  $<$ $\nb$ $<$ $\sta$    & Y  & Y  \cr
mSP4    &   $\na$   $<$ $\cha$ $<$ $\nb$   $<$ $\g$     & Y  & Y  \cr
\hline
mSP5    &   $\na$ $<$ $\sta$  $<$ $\slr$  $<$ $\snl$      & Y  & Y  \cr
mSP6 &   $\na$   $<$ $\sta$  $<$ $\cha$  $<$ $\nb$      & Y  & Y  \cr
mSP7    &   $\na$   $<$ $\sta$  $<$ $\slr$  $<$ $\cha$  & Y  & Y  \cr
mSP8    &   $\na$ $<$ $\sta$  $<$ $A\sim H$             & Y  & Y  \cr
mSP9    &   $\na$   $<$ $\sta$  $<$ $\slr$ $<$ $A/H$    & Y  & Y  \cr
mSP10   &   $\na$   $<$ $\sta$ $<$ $\ta$ $<$ $\slr$     & Y & \cr
 \hline
mSP11   &   $\na$ $<$ $\ta$ $<$ $\cha$  $<$ $\nb$       & Y  & Y  \cr
mSP12 &   $\na$ $<$ $\ta$   $<$ $\sta$ $<$ $\cha$   & Y  & Y  \cr
mSP13   & $\na$   $<$ $\ta$ $<$ $\sta$  $<$ $\slr$      & Y  & Y  \cr
\hline
mSP14   &   $\na$   $<$  $A\sim H$ $<$ $\hc$        & Y  & \cr
mSP15   &   $\na$   $<$ $ A\sim H$ $<$ $\cha$   & Y  & \cr
mSP16   &   $\na$   $<$ $A\sim H$ $<$$\sta$         & Y  & \cr
\hline
mSP17   &   $\na$   $<$ $\sta$ $<$ $\nb$ $<$ $\cha$     & & Y \cr
mSP18   &  $\na$   $<$ $\sta$  $<$ $\slr$  $<$ $\ta$    & & Y \cr
mSP19   &  $\na$ $<$ $\sta$ $<$ $\ta$   $<$ $\cha$  & & Y \cr
 \hline
mSP20  & $\na$ $<$ $\ta$   $<$ $\nb$   $<$ $\cha$   & & Y \cr
mSP21   & $\na$   $<$ $\ta$   $<$ $\sta$  $<$ $\nb$     & & Y \cr
\hline
mSP22   & $\na$   $<$ $\nb$   $<$ $\cha$  $<$ $\g$  & & Y \cr
\hline\hline
 \end{tabular}
\caption{ Hierarchical mass patterns for  the four lightest
sparticles  in mSUGRA when $\mu <0$ and $\mu>0$.
The patterns can be classified according to the next to the lightest
sparticle. For the mSUGRA analysis the next to the lightest sparticle
is found to be either a chargino, a stau, a stop, a CP
even/odd Higgs, or the next lightest neutralino $\nb$. The
notation $A/H $ stands for either $A$ or $H$.
In mSP14-mSP16 it is
possible that the Higgses become
lighter than the LSP. Y stands
for appearance of the pattern for the sub case.}
\label{msptable}
\end{center}
 \end{table}
 \noindent
Since SUGRA models with $\mu>0$ are favored by the  experimental
constraints  much of the analysis presented here  focuses  on this case.
Specifically for the  $\mu>0$ mSUGRA case, we perform
a scan of the parameter space with a total of $2 \times 10^6$
trial parameter points. We delineate the patterns that emerge for the first four
lightest
sparticles.
Here we find that at least  sixteen  hierarchical mass  patterns emerge
which are labeled as mSPs (minimal SUGRA Pattern).
These mSPs can be generally classified according to the type of particle which is next
heavier than the LSP, and we find four classes of patterns in mSUGRA:  the chargino patterns (CP), the stau patterns (SUP),
the stop patterns (SOP),  and the Higgs patterns (HP), as exhibited below
\begin{enumerate}
\item Chargino patterns (CP) : mSP1, mSP2, mSP3, mSP4
\item Stau patterns (SUP) : mSP5, mSP6, mSP7, mSP8, mSP9, mSP10
\item Stop patterns (SOP) : mSP11, mSP12, mSP13
\item Higgs patterns  (HP) : mSP14, mSP15, mSP16.
\end{enumerate}
%%%
%%%
The hierarchical mass patterns  mSP1-mSP16 are defined in Table (\ref{msptable}).
 We note  that the pattern  mSP7 appears in the analyses of \cite{ArnowittTexas,Gounaris:2007gx,Buchmueller:2007zk}.
%%%
%%% --------------- mSPs -------------------- %%%%
\begin{figure}[t]
\begin{center}
\includegraphics[width=9.5 cm,height=8cm]{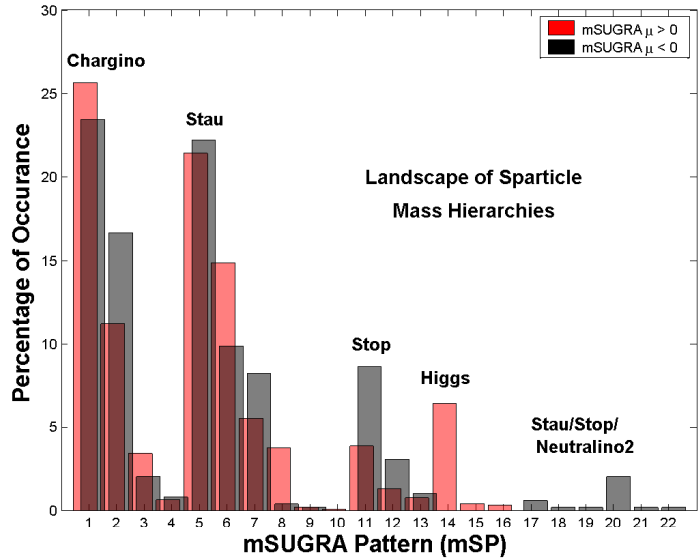}
\caption{Distribution of the surviving hierarchical mass patterns in the landscape
for the mSUGRA model with $\mu >0$ (light) and $\mu <0$ (dark),
under various constraints as discussed in the text.
 } \label{m-land}
\end{center}
\end{figure}
%%% --- table Snowmass & PostWmap and CMS begin--- %%%%
\begin{table}[h]
\begin{center}
%\scriptsize{
\begin{tabular}{|c|c |  }
\hline \hline Snowmass  &  mSP  \\\hline
\hline SPS1a, SPS1b, SPS5 & mSP7 \\
\hline SPS2 & mSP1 \\
\hline SPS3 & mSP5  \\
\hline SPS4, SPS6& mSP3  \\
\hline \hline\end{tabular}
\begin{tabular}{|c|c |  }\hline \hline Post-WMAP3 &  mSP\\\hline
\hline $A',B',C',D',G',H',J', M' $ & mSP5\\
\hline $I', L'$ &  mSP7\\
\hline $E'$ & mSP1 \\
\hline $K'$  & mSP6 \\
\hline \hline\end{tabular}
\begin{tabular}{|c|c |  }\hline \hline CMS LM/HM &  mSP\\\hline
\hline LM1, LM6, HM1 & mSP5\\
\hline LM2, LM5, HM2 &  mSP7\\
\hline LM3, LM7, LM8, LM9, LM10, HM4 & mSP1 \\
\hline LM4, HM3  & mSP3 \\
\hline \hline\end{tabular}
\caption{Mapping between the mSPs and
the Snowmass, Post-WMAP3, and CMS benchmark points. The points $B'
=$~LM1, $I' =$~LM2, $C' =$~LM6. HM1 in SuSpect  has $m_{\na} >
m_{\sta}$, but this is not the case for ISAJET, SPheno, and
SOFTSUSY. Among the CMS benchmarks, only LM1, LM2, LM6, and HM1, HM2 are capable of giving the
correct relic density. Thus the mapping above applies only to the
mass pattern, while all of our mSP and NUSP  benchmark points
satisfy the relic density constraints from MicrOMEGAs with SuSpect.
The CMS test points do a better job of representing mSP1 which is
the dominant pattern found in our analysis. There are no HP test
points or SOP test points in any of the previous works.}
 \label{pattern}\end{center}
\end{table}
We also performed a similar scan for the  mSUGRA  with $\mu<0$ case using
the Monte Carlo simulation with flat priors and the same parameter ranges
as specified in Eq.(\ref{softranges}). Most of the mSP patterns that
appear in the $\mu>0$ case also appear in the $\mu<0$ case (see
Table (\ref{msptable})). However, in addition one finds new
patterns shown below
\begin{enumerate}
\item Stau patterns  (SUP) : mSP17, mSP18, mSP19
\item Stop patterns  (SOP) : mSP20, mSP21
\item Neutralino patterns  (NP) : mSP22.
\end{enumerate}
We note that  the analysis of Ref.\cite{SPM} has a sparticle spectrum  which
corresponds to mSP11 and contains light stops.
Light stops  have also been discussed  recently in \cite{Gladyshev:2007ec,uc}.

While the earlier works which advocated benchmark points
and slopes made good progress in systematizing the search for supersymmetry,
we find that they do not cover the more broad set of possible
mass hierarchies we discuss here. That is, many of the mSP patterns do not appear in the earlier works that
advocated benchmark points for SUSY searches. For example the
Snowmass mSUGRA points (labeled SPS) \cite{Allanach:2002nj} and the
Post-WMAP benchmark points of \cite{Battaglia:2003ab}, make up only
a small fraction of the possible mass hierarchies listed in Table (\ref{msptable}).
The CMS benchmarks classified as  Low Mass (LM) and
High Mass (HM) \cite{Ball:2007zza} (for a recent review see \cite{JL,Spiropulu:2008ug})
does a good job covering the mSP1 pattern which appears as the most dominant
pattern in our analysis, but there are no Higgs patterns or stop patterns discussed
in the CMS benchmarks as well as in SPS or in Post-WMAP benchmarks.
We exhibit the mapping of mSPs with other benchmarks points in a tabular form in Table (\ref{pattern}).

In Fig.~(\ref{m-land}) we give the relative distribution of these
hierarchies found in our Monte Carlo scan. The most common patterns found  are CPs
and SUPs,  especially mSP1 and mSP5.
However there exists a significant region of the parameter space where
SOPs and HPs can be realized.
The  percentages of occurrence
of the various patterns in the mSUGRA landscape  for  both $\mu$ positive and
$\mu$ negative are exhibited in Fig.~(\ref{m-land}).
The analysis of
Fig.~(\ref{m-land}) shows that
the  chargino patterns (CP) are the most  dominant patterns, followed by
the stau patterns (SUP),  the stop patterns (SOP), and the Higgs patterns (HP).
In contrast,  most  emphasis in the literature,  specifically in the context of relic
density analysis,  has focused on the stau patterns, with much less attention
on other patterns. Specifically the Higgs patterns have hardly been investigated
or discussed.  The exceptions to this, in the context of the Higgs patterns, are the  more recent works of
Refs. \cite{Feldman:2007zn,Feldman:2007fq}, and  similar mass ranges for the Higgs bosons have
been studied in  \cite{WP}  (see also \cite{Ellis:2007ss}).

\subsection{The landscape of the 4 lightest  sparticles  in NUSUGRA  \label{A2} }
Next we discuss the landscape of the 4 lightest sparticles for the case of nonuniversal
supergravity models. Here we consider \non in the Higgs sector
(NUH),  in the third  generation sector (NU3), and  in the
gaugino sector (NUG). Such \non appear quite naturally in supergravity models
with a non-minimal K\"{a}hler potential, and  in string and D-Brane models.
The parametrization of the \non
is given by
\begin{equation}
\begin{array}{lcl}
{\rm NUH}&:&M_{H_u}=  m_0(1+\delta_{H_u}),~~ M_{H_d} = m_0(1+\delta_{H_d}),\cr
{\rm NU3}&:&M_{q3} =m_0(1+\delta_{q3}), ~~M_{u3,d3}=m_0(1+\delta_{tbR}),\cr
{\rm NUG}&:&M_{1}=m_{1/2}, ~~~M_{2,3}=m_{1/2}(1+\delta_{M_{2,3}}).
\end{array}\label{nonuni}
\end{equation}
In the above $\delta_{H_u}$ and $\delta_{H_d}$ define the \non for
the up and down Higgs  mass parameters, $M_{q3}$ is the left-handed squark
mass for the 3rd generation, and $M_{u3}$ ($M_{d3}$) are  the
right-handed u-squark (d-squark) masses for the 3rd generation.
The \non  in the gaugino
sector are parameterized here by $\delta_{M_2}$ and $\delta_{M_3}$.
We have carried out a Monte Carlo scan with flat priors using $10^6$ model
points in each of the three types of NUSUGRA models,  taking the same input parameter
ranges
as specified in Eq.(\ref{softranges})  and $-0.9\leqslant\delta\leqslant1$.
Almost all of the mSP patterns seen for the mSUGRA cases were found in
supergravity models with nonuniversal soft breaking,  as the mSUGRA  model is contained
within the nonuniversal supergravity models.  In addition
we find many new patterns labeled NUSPs (nonuniversal SUGRA pattern), and they are
exhibited in  Table (\ref{nusptable}).
%%%% --------------- NUSPs -------------------- %%%%
\begin{table}[htbp]
    \begin{center}
\begin{tabular}{||l||l||c|c||}
\hline\hline NUSP &   Mass Pattern  &     NU3   &   NUG\\\hline\hline
NUSP1   &   $\na$   $<$ $\cha$  $<$ $\nb$   $<$ $\ta$       & Y &   Y \cr
NUSP2   &   $\na$   $<$ $\cha$  $<$ $A\sim H$               &Y     &    \cr
NUSP3   &   $\na$   $<$ $\cha$  $<$ $\sta$  $<$ $\nb$       &    & Y \cr
NUSP4   &   $\na$   $<$ $\cha$  $<$ $\sta$  $<$ $\slr$          & &   Y \cr
\hline
NUSP5   &   $\na$   $<$ $\sta$  $<$ $\snl$  $<$ $\stb$          & Y  &    \cr
NUSP6  &   $\na$   $<$ $\sta$ $<$ $\snl$  $<$ $\cha$        & Y  &\cr
NUSP7   &   $\na$ $<$ $\sta$  $<$ $\ta$   $<$ $A/H$         & & Y \cr
NUSP8   &   $\na$   $<$ $\sta$  $<$ $\slr$  $<$ $\snm$      & &   Y \cr
NUSP9   &   $\na$   $<$ $\sta$  $<$ $\cha$  $<$ $\slr$          & &   Y \cr
\hline
NUSP10  &   $\na$   $<$ $\ta$   $<$ $\g$    $<$ $\cha$          & &   Y \cr
NUSP11  &   $\na$   $<$ $\ta$   $<$ $A\sim H$                       & & Y \cr
\hline
NUSP12  &   $\na$   $<$ $A\sim H$   $<$ $\g$                    && Y \cr
\hline
NUSP13  &   $\na$   $<$ $\g$    $<$ $\cha$ $<$ $\nb$        & &   Y \cr
NUSP14  &   $\na$   $<$ $\g$    $<$ $\ta$ $<$ $\cha$            & &   Y \cr
NUSP15  &   $\na$   $<$ $\g$    $<$ $A\sim H$               &&   Y \cr
  \hline\hline
\end{tabular}
%}
\caption{ New  4 sparticle  mass patterns  that arise in NUSUGRA
over  and above the mSP patterns of  Table (1).
 These  are labeled nonuniversal SUGRA patterns (NUSP) and  at least
 15 new patterns are seen to emerge which are denoted by NUSP1-NUSP15.}
 \label{nusptable}
\end{center}
 \end{table}
 As in the mSUGRA case one finds several pattern classes,
CPs, SUPS, SOPs, and HPs as exhibited below. In addition,
 we  find several  Gluino patterns (GP) where  the gluino is the NLSP.
\begin{enumerate}
\item Chargino patterns (CP) : NUSP1, NUSP2, NUSP3, NUSP4
\item Stau patterns (SUP) : NUSP5, NUSP6, NUSP7, NUSP8, NUSP9
\item Stop patterns (SOP) : NUSP10, NUSP11
\item Higgs patterns (HP) : NUSP12
\item Gluino patterns (GP) : NUSP13, NUSP14, NUSP15.
\end{enumerate}
It is  interesting to note that for the  4 sparticle landscape we find saturation in the number of
mass hierarchies that are present.   For example, for the case $\mu>0$ in
mSUGRA , increasing the soft parameter scan from $1\times 10^6$ parameter
model points to $2\times 10^6$ model points does not increase the
number of 4 sparticle patterns. In this context it becomes relevant to
examine as to what degree the relic density and other experimental
constraints play a role in constraining the parameter space and  thus
reducing the number of patterns. This is exhibited in
Table (\ref{tab:spnum}) where we demonstrate how the relic density
and the other experimental constraints  decrease the number of
admissible model points in the allowed parameter space  for the mSUGRA
models with both $\mu>0$ and $\mu<0$, and also for the cases with \non
 in the Higgs sector, \non in the third generation sector,
and with \non in the gaugino sector. In each case we start with
$10^6$ model points at the GUT scale, and find that the electroweak symmetry
breaking constraints reduce the number of viable models to about
$1/4$ of what we started with. We find that the allowed number of models
translates into SUGRA mass patterns which are typically less than 100.  The
admissible set of parameter points reduces drastically when the
relic density constraints are imposed and are then found to typically reduce
the number of models
by a factor of about 200 or more, with a reduction in the number of
allowed patterns by a factor of 2 or more. Inclusion of all other
experimental constraints further reduces the  number of admissible points by
a factor between 30\% and 50\%,
with a corresponding
reduction in the number of patterns by up to 40\%.
The above
analysis shows that there is an enormous reduction in the number of
admissible models and the corresponding number of  hierarchical mass patterns  after the
constraints of radiative breaking of the electroweak symmetry, relic
density constraints, and other experimental constraints are imposed.
\begin{table}[htbp]
    \begin{center}
 \scriptsize{
\begin{tabular}{|c|c|c|c|c|c|c|c|}
\hline
Model  &   Trial      &   Output  &   No. of    &   Relic Density          &  No. of  &   All          & No. of  \\
Type    &   Models  &   Models   &  Patterns  &   Constraints    & Patterns & Constraints & Patterns\\
\hline mSUGRA($\mu>0$) &   10$^6$   & 265,875 &   55  &   1,360   &   22  & 902 & 16  \\
\hline mSUGRA($\mu<0$) & 10$^6$   &   226,991 &   63  & 1,000   & 31  &   487 & 18  \\
\hline NUH($\mu>0$)   &   10$^6$   & 222,023 &   59 &   1,024   & 24  &   724 &   15  \\
\hline NU3($\mu>0$) & 10$^6$  &   229,928 &   73  &   970 &   28  &   650 &   20  \\
\hline NUG($\mu>0$) &   10$^6$   &   273,846 &   103 &   1,788   & 36 & 1,294   &   28  \\  \hline
\end{tabular}
}
\caption[]{
An analysis of  mass patterns  for the four lightest sparticles. Exhibited  in the table
are  the model type, the number of trial input points
for each model, the number surviving the radiative  electroweak
symmetry breaking scheme as given by SuSpect (column 3), the
number surviving when the relic density  constraints  are applied with MicrOMEGAs
(column 5),  the number surviving with inclusion of all experimental  collider
constraints (column 7), along with the corresponding number of  hierarchical mass
patterns in each case  (column 8).} \label{tab:spnum}
    \end{center}
 \end{table}

 %%%%% ----
\subsection{Hierarchical patterns for the full sparticle spectrum \label{A4}}

\begin{figure*}[htbp]
\centering
\hspace*{-.1in}\includegraphics[width=6.5cm,height=5cm]{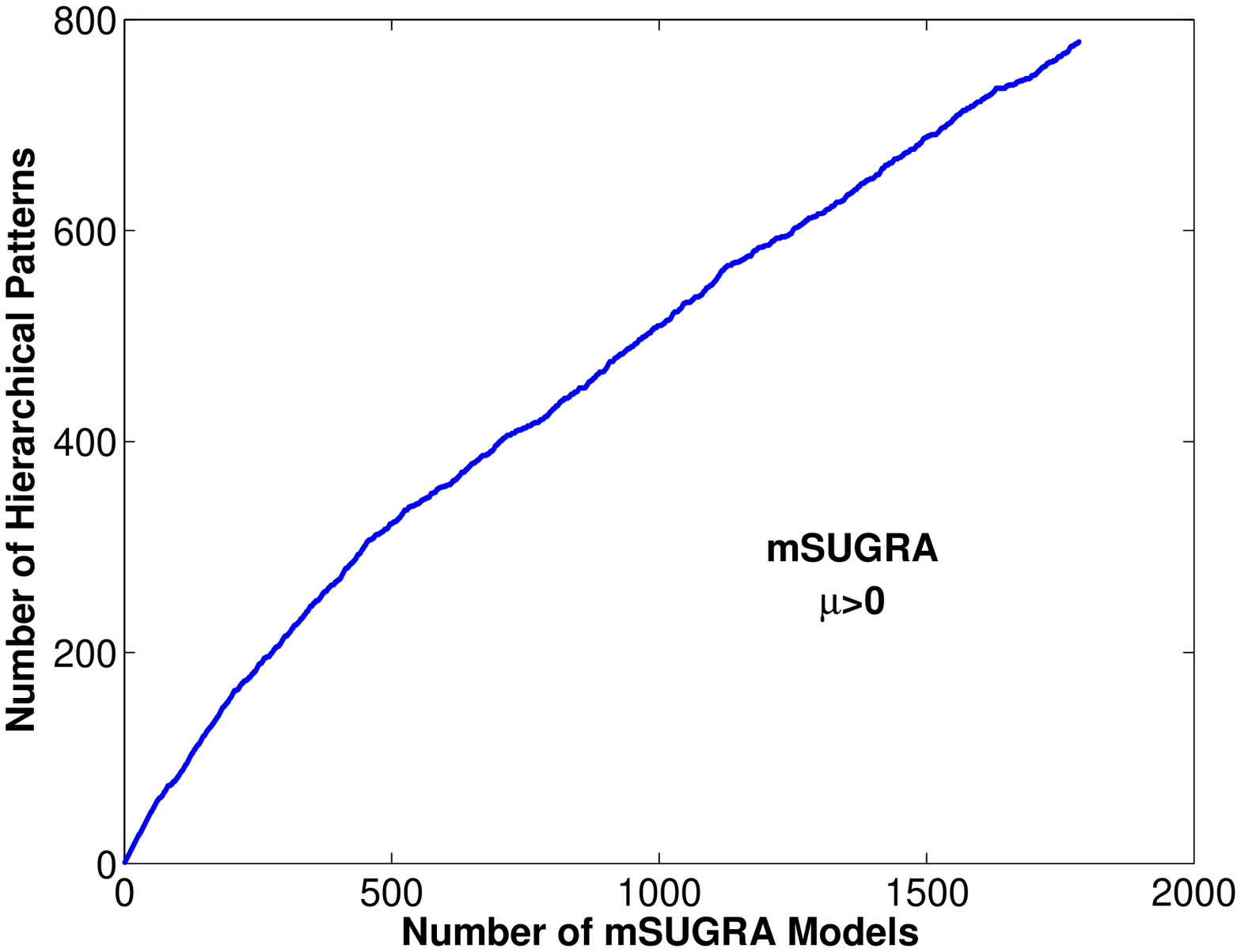}
\hspace*{.2in}\includegraphics[width=6.5cm,height=5cm]{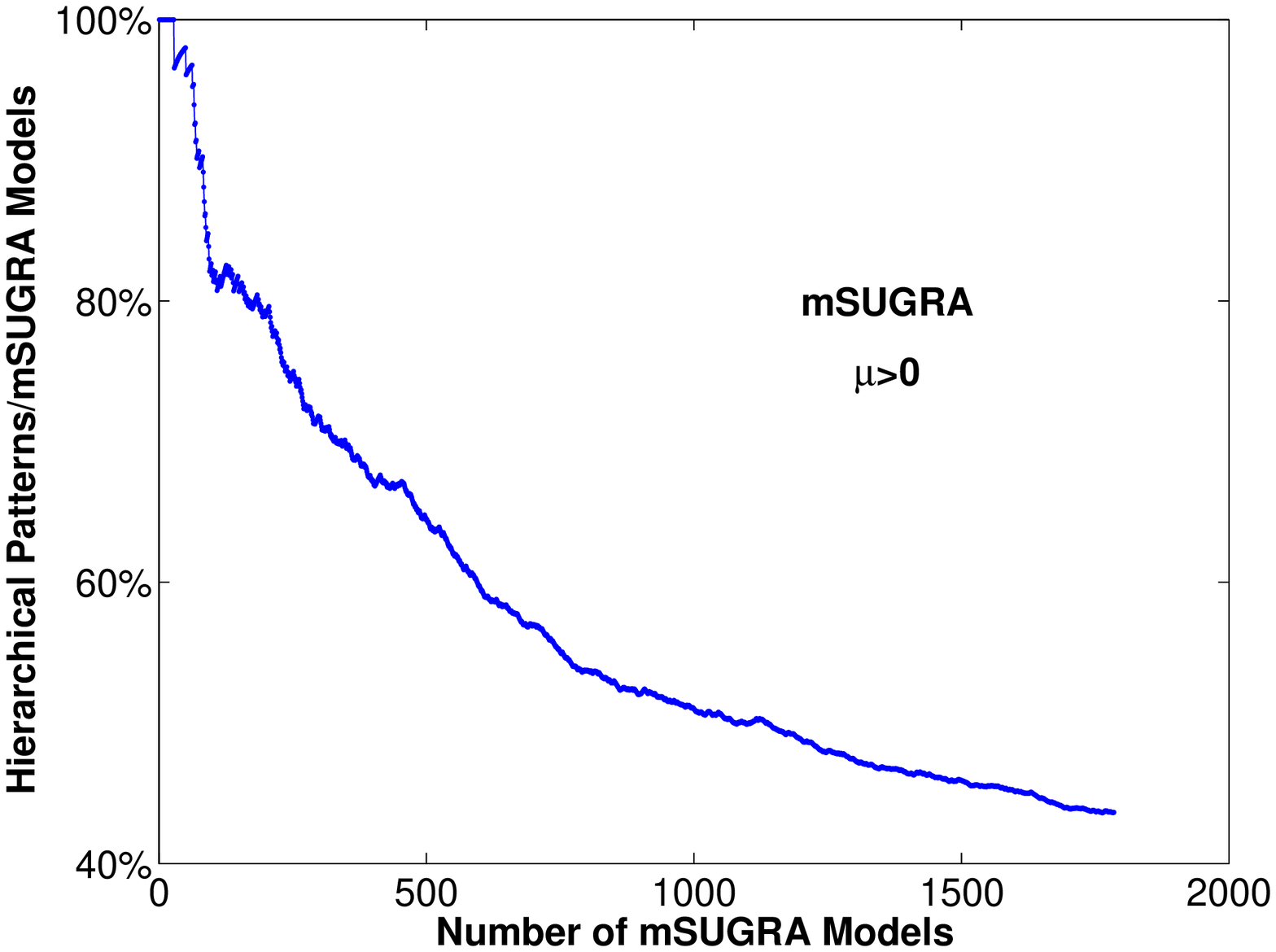}
\caption{ Left panel: The number of  hierarchical mass
patterns for 32 sparticles vs the number of trial  points for mSUGRA
models which survive the electroweak symmetry breaking constraints,
the relic density and all other experimental constraints. The number
of  hierarchical mass patterns show a trend towards saturation.
Right panel :  A similar phenomenon is seen in the ratio between the
number of patterns over the number of surviving trial points  in
mSUGRA models.}
 \label{fig:saturation}
\end{figure*}

We discuss now the number of hierarchical mass patterns  for the full  set of
 32 sparticles
in  SUGRA models when the constraints of electroweak symmetry, relic
density, and other experimental constraints are imposed.  The result of
the analysis is given in Fig.(\ref{fig:saturation}) and Table (\ref {tab:wholespectrum}).
%%%%%%%%%%%%%%%%%%%%%%%%%
\begin{table}[htbp]
    \begin{center}
    \scriptsize{
\begin{tabular}
{|r|c|c|}
\hline
Models [No.]    & No. after constraints  &  No. of patterns\\
\hline
mSUGRA($\mu>0$) [$10^6$]    &   902 &   505    \\\hline
mSUGRA($\mu<0$) [$10^6$]   &   487 &   268    \\\hline
NUH($\mu>0$) [$10^6$]  &   724 &   517   \\ \hline
NU3($\mu>0$) [$10^6$]    &   650 &   528     \\\hline
NUG($\mu>0$) [$10^6$]    &   1294    &   1092      \\\hline
All Above[$5 \times 10^6$]    &   4057    &   2557      \\
\hline
\end{tabular}
} \caption[]{The table exhibits a dramatic reduction of the
landscape from upward of $\sim O(10^{28})$
hierarchical mass patterns for
the 32 sparticle masses  to a much smaller number when the
electroweak symmetry breaking constraints, the relic density
constraints, and other experimental constraints are applied.
Column 1 shows  one million input parameter points for each
of the models investigated,  and the
number surviving all the constraints are exhibited in column 2,
while column 3 gives the number of  hierarchical patterns.}
\label{tab:wholespectrum}
    \end{center}
 \end{table}
Here one finds that increasing the number of model points in the scan
does increase the number of patterns. However, the ratio of the
number of  patterns to the total number of models that survive all
the constraints  from  the scan decreases sharply as shown in the right
panel of Fig.(\ref{fig:saturation}). This means  that although
saturation is not yet  achieved one is moving fast  towards
achieving saturation with a relatively small number of allowed
patterns for all the 32  sparticles within SUGRA models consistent
with the various experimental constraints.
The analysis of Table (\ref{tab:wholespectrum}) shows
that the number of allowed patterns for the 32 sparticles,
which in the MSSM without the SUGRA framework  can be as large as $O(10^{28})$
or larger, reduces rather drastically when
various constraints are applied in supergravity models.
We note that some patterns are repeated as we move across different model types listed in the
first column  of Table (\ref{tab:wholespectrum}). Thus the total number of patterns listed at the
bottom of the last column of this table is smaller than the sum of patterns listed above in that
column.  We note that the precise number and nature of the patterns are dependent on the
input parameters such as the top mass and a significant shift in the input values could modify the
pattern structure.

%%%%%%%%%%%%%%%%%%%%%%%%%%%%%%%
% ---------------------------- Monte Carlo mSUGRA scans  ----------------------------
\begin{figure*}[htb]
\centering
\hspace*{-.1in}\includegraphics[width=7.0cm,height=6.0cm]{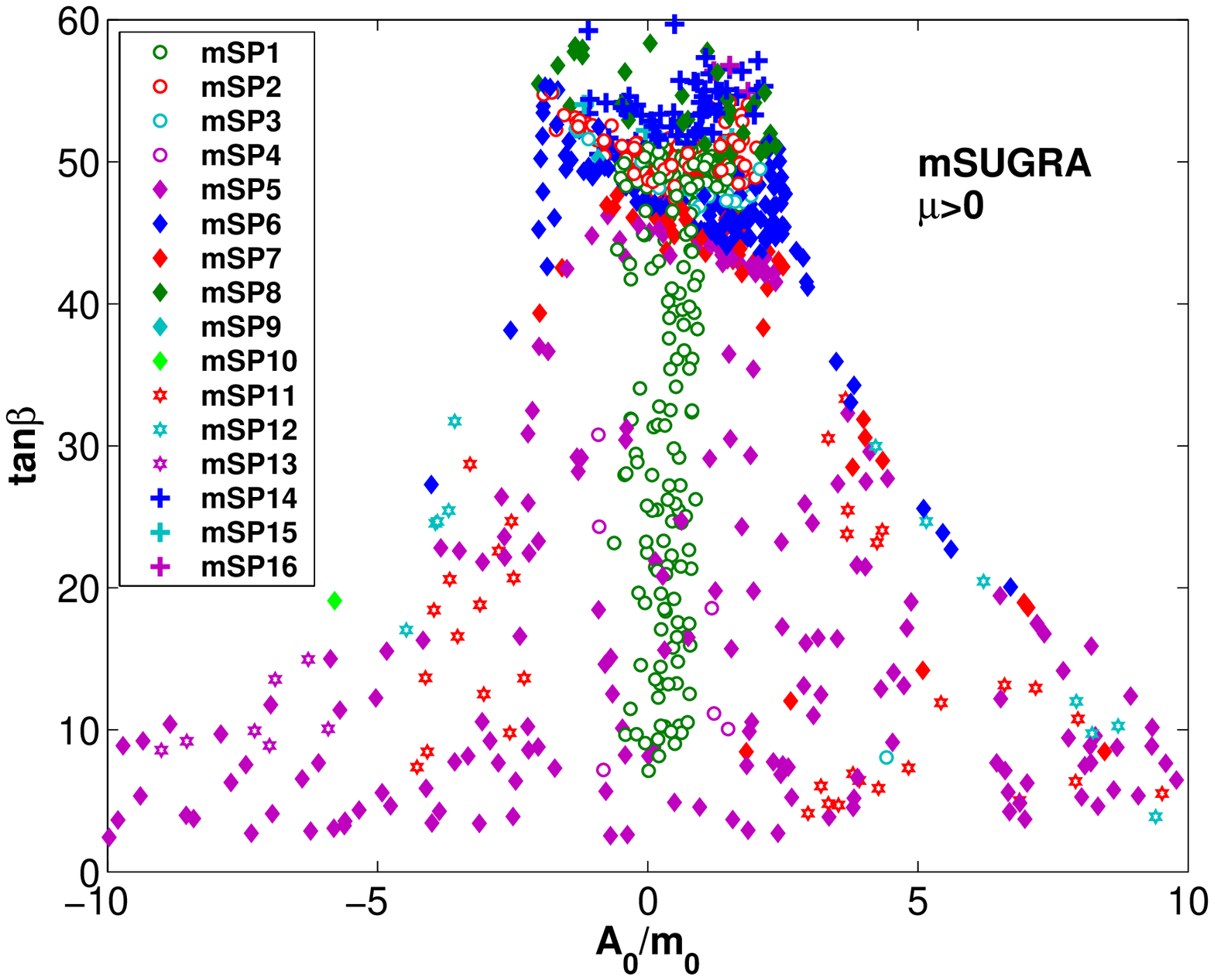}
\hspace*{.2in}\includegraphics[width=7.0cm,height=6.0cm]{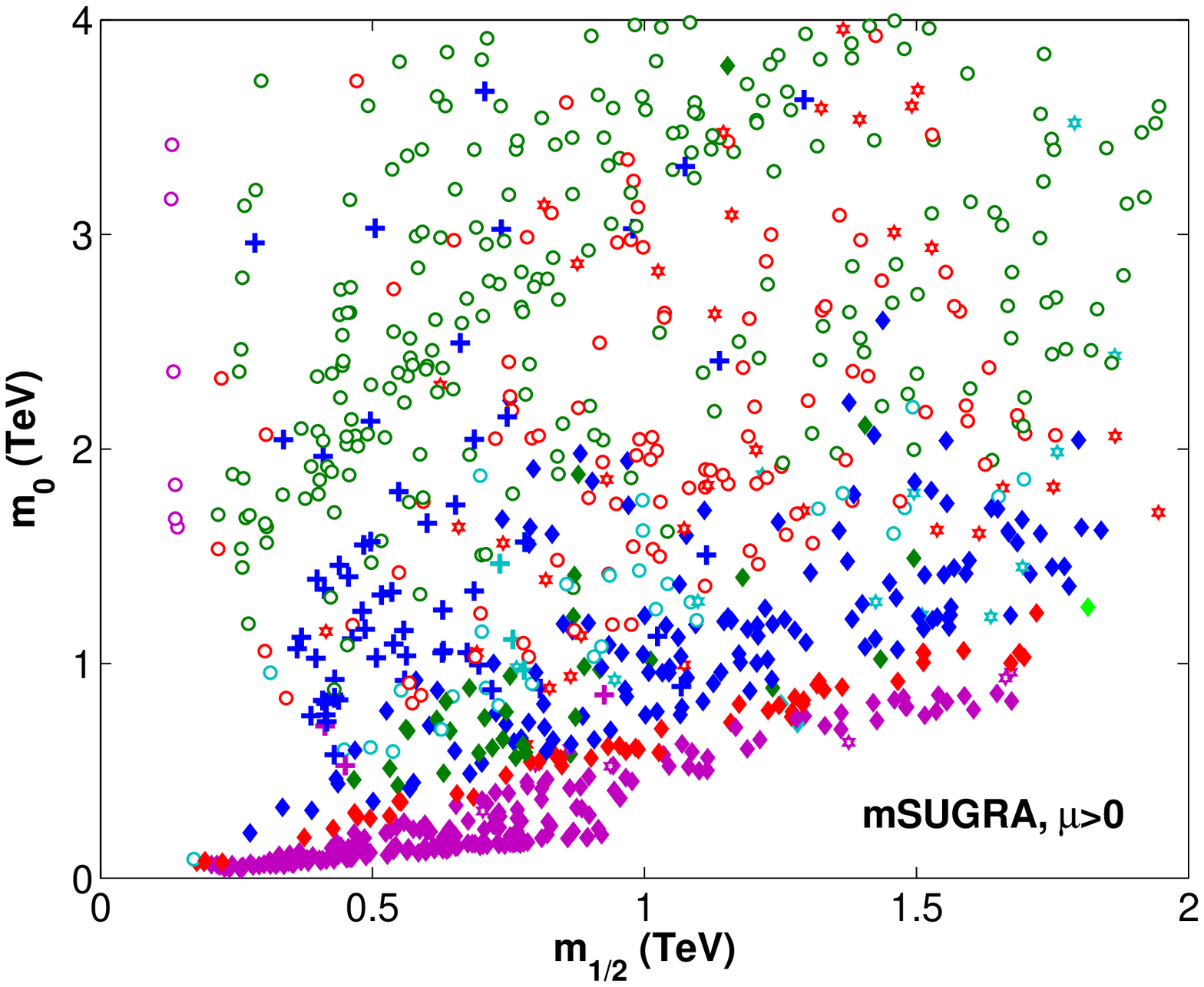}
\hspace*{-.1in}\includegraphics[width=7.0cm,height=6.0cm]{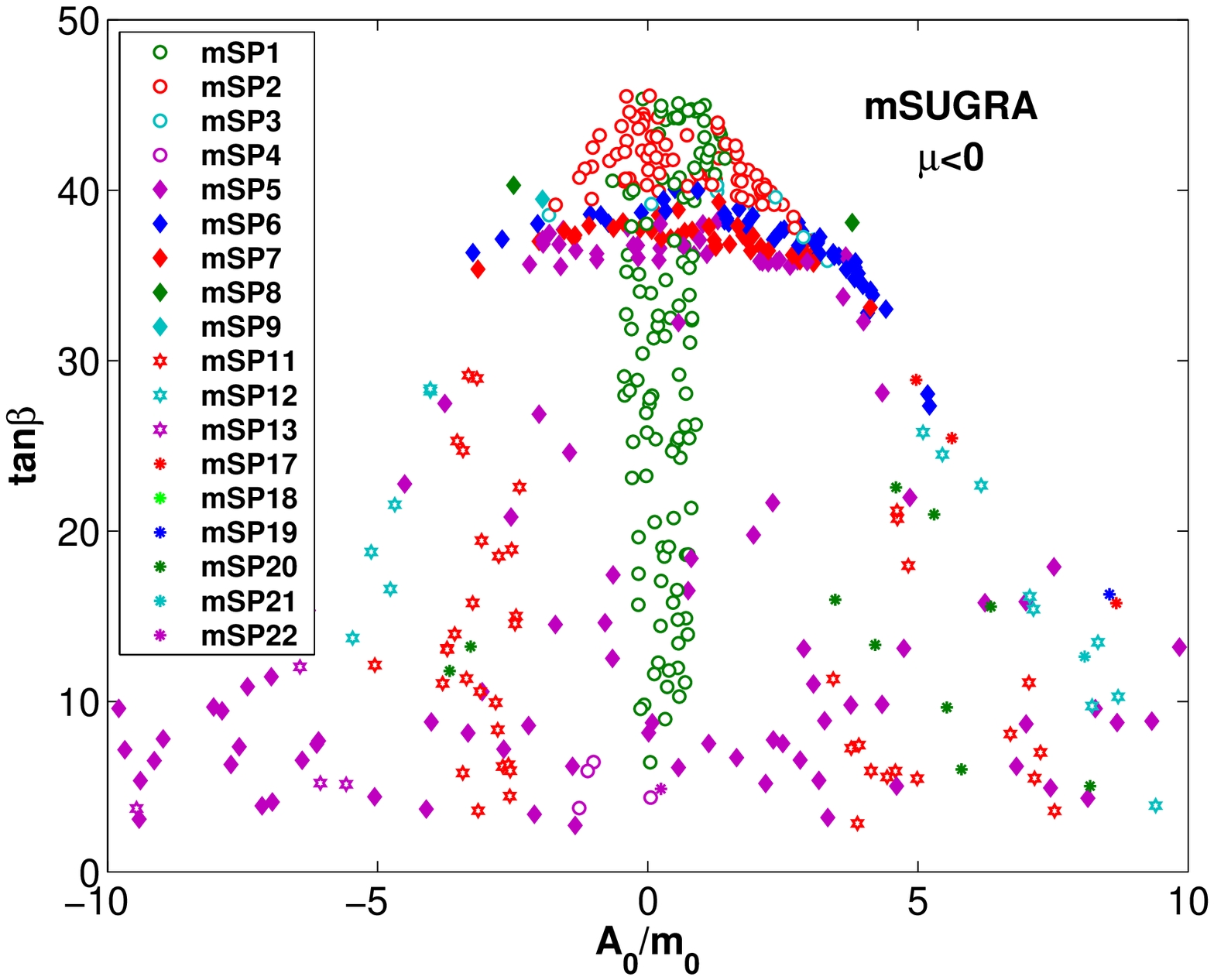}
\hspace*{.2in}\includegraphics[width=7.0cm,height=6.0cm]{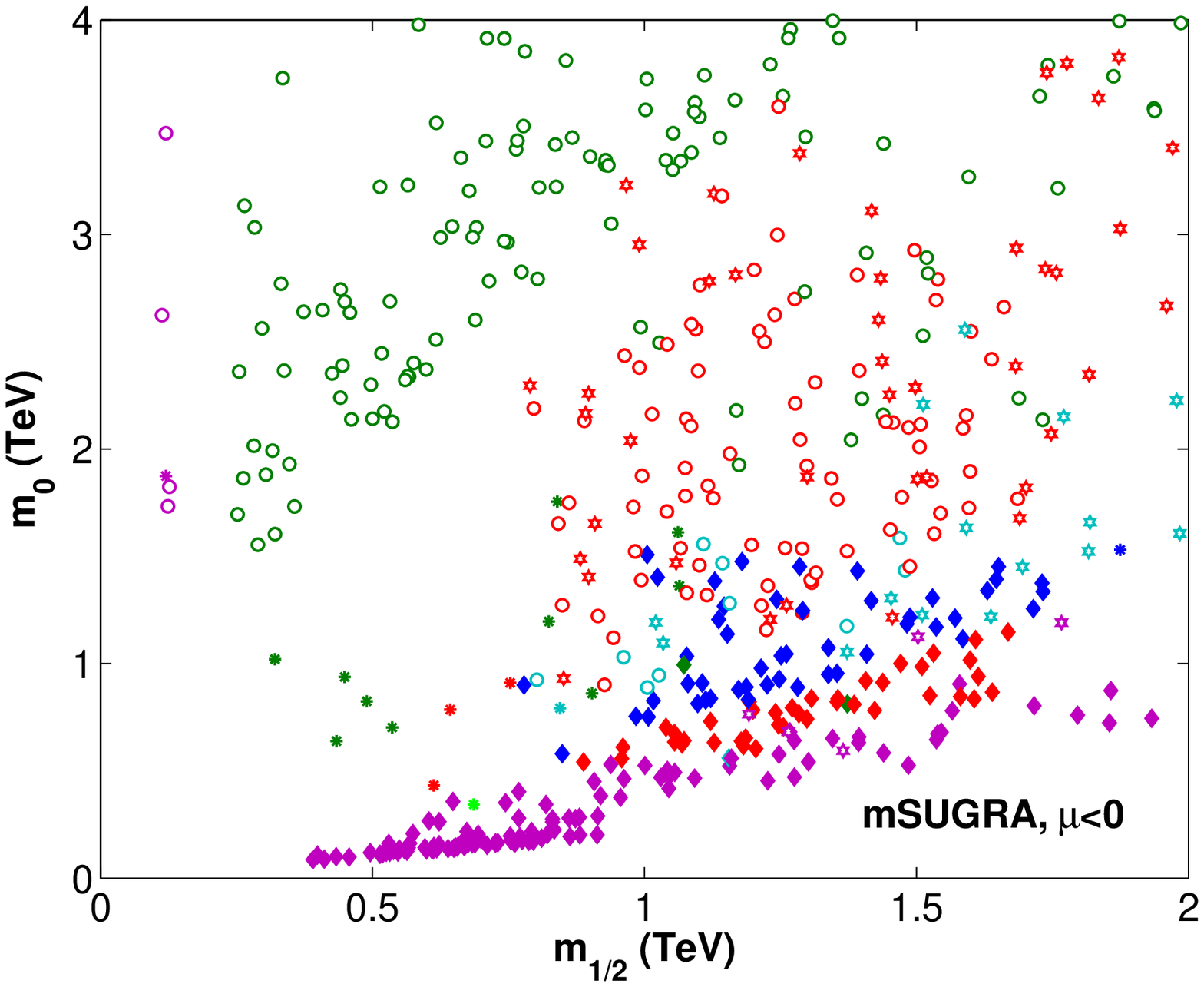}
\caption[]{The dispersion of mSPs arising in mSUGRA
in the $\tan\beta$ vs $A_0/m_0$ plane (left panels), and in the
$m_0$ vs $m_{{1}/{2}}$ plane (right panels) for the $\mu>0$ case
(upper panels) and $\mu<0$ case (lower panels). The analysis is
based on a  scan of $10^6$ trial model points with flat priors in
the ranges $m_0<4{\rm~TeV}$, $m_{1/2}<2 {\rm~TeV}$, $1<\tan\beta<60$, and
$|A_0/m_0|<10$.
 mSP1 is confined to the region where $|A_0/m_0|<2$.
For the case $\mu<0$, no HPs are seen, and also, no model points
survive in the region where $\tan\beta > 50$ in contrast to the
$\mu>0$ case where there is a significant number for $\tan\beta
\gtrsim 45$.} \label{fig:spec}
\end{figure*}

% ---------------------------- special mSUGRA scans  ----------------------------

\begin{figure*}[htb]
\centering
\hspace*{-.1in}\includegraphics[width=7.0cm,height=6.0cm]{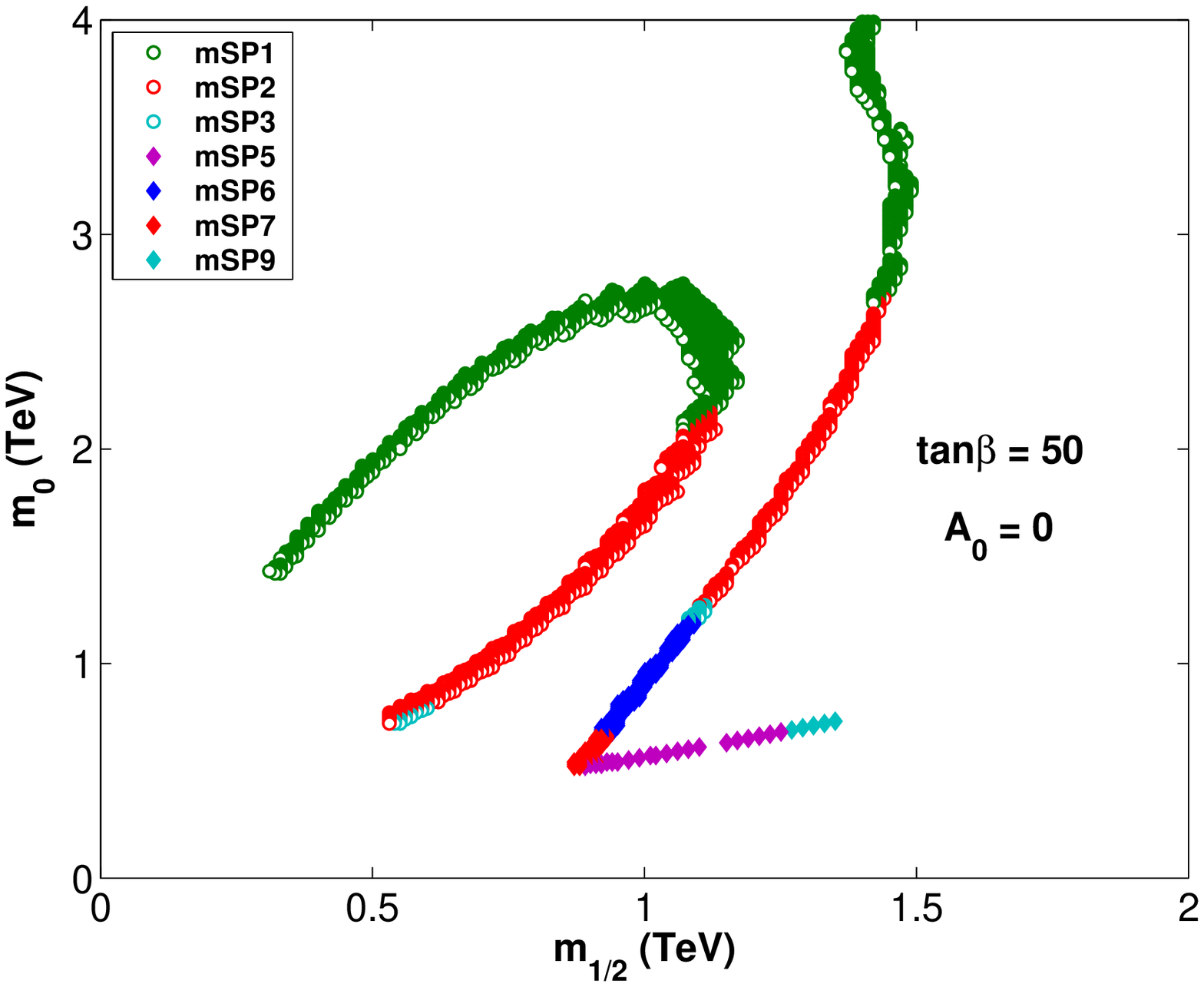}
\hspace*{.2in}\includegraphics[width=7.0cm,height=6.0cm]{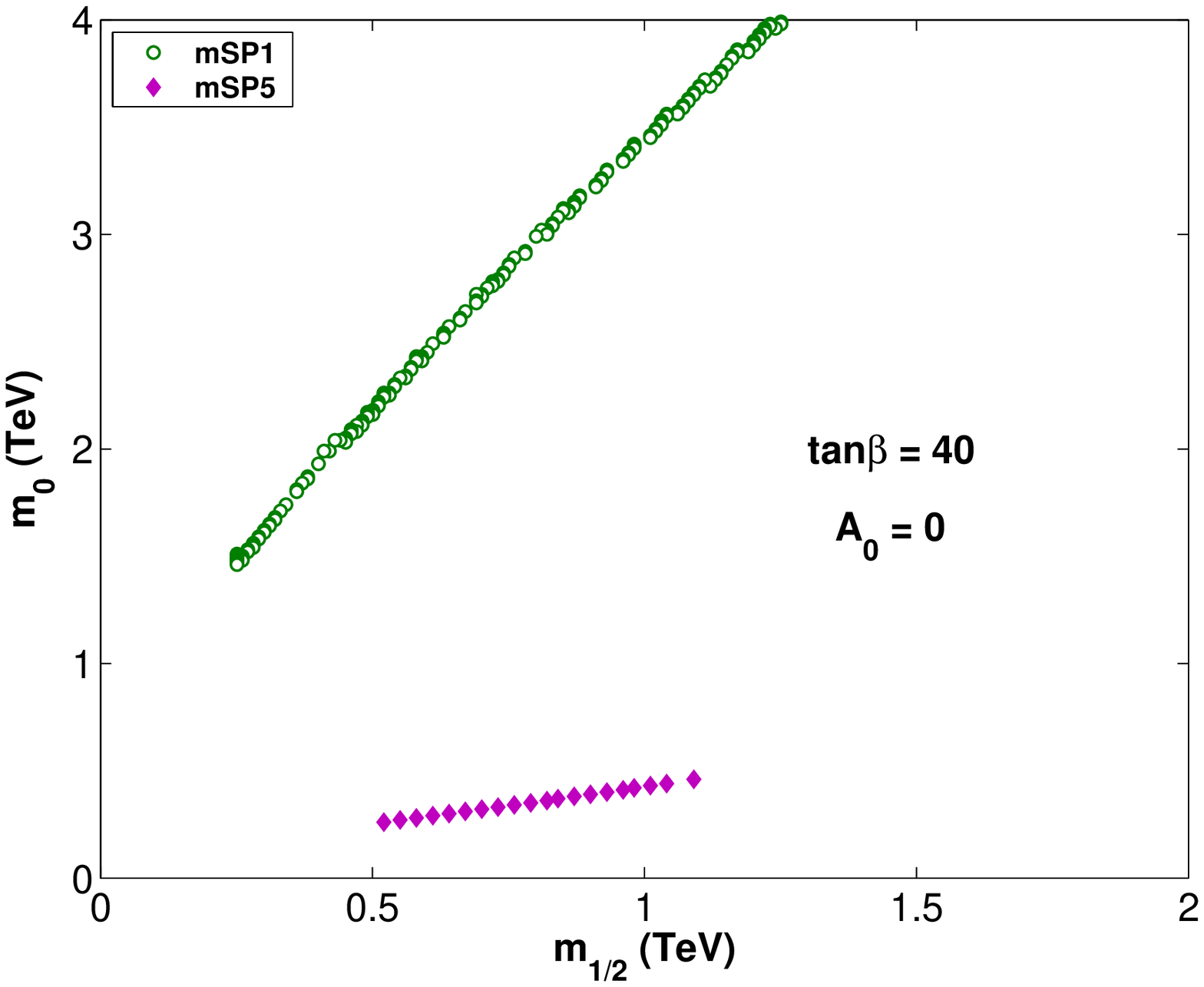}
\hspace*{-.1in}\includegraphics[width=7.0cm,height=6.0cm]{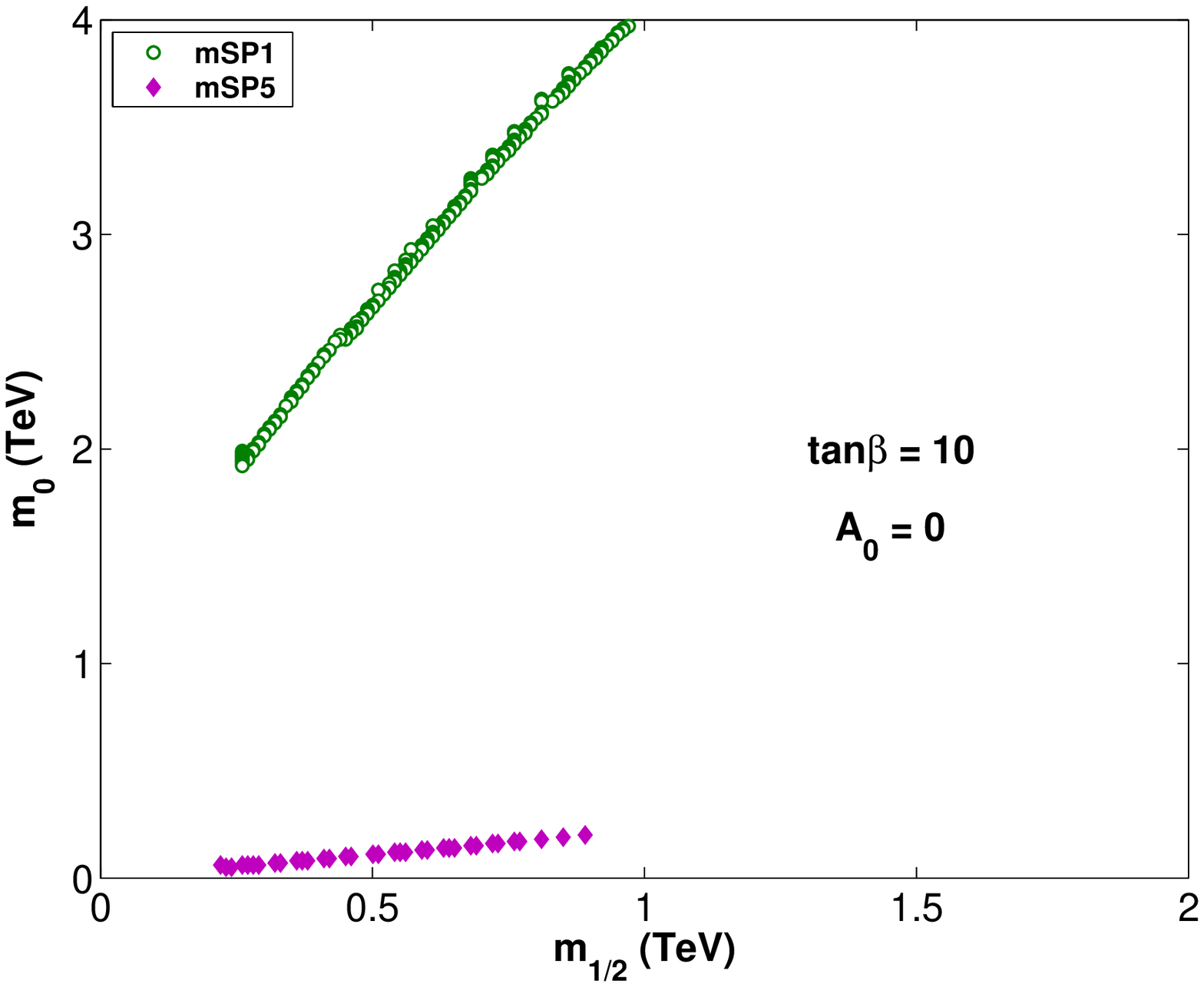}
\hspace*{.2in}\includegraphics[width=7.0cm,height=6.0cm]{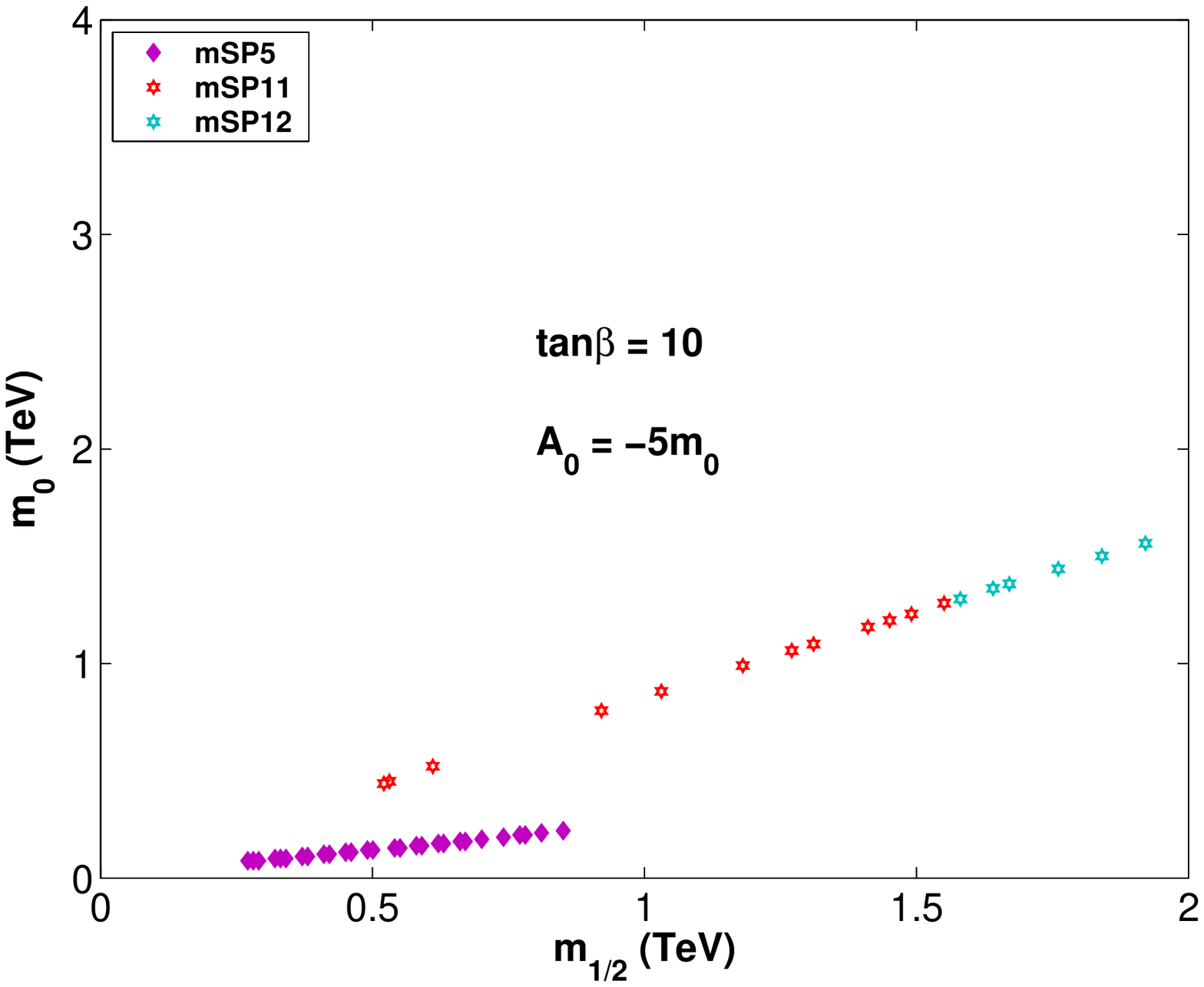}
\caption[]{
Dispersion of patterns in the $m_0$ vs $m_{1/2}$ plane for fixed values of  $\tan\beta$
and $A_0/m_0$. The region scanned is  in the range $m_0 < 4$~{\rm TeV} and
$m_{1/2} < 2$~{\rm TeV}  with a 10~{\rm GeV} increment for each mass.
Only a subset of the allowed parameter points relative to Fig.(\ref{fig:spec}) remain,
since the scans are on constrained surfaces in the mSUGRA parameter space.
}
\label{fig:grid}
\end{figure*}
\section{Sparticle Patterns and the Nature of Soft Breaking\label{B}}
\subsection{Correlating mass hierarchies with the soft parameter space}
It is interesting to ask if the patterns can be traced back to some specific regions
of the parameter of soft breaking from where they originate. This indeed is the case,
at least, for some of the patterns. The analysis illustrating the origin of the patterns
in the parameter space is given in  Fig.(\ref{fig:spec}).
Exhibited are the landscape of sparticle mass spectra in the planes  (I)
$\tan\beta$ vs $A_0/m_0$ and (II) $m_0$ vs $m_{1/2}$, when the soft
parameters  are allowed to vary in the  ranges given in
Eq.(\ref{softranges}). Many interesting observations can be made
from these spectral decompositions. For example, a significant set
of the mSP1 (CP) models lie in the region $|A_0/m_0|<2$ and correspond to
the Hyperbolic branch(HB)/Focus Point
(FP) \cite{hb} regions, while most of the
SOPs have a rather large ratio of $A_0/m_0$ with the satisfaction of REWSB.
In this analysis we require
that there be no charge or color breaking (CCB)\cite{Frere:1983ag,Casas:1995pd}
at the electroweak scale.
We note in passing that it has been argued that even if the true minimum is not
color or charge preserving, the early universe is likely to occupy the CCB preserving
minimum and such minima may still be acceptable if the tunneling lifetime from the false
to the true vacuum is much greater than the present age of the universe\cite{Kusenko:1996jn}.
Next, we note that for the mSUGRA $\mu>0$ case,
the region around
$\tan\beta= 50$  has a large number of models that can be realized, while
the region around $\tan\beta= 30$  has far less model points. We also note
that most of the HPs reside only in the very high $\tan\beta$
region in mSUGRA, but this situation can be changed significantly
in
the NUH case where HP points can be realized in the $\tan\beta$
region as low as $\tan\beta \sim 20$.
In the $m_0$ vs $m_{1/2}$ plane,
one finds  that most of CPs and HPs
have a larger universal scalar mass  than most of the
SUPs and  SOPs.

Often in the literature one limits the analysis by fixing specific values of
$A_0$ and $\tan\beta$.  For $A_0$ the value most investigated is
$A_0=0$.
However, constraining the values of $A_0$ or of $\tan\beta$
 artificially eliminates a very significant part of the allowed
 parameter space where all the relevant constraints (the REWSB constraint as well as
 the relic density and the experimental constraints) can be satisfied
 as seen in Fig.(\ref{fig:spec}).
One can extract the familiar plots one finds in the literature
where $A_0$ and $\tan\beta$ are constrained from a reduction of the
top-right  panel of  Fig.(\ref{fig:spec}). The results of this reduction
are  shown in  Fig.(\ref{fig:grid}) with a focused scan in specific regions
of the soft parameter space.
Specifically the bottom-left and  top-right panels of Fig.(\ref{fig:grid})
show the familiar stau coannihilation\cite{Ellis:1999mm,Gomez:2000sj,ArnowittTexas}
regions and the HB/FP branch,
 the bottom-right panel gives the stau coannihilation region and
the stop coannihilation region because of the relatively large $A_0$ value,
and the top-left panel is of the form seen in the works of Djouadi
et al.~\cite{Djouadi:2006be} where the Higgs funnel plays an important
role in the satisfaction of the relic density.

\begin{figure*}[htb]
\centering
\hspace*{-.1in}\includegraphics[width=7.0cm,height=6.0cm]{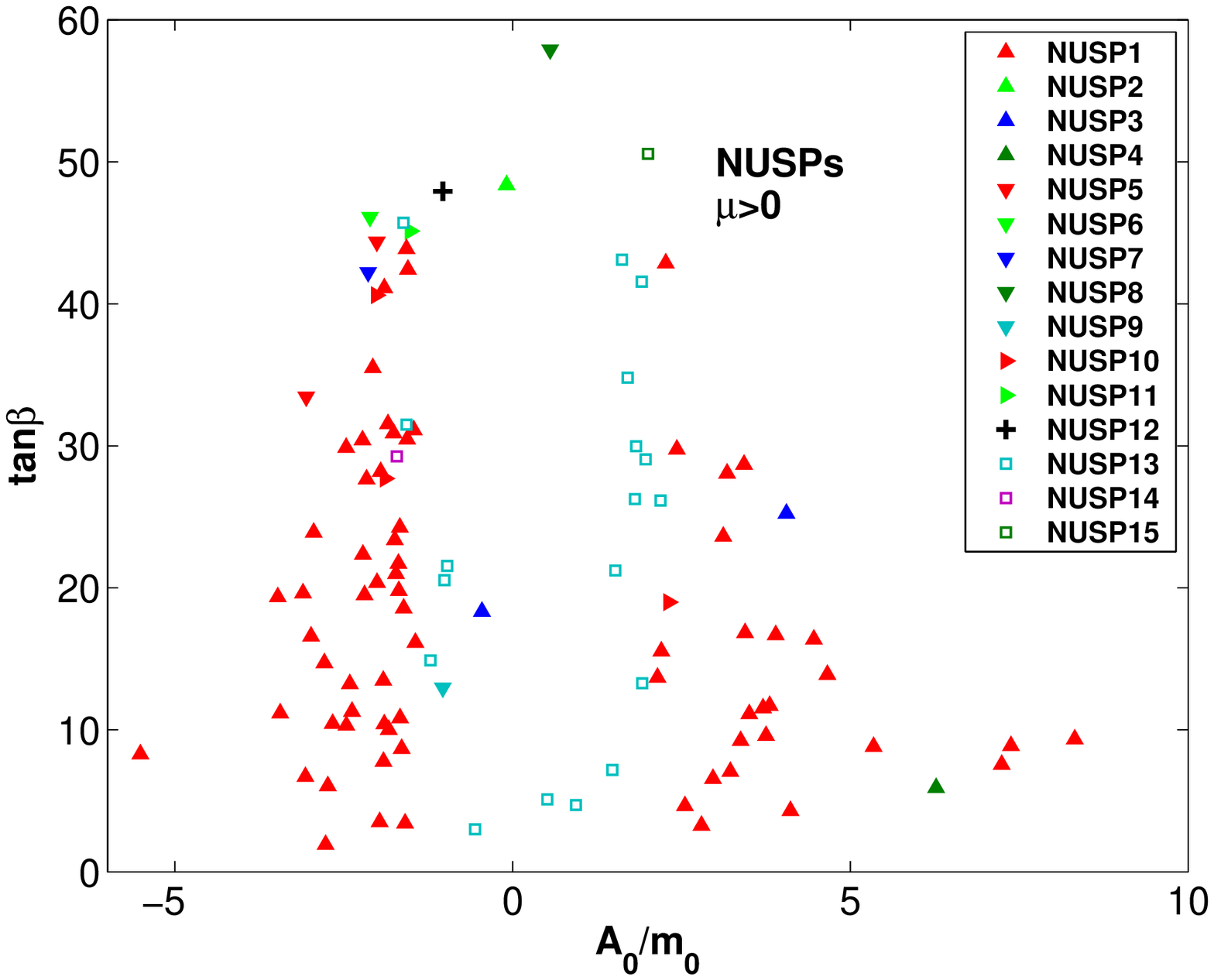}
\hspace*{.2in}\includegraphics[width=7.0cm,height=6.0cm]{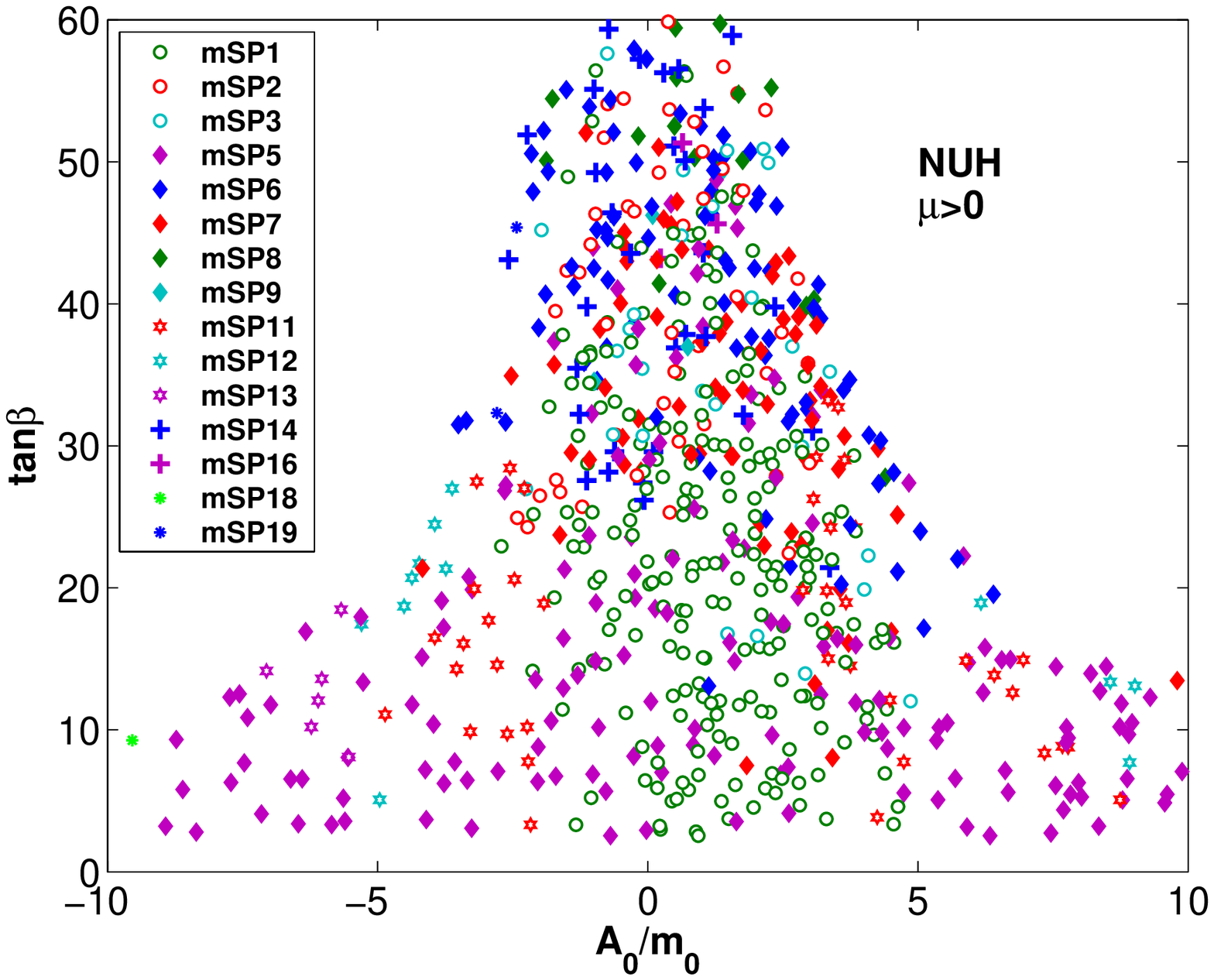}
\hspace*{-.1in}\includegraphics[width=7.0cm,height=6.0cm]{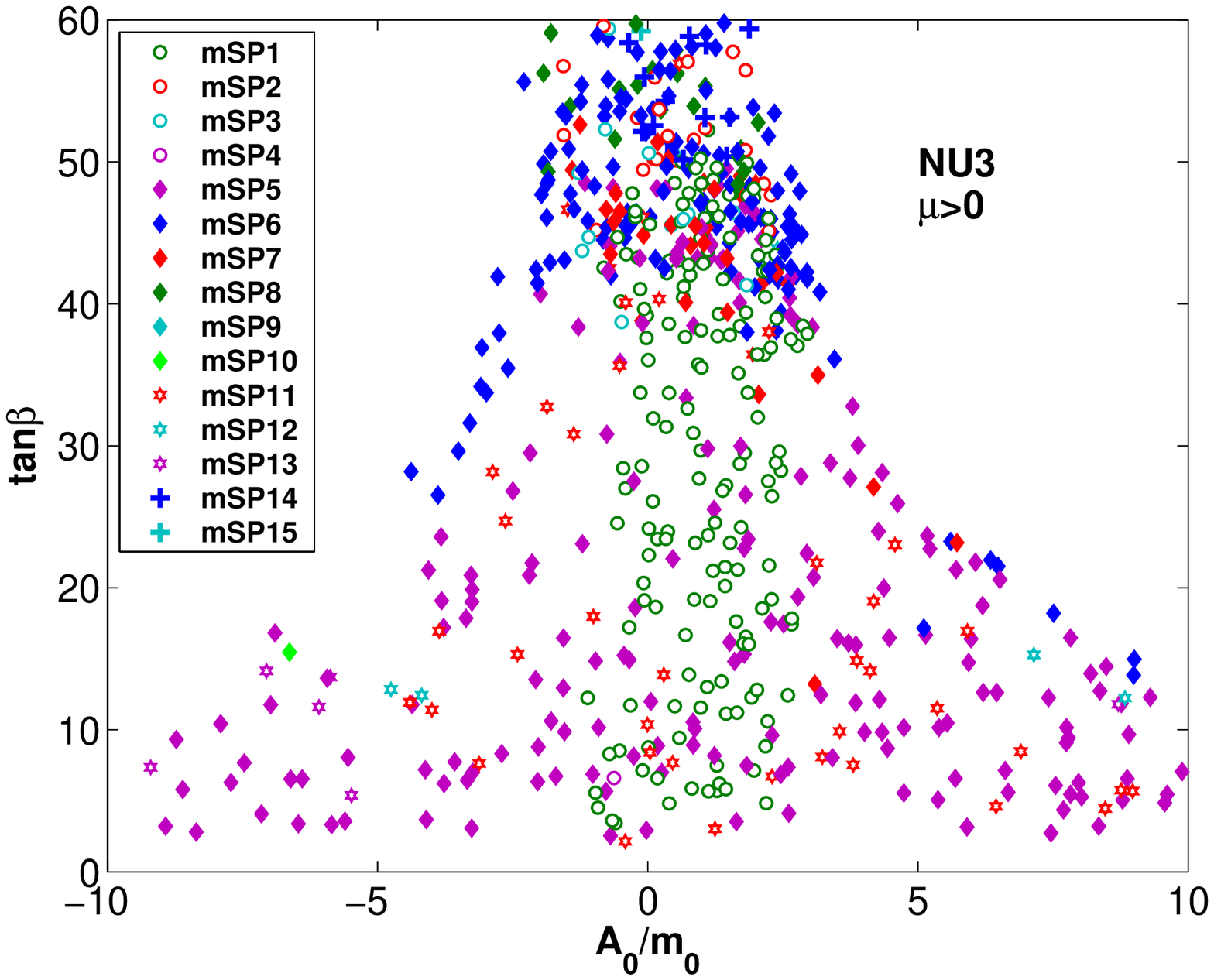}
\hspace*{.2in}\includegraphics[width=7.0cm,height=6.0cm]{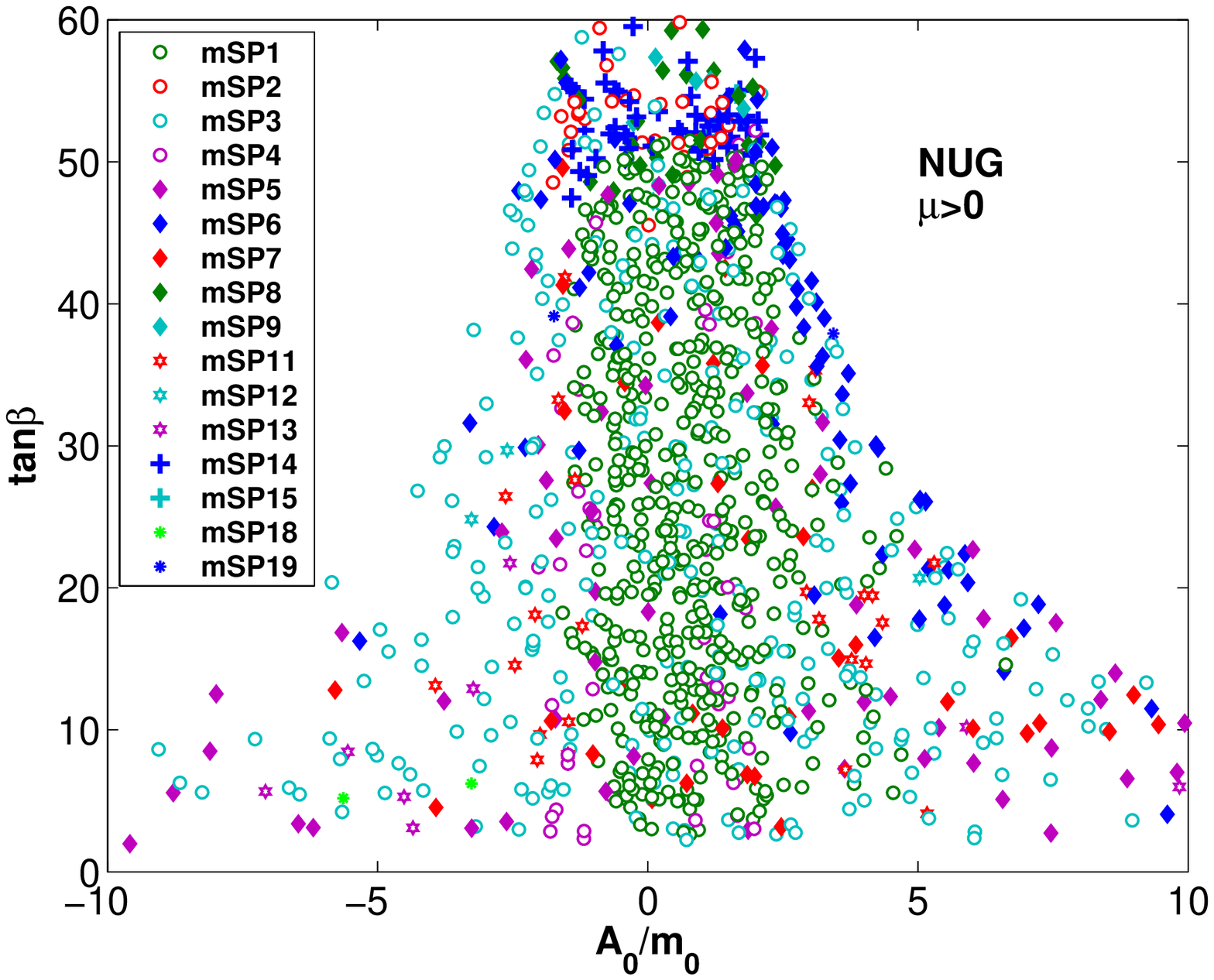}
\caption[]{
An exhibition of the
 NUSPs and mSPs for the NUH, NU3, and NUG models in the
$\tan\beta$ vs $A_0/m_0$ plane. The range of SUGRA parameters are the
same as the case mSUGRA ($\mu>0$). One may notice that the mSP1 points
arising from NU models lie in a relatively larger $A_0/m_0$ region.
Most of the models in NU cases are still mSPs, and among the NUSPs, only
two patterns have a relatively large population,  these being  NUSP1 and NUSP13.
One may also notice that in NUH case, the HPs can exist in a low
$\tan\beta$ region as opposed to the mSUGRA case where
HPs can either exist in the large $\tan\beta$ region ($\mu>0$)
or are totally eliminated  ($\mu<0$).}
 \label{fig:nuspec}
\end{figure*}
An analysis similar to that of Fig.(\ref{fig:spec}) for the
nonuniversal case is given in Fig.(\ref{fig:nuspec}). Here in
addition to the mSPs new patterns emerge which we label as
nonuniversal sugra patterns or NUSPs. Among the NUSPs  the  dominant
patterns are NUSP1 (CP) and NUSP13 (GP), which
 are seen to arise the model with \non in the gaugino sector, i.e., the NUG model.
In general, the NUG is  dominated by the  CP patterns
whereas the NUH case is rather diverse  offering
the possibility of Higgs patterns
at lower, less fine tuned values of $\tan \beta$.

\subsection{Benchmarks for sparticle patterns   \label{B3}  }
As discussed in Sec.(\ref{A1}), many of the sparticle mass patterns
discussed in this analysis do not appear in the Snowmass, Post-WMAP,
and CMS benchmark points. With some of these mSP and NUSP having a
significant
 probability of occurrence, we therefore provide a larger set of
benchmark points for the various patterns in different SUGRA
scenarios. These benchmark points are exhibited in
Tables (\ref{b1},\ref{b4},\ref{b2},\ref{b3},\ref{b5}) of  the
Appendix.   Each of these benchmarks satisfies the relic density and
other experimental constraints with SuSpect  linked to MicrOMEGAs. We
have explicitly checked that the first mSP benchmark point in each
of the tables can be reproduced by using  SPheno, and SOFTSUSY
by allowing minor variations on the input parameters.
The benchmarks are chosen to  cover wide parts  of the SUGRA
parameter space.
We give these benchmarks, several for each mass pattern,
as the search for SUSY from the point of view of mass patterns
has important consequences for LHC experimental searches.
 Some of the patterns are correlated  with certain
well investigated phenomena such as the HB/FP branches of REWSB
and the stau-neutralino co-annihilation regions.
However, many of
the patterns arise from multiple annihilation processes.

%% ---------------------------- LHC Signatures  ----------------------------

\section{LHC Signatures for Mass Patterns\label{D}}

\subsection{Event generation and detector simulation \label{Dsub11}}
Before moving to the discussion of the LHC signatures arising from
various mSPs and NUSPs, we first give a detailed description of our
LHC simulation procedure.

After the imposition of all the
constraints mentioned in the previous sections, such as the relic
density constraints from WMAP data, the constraints on the FCNCs, as
well as mass limits on the sparticle spectrum,
we are left with the candidate model points for the signature analysis.
For each of these model points, a SUSY
Les Houches Accord (SLHA) file \cite{SKANDS} is interfaced to
PYTHIA 6.4.11 \cite{PYTHIA} through  PGS4 \cite{PGS}
for the computation of SUSY production cross sections and branching fractions.
 In this analysis, for signals, we have generated all
of PYTHIA's $2 \to 2$ SUSY production modes using MSEL~$=$~39\footnote
 {More specifically this choice
  generates 91 SUSY production modes including gaugino, squark, slepton, and SUSY Higgs
pair production but leaves out singly produced Higgs production.
For further details, see \cite{PYTHIA}.
A treatment of singly produced Higgs production in the context
of sparticle mass hierarchies was included in the analysis of Ref. \cite{Feldman:2007fq}.}.
Leading
order  cross sections from PYTHIA and leading order cross sections from
PROSPINO 2.0 \cite{PROSPINO} were cross checked against one another for consistency
over several regions of the soft parameter space. TAUOLA \cite{TAUOLA} is called by
PGS4 for the calculation of tau branching fractions as controlled in
the PYTHIA parameter card (.pyt) file.
With PGS4 we use the Level 1 (L1) triggers based on  the Compact Muon Solenoid detector
(CMS) specifications \cite{CMS,Ball:2007zza} and the LHC detector card.
 Muon isolation is controlled by employing the cleaning script in PGS4.
 We take the
experimental nomenclature of lepton being defined only as electron
or muon and thus distinguish electrons and muons from tau leptons.
SM backgrounds have been generated with
QCD multi-jet production due to light quark flavors,
 heavy flavor jets ($b \bar b$,  $t \bar t$), Drell-Yan,
single $Z/W$ production in association with quarks and gluons ($Z$+ jets / $W$+ jets),
and  $ZZ$, $WZ$, $WW$ pair production resulting in multi-leptonic backgrounds.
Extraction of final state particles from the PGS4 event record is
accomplished with a code SMART (~=~SUSY Matrix Routine)
written by us \cite{Feldman:2007zn} which provides an optimized processing of PGS4
event data files.
The standard criteria for the discovery limit of new signals is that the SUSY signals
should exceed either $5\sqrt{N_{\rm SM}}$ or 10 whichever is larger,  i.e.,
${\rm N_{SUSY}}>{\rm Max}\left\{5\sqrt{N_{\rm SM}},10\right\}$ and such a criteria is imposed where relevant.
We have also  cross checked various results of our analysis with three CMS notes \cite{CMSnote1,CMSnote2,CMSnote3}
and we have found agreement with these works using SMART and PGS4 for signal and backgrounds.

We note that several works where sparticle signatures are discussed have appeared
recently \cite{chameleon,Kane:2006yi,Conlon:2007xv,Baer:2007ya,Mercadante:2007zz,Kitano:2006gv}.
However, the issue of  hierarchical mass patterns and the correlation of signatures
with such patterns has not been discussed which is what the
analysis of this work investigates.

\subsection{Post trigger level cuts and LHC signatures \label{D1}}

Generally speaking, there are two kinds of LHC signatures: (i) event  counting
signatures,  and (ii) kinematical signatures. We have investigated both of
these for the purpose of discriminating the sparticle mass patterns.
We list our  event counting signatures in Table (\ref{tab:counting}), where we have
carried out analyses of a large set of lepton + jet signals.
In our counting procedure, only electron and muon are
counted as leptons, while tau jets are counted independently.
For clarity, from here on, our use of `jet(s)' will exclude tau jets. Thus, for jet
identification, we divide jets into two categories: b-tagged jets
and jets without b-tagging, which we simply label as b-jets and
non-b-jets (see also \cite{chameleon}).
There are some counting signatures that only concern one
class of measurable events, for example,
the number of events containing one tagged b-jet and any other final state
particles.
There are also types of signatures of final state particles with
combinations of two or three different species.
For instance, one such example would be the number of events in which there is
a single lepton and a single tau.

When performing the analysis of event counting, for  each  SUGRA model point,
we impose  global post
trigger cuts to analyze most of our PGS4 data.   Below we give
our default post trigger cuts which are used
throughout the paper unless stated otherwise.
\begin{enumerate}
\item In an event, we only select photons, electrons, and muons
that have transverse momentum $P^{p}_T>10$ GeV and $|\eta^{p}|<2.4$, $p=(\gamma,e,\mu)$.
\item Taus which satisfy $P^{\tau}_T>10$ GeV and $|\eta^{\tau}|<2.0$ are selected.
\item For hadronic jets, only those satisfying $P^{j}_T>60$ GeV and $|\eta^{j}|<3$ are selected.
\item We require a large amount of missing transverse momentum, $P^{miss}_T>200$ GeV.
\item There are at least two jets that satisfy the $P_T$ and $\eta$ cuts.
\end{enumerate}
Our default post trigger level cuts are standard
and are designed to  suppress the
Standard Model background, and highlight the SUSY events over a broad class of models.

 % ---------------------------- Signature TABLE  ---------------------------
\begin{table*}[htb]
    \begin{center}
\begin{tabular}{|c|l||c|l|}
\hline Signature   &   Description &   Signature   &
Description \\  \hline 0L  &   0 Lepton    &   0T  &
0 $\tau$    \\  \hline 1L  &   1 Lepton    &   1T  &   1 $\tau$
\\  \hline 2L  &   2 Leptons    &   2T  &   2 $\tau$    \\  \hline 3L
&   3 Leptons    &   3T  &   3 $\tau$    \\  \hline 4L  &   4 Leptons
and more   &   4T  &   4 $\tau$ and more   \\  \hline 0L1b    &   0
Lepton + 1 b-jet  &   0T1b    &   0 $\tau$ + 1 b-jet  \\  \hline
1L1b    &   1 Lepton + 1 b-jet  &   1T1b    &   1 $\tau$ + 1 b-jet
\\  \hline 2L1b    &   2 Leptons + 1 b-jet  &   2T1b    &   2 $\tau$
+ 1 b-jet  \\  \hline 0L2b    &   0 Lepton + 2 b-jets  &   0T2b    &
0 $\tau$ + 2 b-jets  \\  \hline 1L2b    &   1 Lepton + 2 b-jets  &
1T2b    &   1 $\tau$ + 2 b-jets  \\  \hline 2L2b    &   2 Leptons + 2
b-jets  &   2T2b    &   2 $\tau$ + 2 b-jets  \\  \hline ep  &   $e^+$
in 1L &   em  &   $e^-$ in 1L \\  \hline mp  &   $\mu^+$ in 1L   &
mm  &   $\mu^-$ in 1L   \\  \hline tp  &   $\tau^+$ in 1T  &   tm  &
$\tau^-$ in 1T  \\  \hline OS  &   Opposite Sign Di-Leptons &   0b  &
0 b-jet \\  \hline SS  &   Same Sign Di-Leptons &   1b  &   1 b-jet
\\  \hline OSSF    &   Opposite Sign Same Flavor Di-Leptons &   2b  &
2 b-jets \\  \hline SSSF    &   Same Sign Same Flavor Di-Leptons &
3b  &   3 b-jets \\  \hline OST &   Opposite Sign Di-$\tau$ &   4b  &
4 b-jets and more    \\  \hline SST &   Same Sign Di-$\tau$ & TL  &
1 $\tau$ plus 1 Lepton  \\  \hline
\end{tabular}
\begin{tabular}{|l|}
\hline Kinematical signatures\\  \hline 1. $P_T^{miss}$ \\  \hline
2. Effective Mass = $P_T^{miss}$ + $\sum_j P_T^j$\\  \hline 3.
Invariant Mass of all jets\\  \hline 4. Invariant Mass of $e^+e^-$
pair\\  \hline 5. Invariant Mass of $\mu^+\mu^-$ pair\\  \hline 6.
Invariant Mass of $\tau^+\tau^-$ pair\\  \hline
\end{tabular}
\caption[]{ The tables give  a list of 40 counting signatures along with the kinematical signatures
 analyzed for each  point in the SUGRA model
parameter space. $L=e,\mu$ signifies only electrons and
muons.} \label{tab:counting}
    \end{center}
 \end{table*}

The different kinematical signatures we investigated for the purpose
of discriminating among sparticle mass patterns are also exhibited in Table
(\ref{tab:counting}). One may further divide the kinematical
signatures into two classes: namely those involving transverse momentum $P_T$ and
those which involve invariant
mass. For those involving  $P_T$, we have investigated missing $P_T$ distributions
and the effective mass, the latter being the sum of missing
$P_T$ and $P_T$ of all jets contained  within an event.
For the kinematical variables using  invariant
mass, we reconstruct such quantities for four different cases, i.e.,
 the invariant mass for all jets, for $e^+e^-$ pair, for
$\mu^+\mu^-$ pair, and for $\tau^+\tau^-$ pair. The reconstruction of the
invariant mass of $\tau^+\tau^-$ pair is  based on hadronically decaying
taus (for recent analyses see \cite{ArnowittTexas}).

 %---------------------------- Signature for mSPs  ----------------------------

\subsection{Discrimination among mSPs in mSUGRA\label{D2}}

We turn now to a discussion of how one may distinguish among different
patterns. The analysis begins by considering the 902  model points that survive our
mSUGRA scan with $10^6$ trial points, and simulating  their LHC signals
with PGS4 using, for illustration, 10 fb$^{-1}$ of integrated luminosity at the LHC.
In our analysis we will focus mostly on the counting signatures.
Here the most useful counting signature is the total number of SUSY
events after trigger level cuts and post trigger level cuts are imposed.
All other counting signatures are normalized with respect to the
total number of SUSY events passing the cuts
and thus appear as fractions lying
between (0,1) in our figures.
To keep the analysis statistically significant, we admit only those points in the
parameter space that generate at least 500 total SUSY events.

\begin{figure*}[htb]
\centering
\includegraphics[width=7.0cm,height=6.0cm]{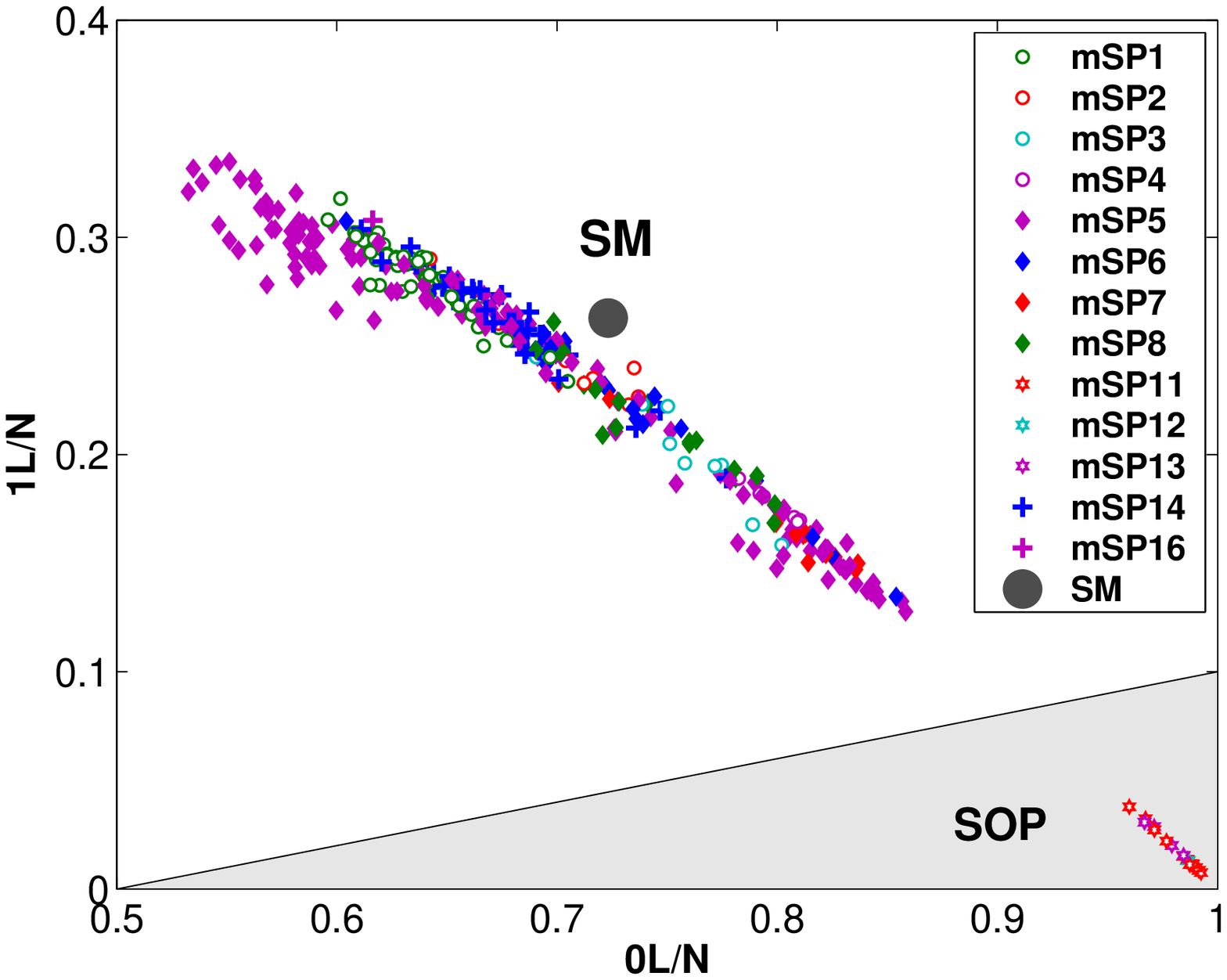}
\includegraphics[width=7.0cm,height=6.0cm]{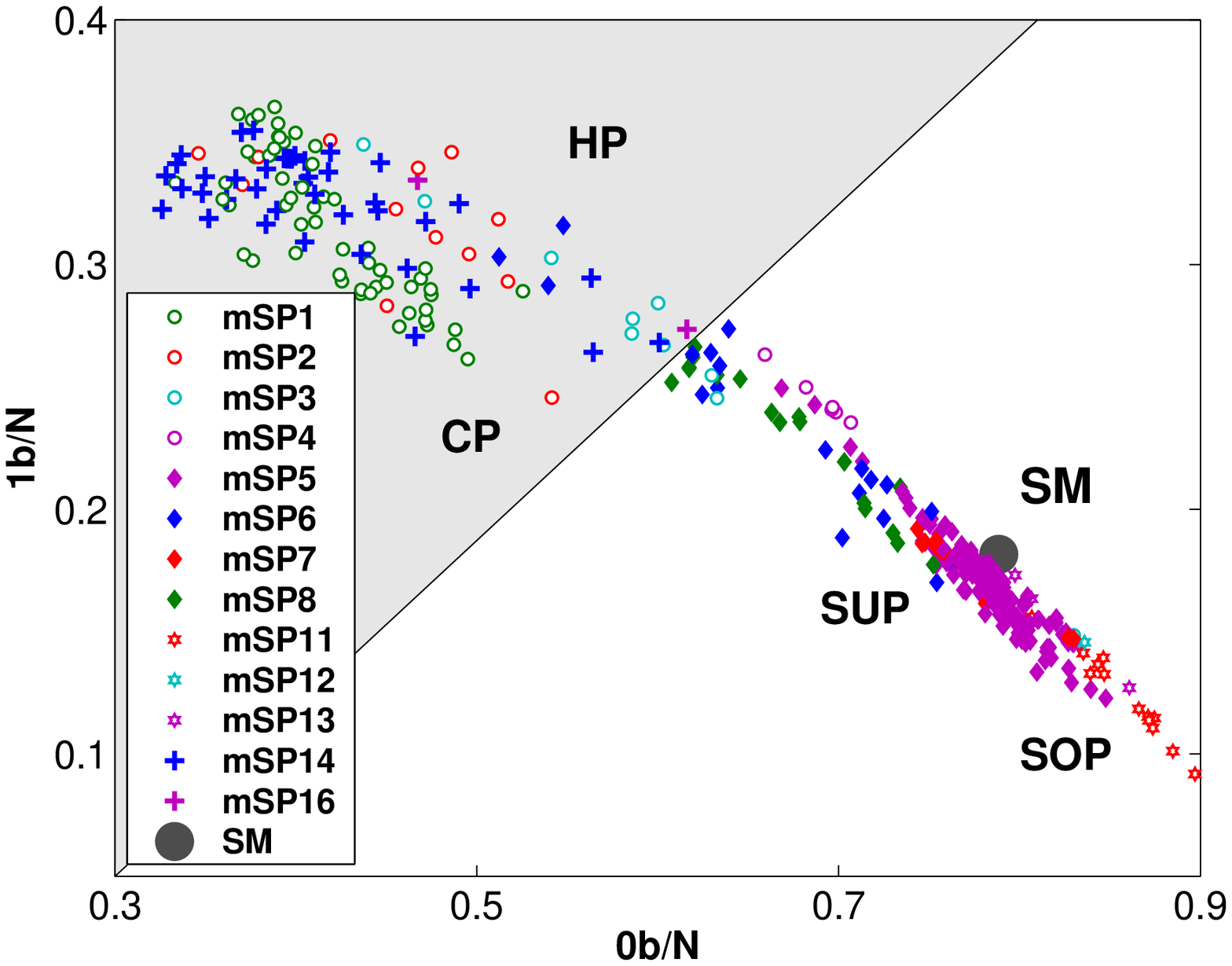}
\includegraphics[width=7.0cm,height=6.0cm]{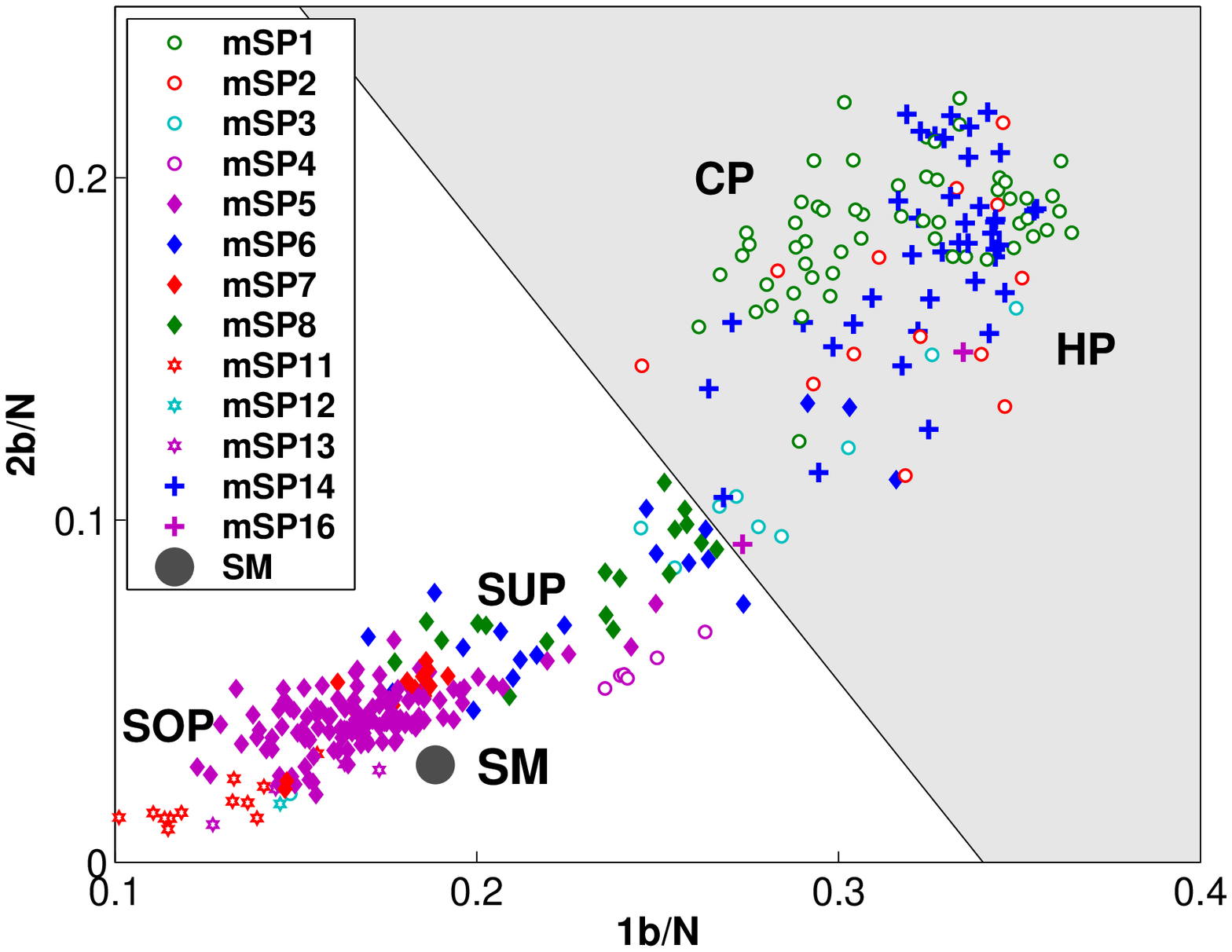}
\includegraphics[width=7.0cm,height=6.0cm]{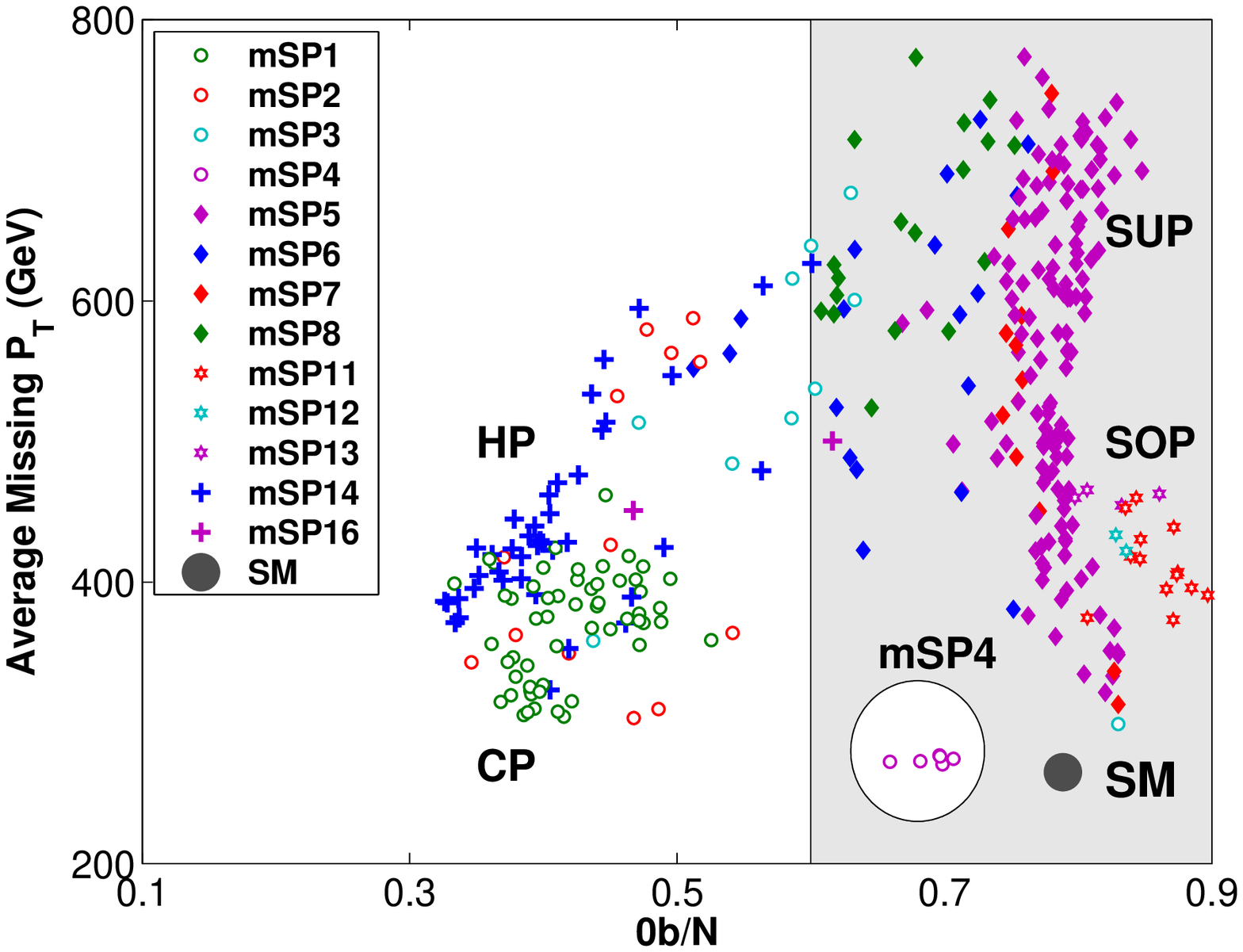}
\caption{
Top Left: An exhibition of the mSPs in the 1L
vs 0L where the fraction of events to the
total number of events  in each case  is plotted.
 The analysis shows that the Stop Patterns (SOP)
appearing on the right-bottom corner are easily distinguished from
other patterns. The analysis shows that SOP has few lepton signals.
Top Right and Bottom Left: Plots in the signature space with fraction of events with
 1b vs 0b and 2b vs 1b  exhibiting the separation of CPs and HPs from SOPs
and SUPs, with CPs and HPs occupying one region, and SOPs and SUPs occupy another in
this signature space except for a very small overlap.
Bottom Right: An exhibition of the mSPs in the signature space with the average
missing $P_T$ for each parameter point in the mSUGRA parameter
space along the y-axis and the fraction of events with $0b$ along the
x-axis. The plot shows a separation of the CPs and HPs from SOPs and
SUPs. Further, mSP4 appears isolated in this plot.
Most of the CPs and HPs have less than 60\% events without b-jet content.
The ratios for the SUSY models refer to the SUSY signal only.  The SM point is purely
background.
}
\label{fig:bjet}
\end{figure*}

%%%%  ---- SUP and CP ---- %%%%

\begin{figure*}[htb]
\centering
\includegraphics[width=7.0cm,height=6.0cm]{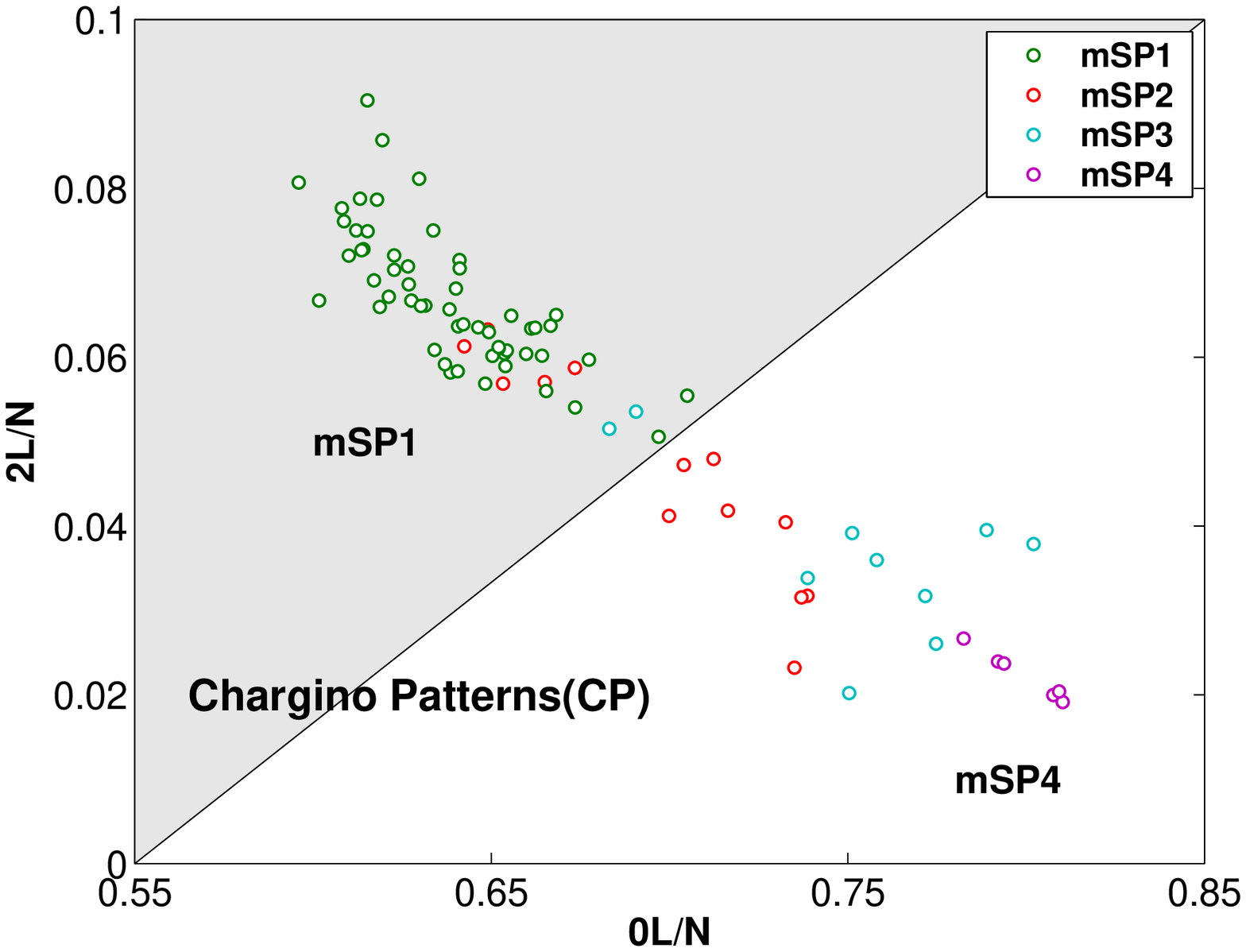}
\includegraphics[width=7.0cm,height=6.0cm]{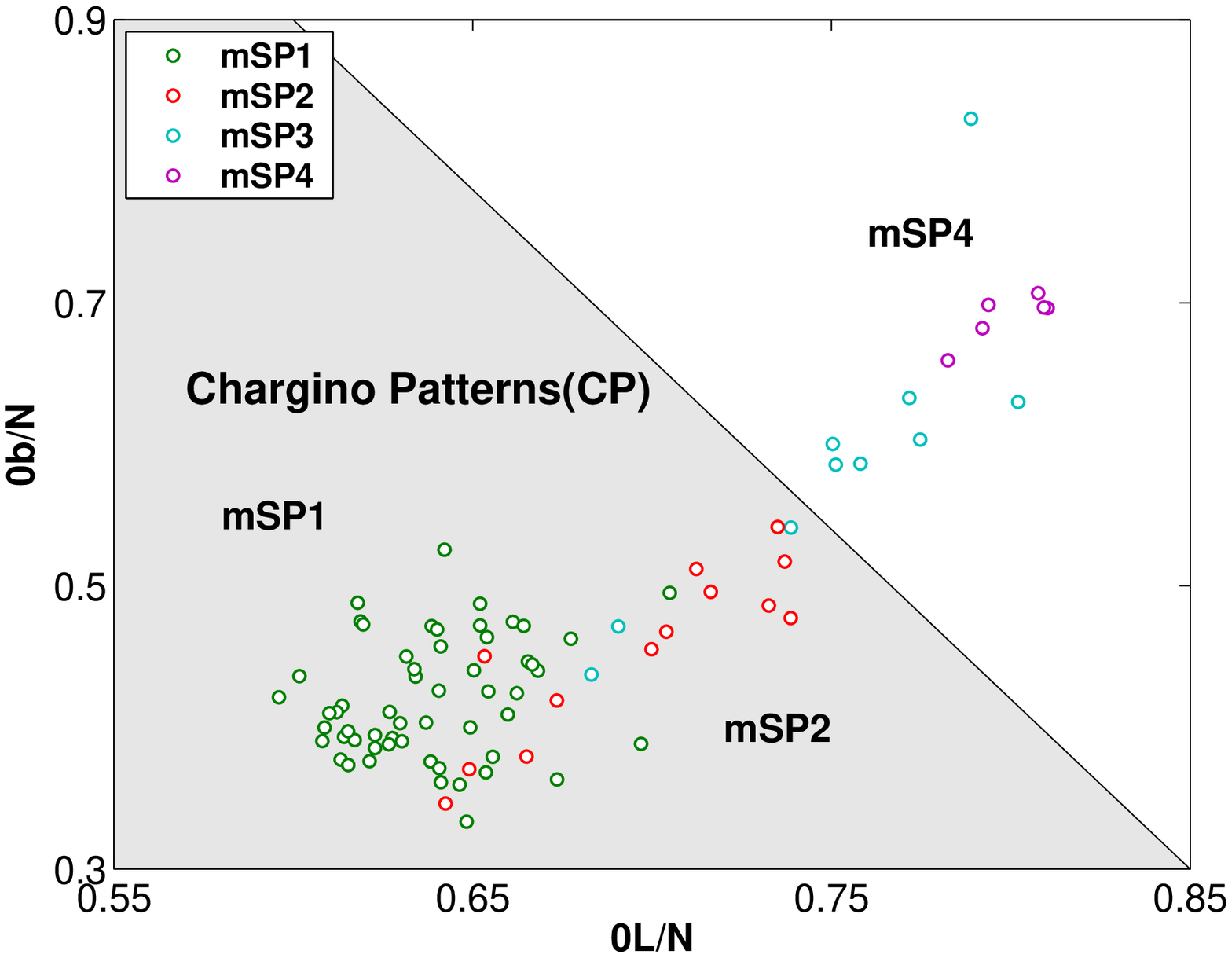}
\includegraphics[width=7.0cm,height=6.0cm]{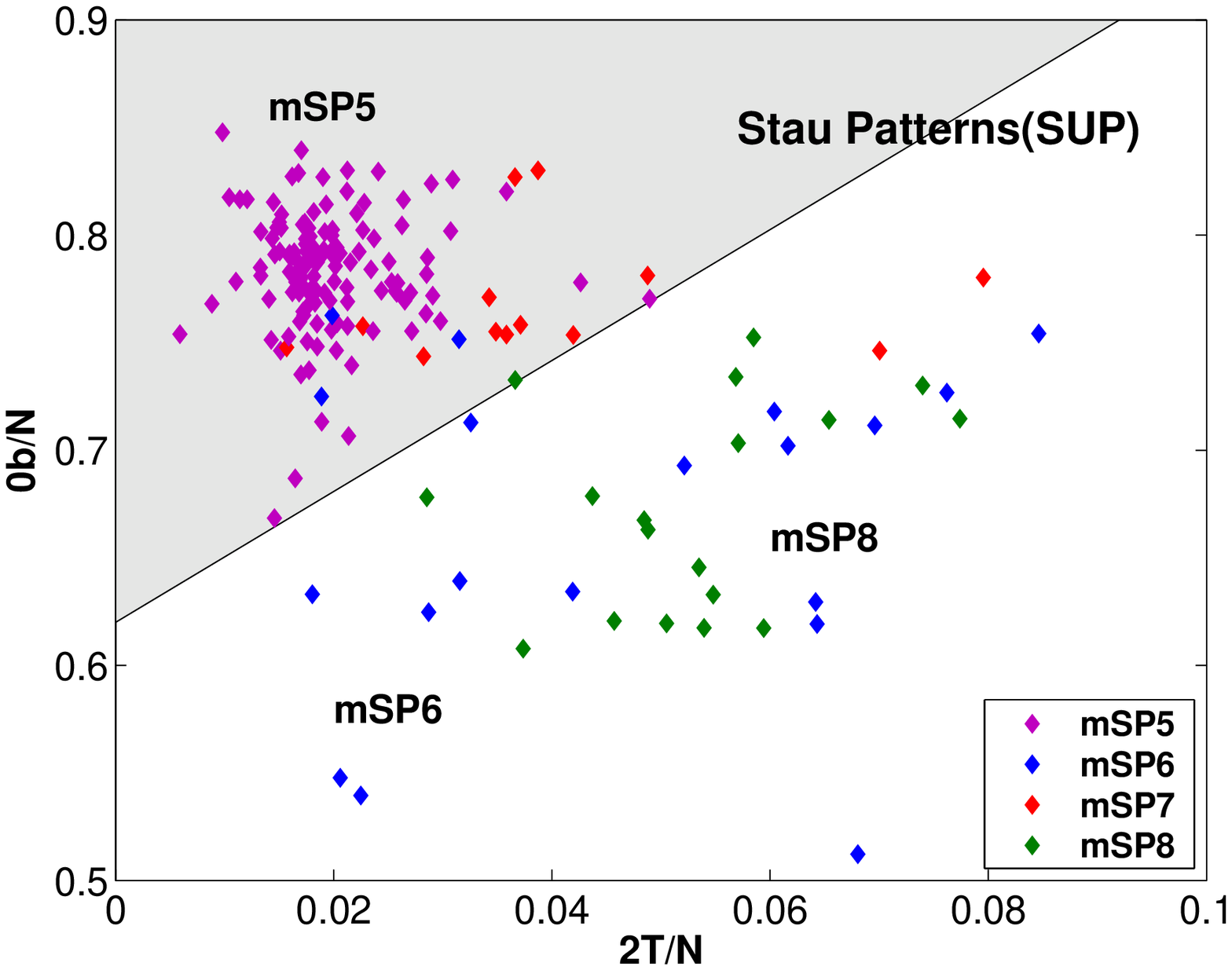}
\includegraphics[width=7.0cm,height=6.0cm]{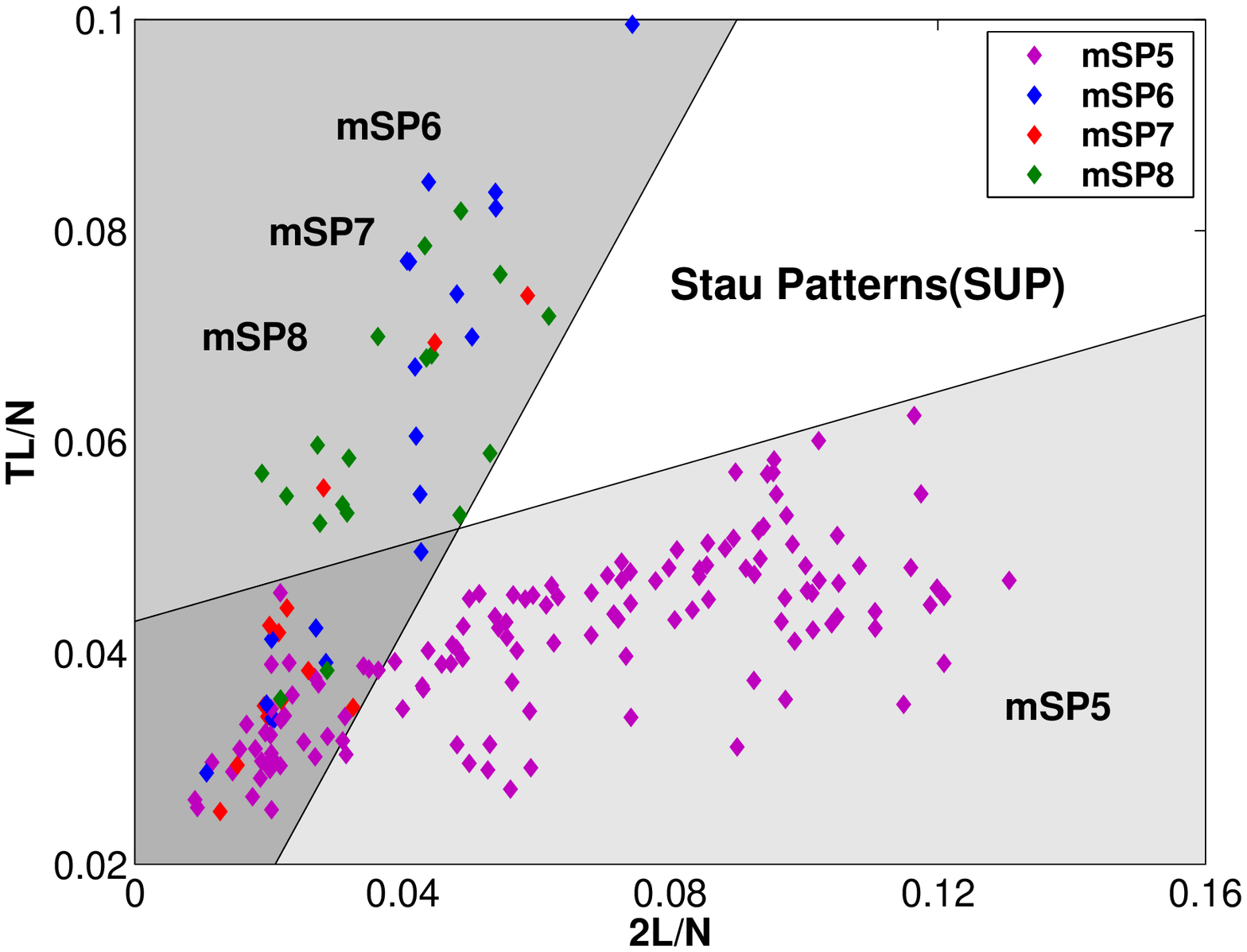}
\includegraphics[width=7.0cm,height=6.0cm]{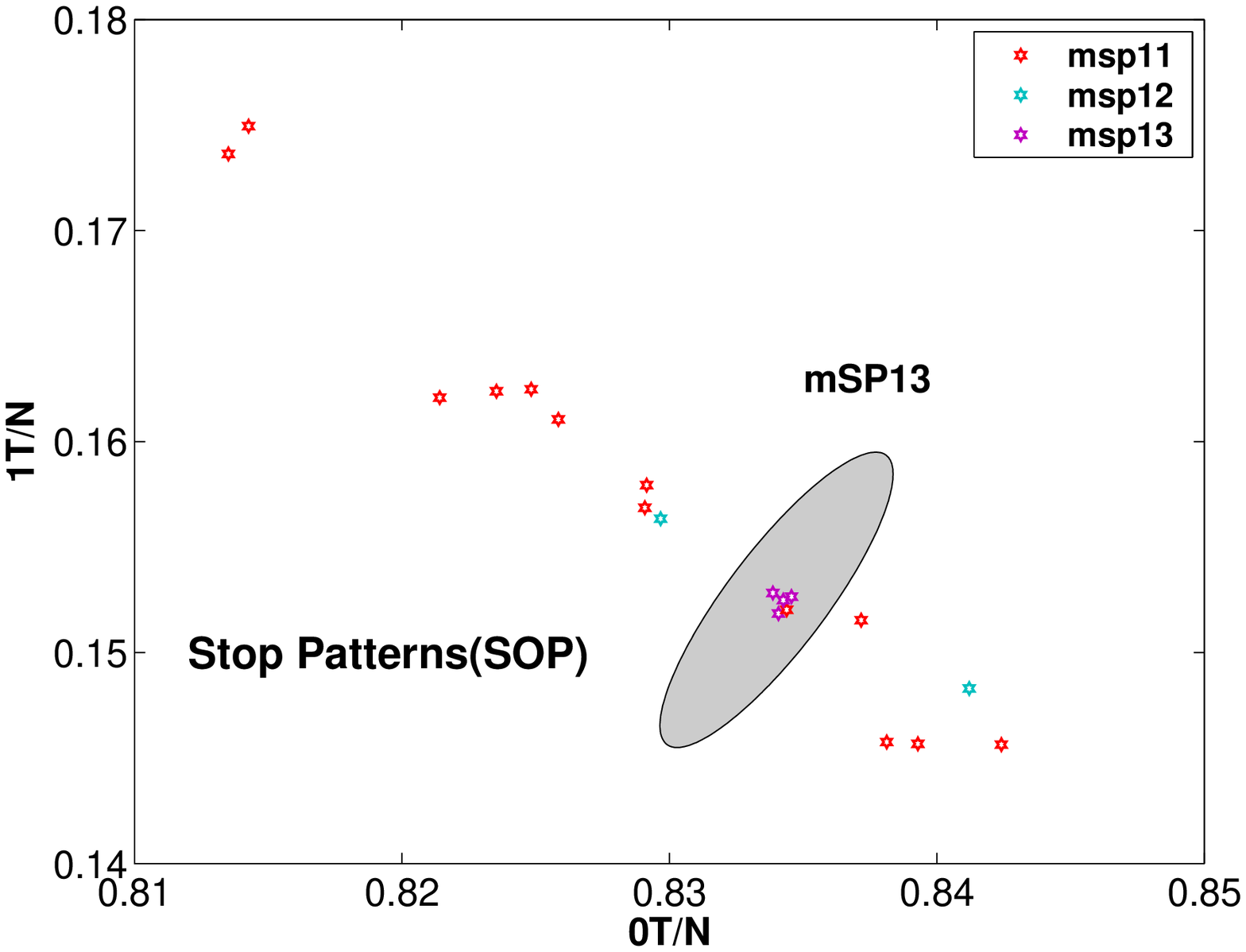}
\includegraphics[width=7.0cm,height=6.0cm]{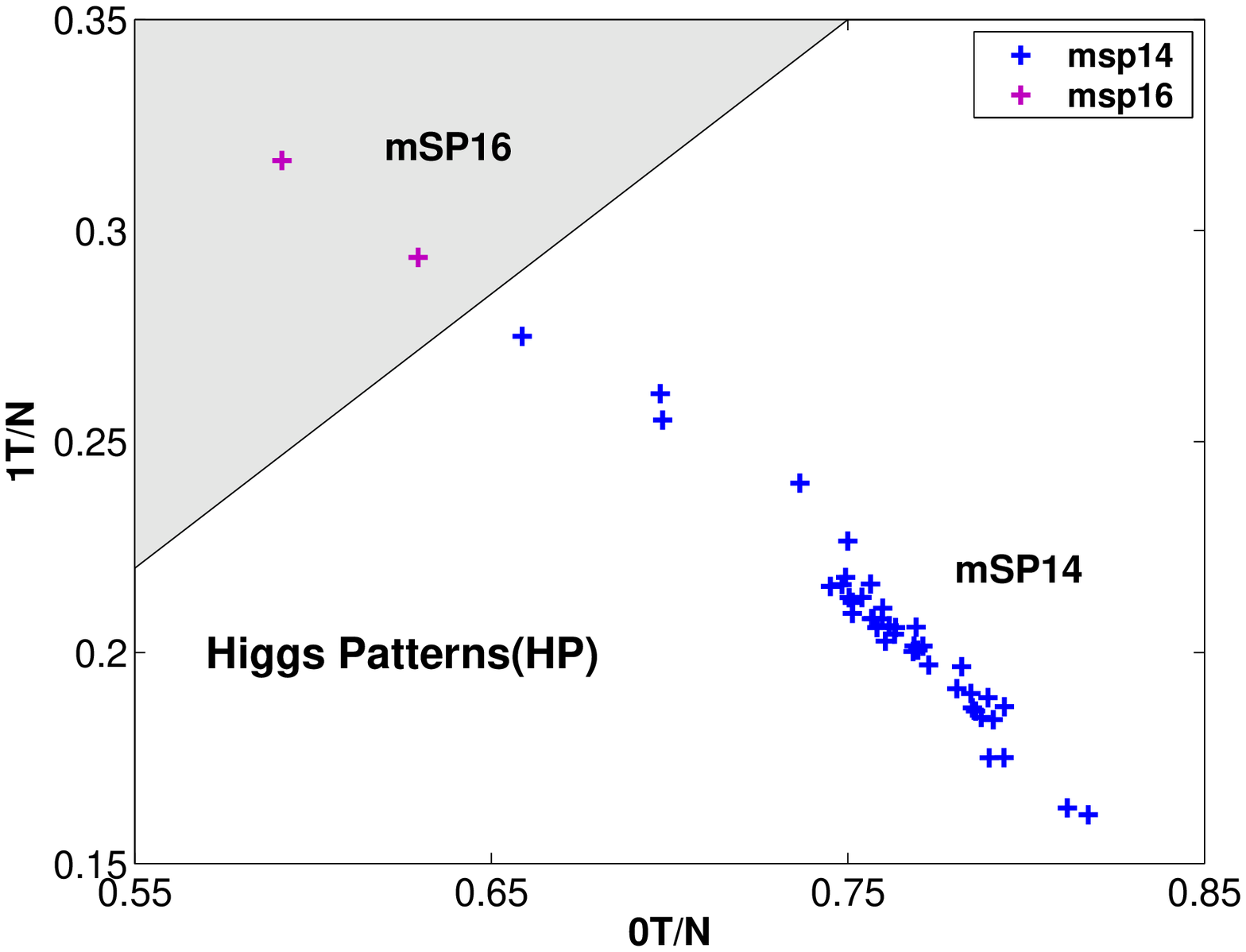}
\caption{
An exhibition of how the mSPs can be  discriminated within a given
class, i.e., within CPs, SUPs, SOPs, and HPs.  The analysis shows
that  patterns within a given class can be discriminated.
}
\label{fig:sups}
\end{figure*}

We give now the details of the analysis. In Fig.(\ref{fig:bjet}), we
investigate the signature space spanned by a variety of signature
channels. The top left panel gives a plot with one signature
consisting of events with one lepton and the second signature
consisting of events with no leptons.  It is seen that the stop
patterns (SOPs) that survive the cuts are confined in a  small
region at the right-bottom corner and have a significant separation
from all other mSPs. The panel illustrates the negligible leptonic
content in stop decays. The top-right panel is a plot between two
signatures where one signature contains a tagged b-jet while the
other signature has no tagged b-jets.  In this case one finds a
significant separation of the CPs and HPs  from SUPs and SOPs. The
lower-left panel gives a plot where one signature has two tagged
b-jets and the other signature has only one tagged b-jet. One again
finds that the CPs and HPs are well separated from the SOPs and the
SUPs for much the same reason as in upper-right panel.  Finally,  a
plot is given in the lower-right panel where one signature is the
average missing $P_T$ while the other signature involves events with
no tagged b-jets. Again in this plot the CPs (which include mSP4)
and HPs are well separated from the SOPs and SUPs.

The analysis of  Fig.(\ref{fig:bjet}) exhibits that
for some cases, e.g., for the patterns CP and HP in the upper right
hand corner of Fig.(\ref{fig:bjet}),  the separation between the SUGRA
prediction and the Standard Model background is strikingly clear, allowing for the
 identification not only of new physics but also of the nature of the pattern that
leads to such a signature.

We discuss now the possibility of discriminating sub-patterns within a given pattern class.
 An analysis illustrating this possibility
is given in Fig.~(\ref{fig:sups}). Here the top two panels illustrate
how  the sub-patterns mSP1, mSP2, mSP4 within the chargino class (CP)
are distinguishable  with appropriate choice of the signatures.
A similar analysis regarding the discrimination for the sub-patterns in the stau class  (SUP)
is given in the two middle panels. The lower-left panel gives an analysis of how one may
discriminate the stop sub-patterns mSP11, mSP12, mSP13 in the stop class (SOP),
and finally the  lower-right panel shows the plots that allows one to discriminate
the Higgs patterns mSP14 and mSP16 from each other.
There are a variety of other plots which allow one to discriminate
among patterns.
 With 40 counting signatures one  can have 780
%$^{40}C_{2}$
such plots and it is not possible to display all of them.
A global analysis where the signatures are simultaneously considered for
a large collection of mSPs and NUSPs is discussed in Sec.(\ref{D7}).

 As mentioned in the above analysis we have included models which can produce at
least 500 SUSY events with 10 fb$^{-1}$  which is  lower than
our estimated discovery limits for total SUSY events which are about 2200 in this case. The reason for
inclusion of points below the discovery limit in the total SUSY events is
 that some of them can be detected in other
channels such as in the trileptonic channel while others will be
detectable as the luminosity goes higher. We note in passing that
reduction of admissible points makes separation of patterns easier.

\subsection{Sparticle signatures including nonuniversalities   \label{D3}   }

%%% ------ SUGRA (m+NU) Signals ----
\begin{figure}[htb]
\centering
\includegraphics[width=7.0cm,height=6.0cm]{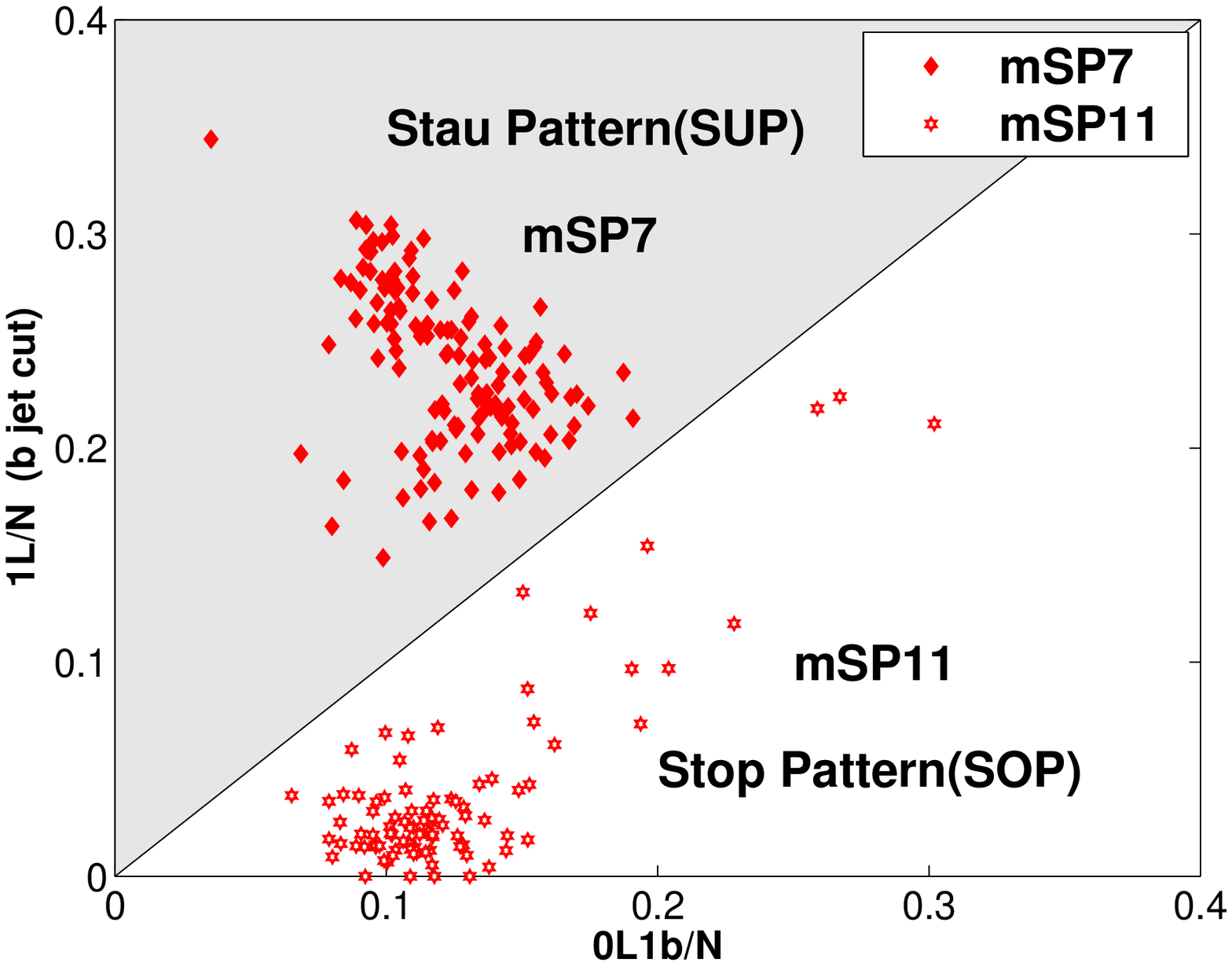}
\includegraphics[width=7.0cm,height=6.0cm]{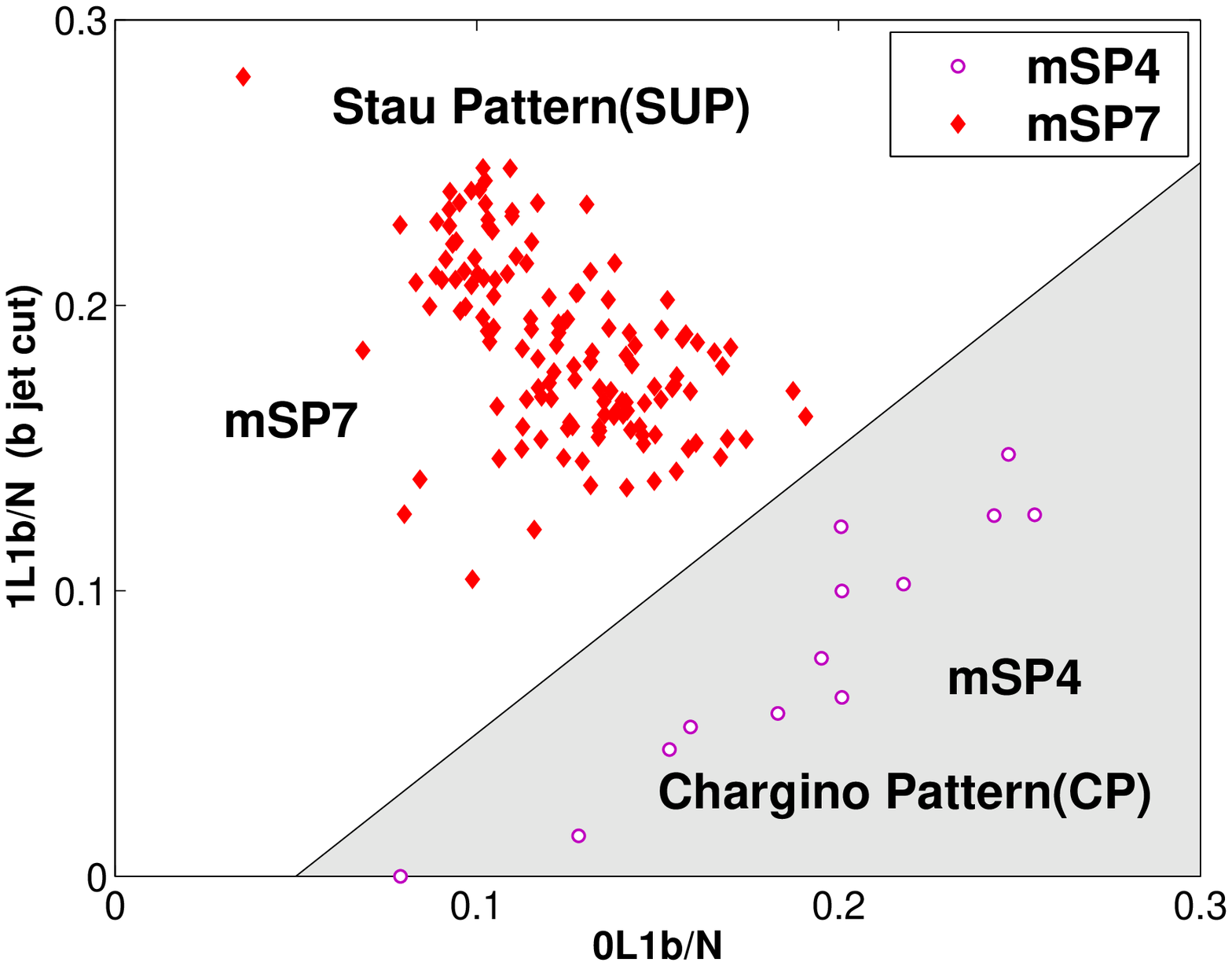}
\includegraphics[width=7.0cm,height=6.0cm]{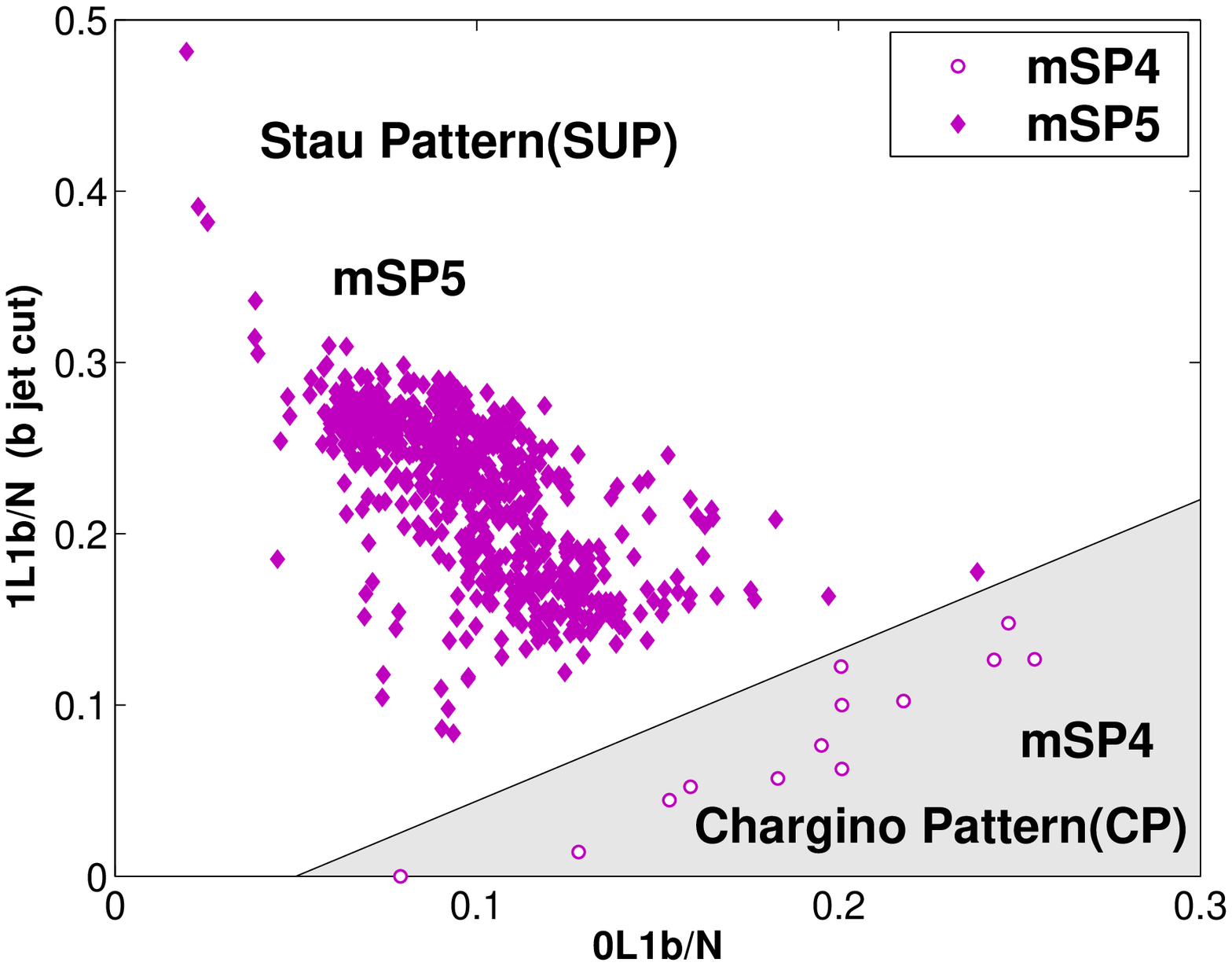}
\includegraphics[width=7.0cm,height=6.0cm]{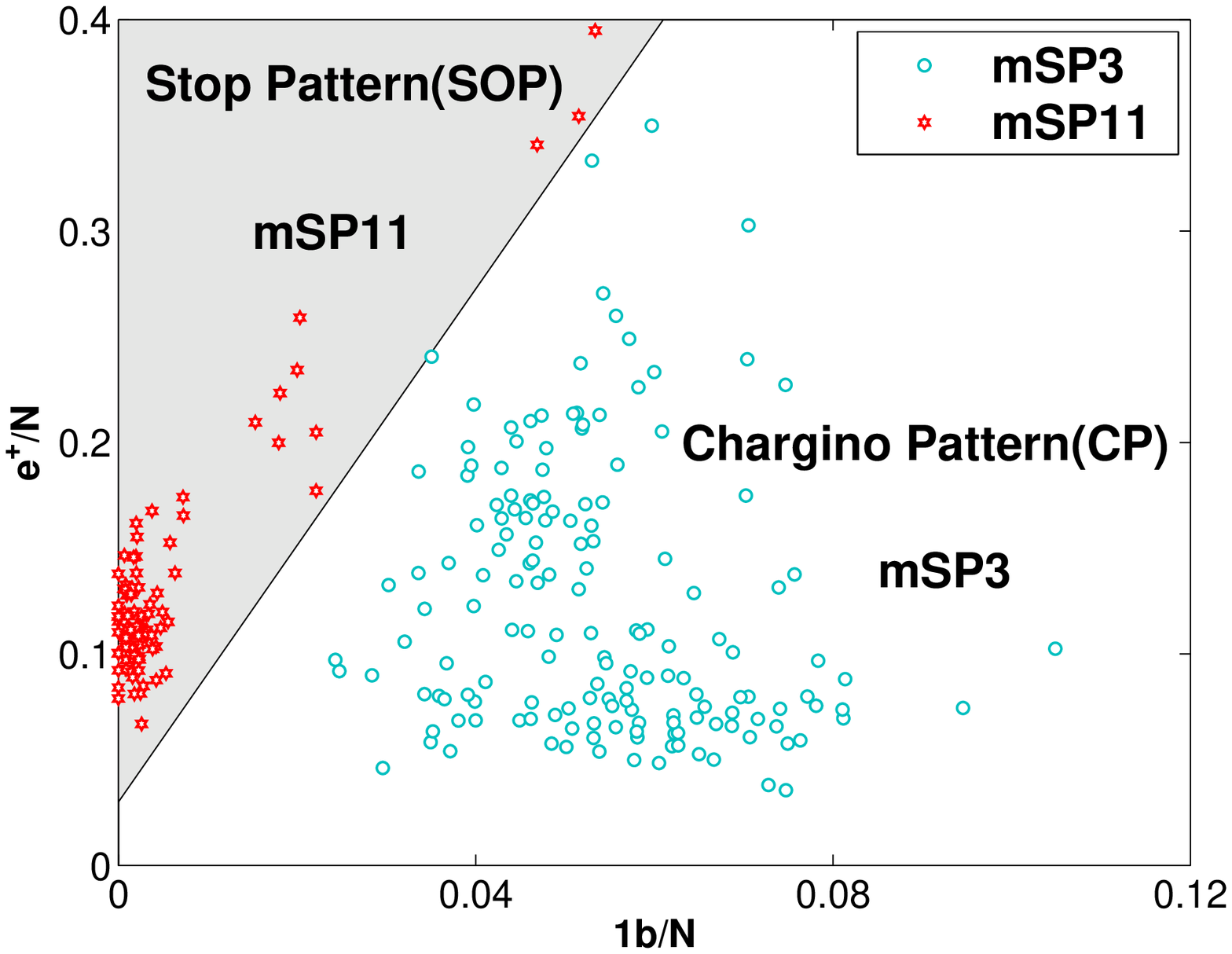}
\caption[]{
Discrimination among mSPs within both mSUGRA and
NUSUGRA models.  Two  mSPs are presented in each figure in
different signature spaces to show the separation for each case.
Signals are simulated with constant number of events in PGS4
for each pattern.
} \label{fig:sugrafig}
\end{figure}

In this subsection, we give an analysis including \non in three
different sectors: NUH, NU3, and NUG.  In our analysis we simulate
various models with the same constant number of events N which we
take as an example to be N$= 10^4$. To discriminate among the
patterns in the signature space, we introduce another set of post
trigger cuts, which we denote as 'b jet cuts', in addition to  the
default post trigger cuts specified in Sec.(\ref{D1}). The criteria
in the b-jet cuts are the same as the default post trigger cuts,
except that we change the condition `at least two hadronic jets in
the event' to `specifically at least one b-tagged jet in the event'.
We exhibit our analysis utilizing both the default cuts and the b
jet cuts in  Fig.(\ref{fig:sugrafig}). One can see that even with
inclusion of a variety of soft breaking scenarios, some mSPs still
have very distinct signatures in some specific channels.

Thus in the top-left panel  we give a plot of mSP7 (SUP) and
mSP11 (SOP) in the signature space 1L/N (b jet cuts) vs 0L1b/N,
where 0L1b/N is obtained with the default post trigger cuts.
Here we find that these two model types are clearly distinguishable
as highlighted by shaded and unshaded regions. A similar analysis with
signatures consisting of 1L1b/N (b jet cuts) vs 0L1b/N for
mSP4 (CP) and mSP7 (SUP) is given in the top-right panel.
The lower-left panel gives an analysis of mSP4 (CP) and mSP5 (SUP)
also in the signature space consisting of 1L1b/N (b jet cuts) vs 0L1b/N.
Finally, in the lower-right panel we give an analysis of mSP3 (CP) and mSP11 (SOP)
in the signature plane $e^+$/N vs 1b/N.  These analyses
illustrate that the patterns and often even the sub-patterns can be discriminated
 with the appropriate choice of signatures for a general class of SUGRA models
including \non.

%%%%%%%%%%%%%%%%

\subsection{The trileptonic signal as a pattern discriminant \label{D6} }

\begin{figure*}[htb]
  \begin{center}
\includegraphics[width=7.0cm,height=6.0cm]{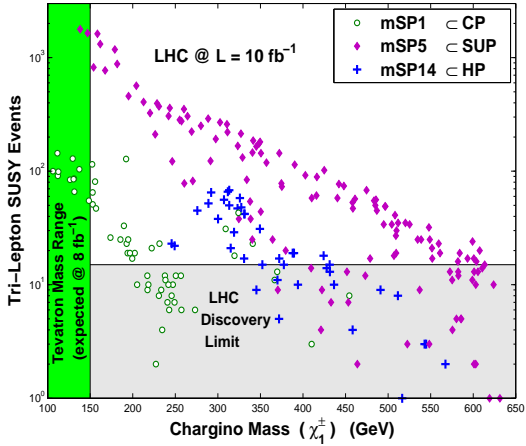}
\caption[]{
A plot of the number of trilepton events versus  the light chargino mass for three patterns,
one from each class, CP, SUP and HP. The SUP pattern gives the largest trileptonic signal
followed by the HP and CP patterns.
}
\label{fig:sigmass2}
  \end{center}
\end{figure*}

 \begin{figure*}[htb]
  \begin{center}
\includegraphics[width=7.0cm,height=6.0cm]{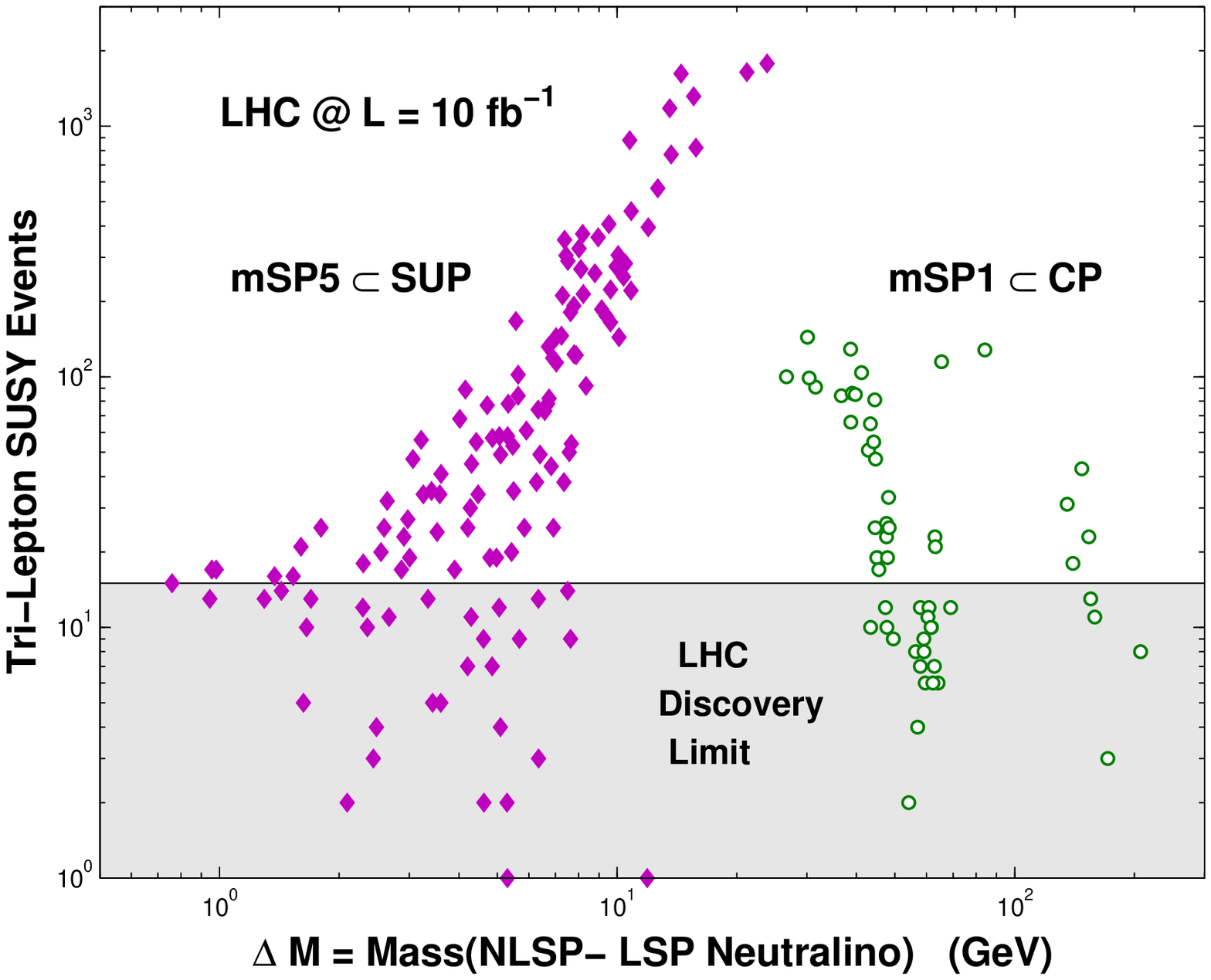}
\includegraphics[width=7.0cm,height=6.0cm]{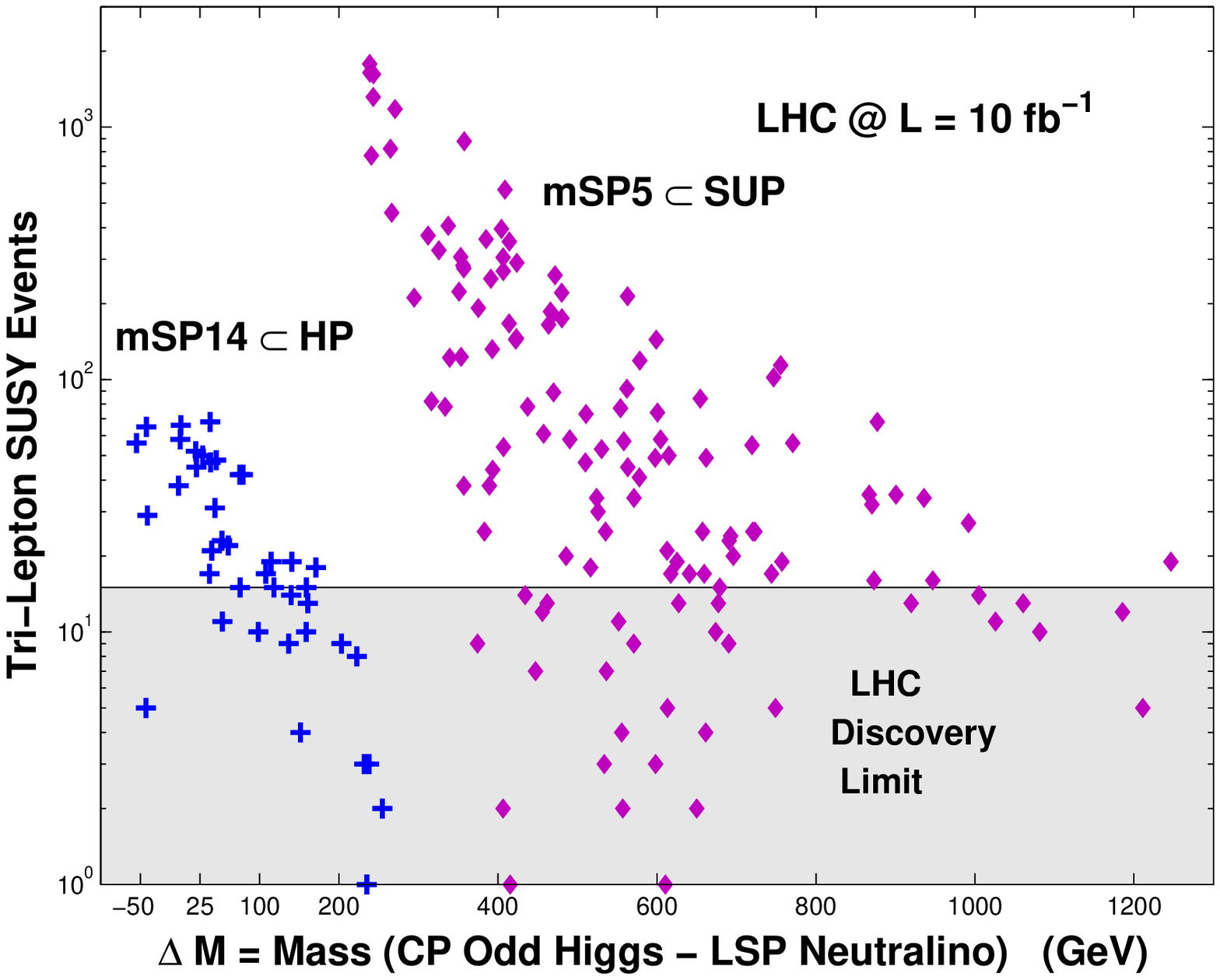}
\caption[]{The number of tri-lepton events versus  the
sparticle mass splittings. The left panel shows clear separations
for hierarchical mass patterns in the number of trilepton events
produced with  $10~\rm fb^{-1}$ as a function of the NLSP and the  LSP
mass splitting for the chargino (CP) pattern mSP1 and Stau (SUP) mSP5.
The plot on the right shows a similar effect for the case where the mass splitting is taken
to be the difference of the CP odd Higgs boson mass and the LSP
for both the Higgs pattern mSP14 and the stau pattern mSP5.
The Standard Model background is highly suppressed in this channel.
}
\label{fig:sigmass1}
  \end{center}
\end{figure*}

The trileptonic signal is an important signal for the discovery of supersymmetry.
For on-shell decays the trileptonic signal was discussed in the early days in
\cite{earlypheno,Baer:1986vf} and for off-shell decays in \cite{Nath:1987sw}.
(For a recent application see \cite{CMSnote3}).
Here we discuss the trileptonic signal in the context of discrimination
of hierarchical  patterns.
In Fig.~(\ref{fig:sigmass2}) we exhibit the dependency of the trilepton
signal on the chargino mass. It is seen  that mSP5 gives
the largest number of events in this channel while
 the CP pattern (mSP1) and the HP pattern (mSP14)
can also produce a large number of trilepton events above the
discovery limit,  while the chargino mass reach is extended for the
mSP5 as opposed to the mSP1 and mSP14. The above observations hold
for some of the other SUP patterns as well. Thus the trileptonic
signal is strong enough to be probed up to  chargino masses of about
600~\rm GeV in the SUP pattern. Another interesting display of the
trileptonic signal is when this signal is plotted against some
relevant mass splittings.
 Thus  the left-panel of Fig.~(\ref{fig:sigmass1}) gives an analysis for the trileptonic signal for
two patterns:  the Chargino pattern mSP1
and the Stau pattern mSP5 plotted against the  NLSP-LSP mass
 splitting  with  $10~\rm fb^{-1}$ of data.
The analysis of  the left-panel of Fig.~(\ref{fig:sigmass1}) shows that
the SUP pattern
presents an excellent opportunity for discovering SUSY through the 3
lepton mode.  The analysis also shows a clear separation  among mass patterns
and further a majority of the model points stand above the discovery limit which
in this channel is  $\approx 15$  events under the post trigger level
cuts discussed in Sec.(\ref{D1}).
The right-panel of
Fig.~(\ref{fig:sigmass1}) gives an analysis of the trileptonic signal
 vs the mass splitting of the CP odd Higgs and the
lightest neutralino LSP for patterns mSP5  and mSP14.
Again, we see a clear separation of model points.
We note that CP odd Higgs can sometimes be even
 lighter than the LSP, and thus the quantity $\Delta M= M_A-M_{\tilde \chi_1^0}$ plotted
 on the x-axis can sometimes become  negative.

\subsection{Kinematical distributions \label{D5} }

\begin{figure*}[htb]
  \begin{center}
\includegraphics[width=7.0cm,height=6.0cm]{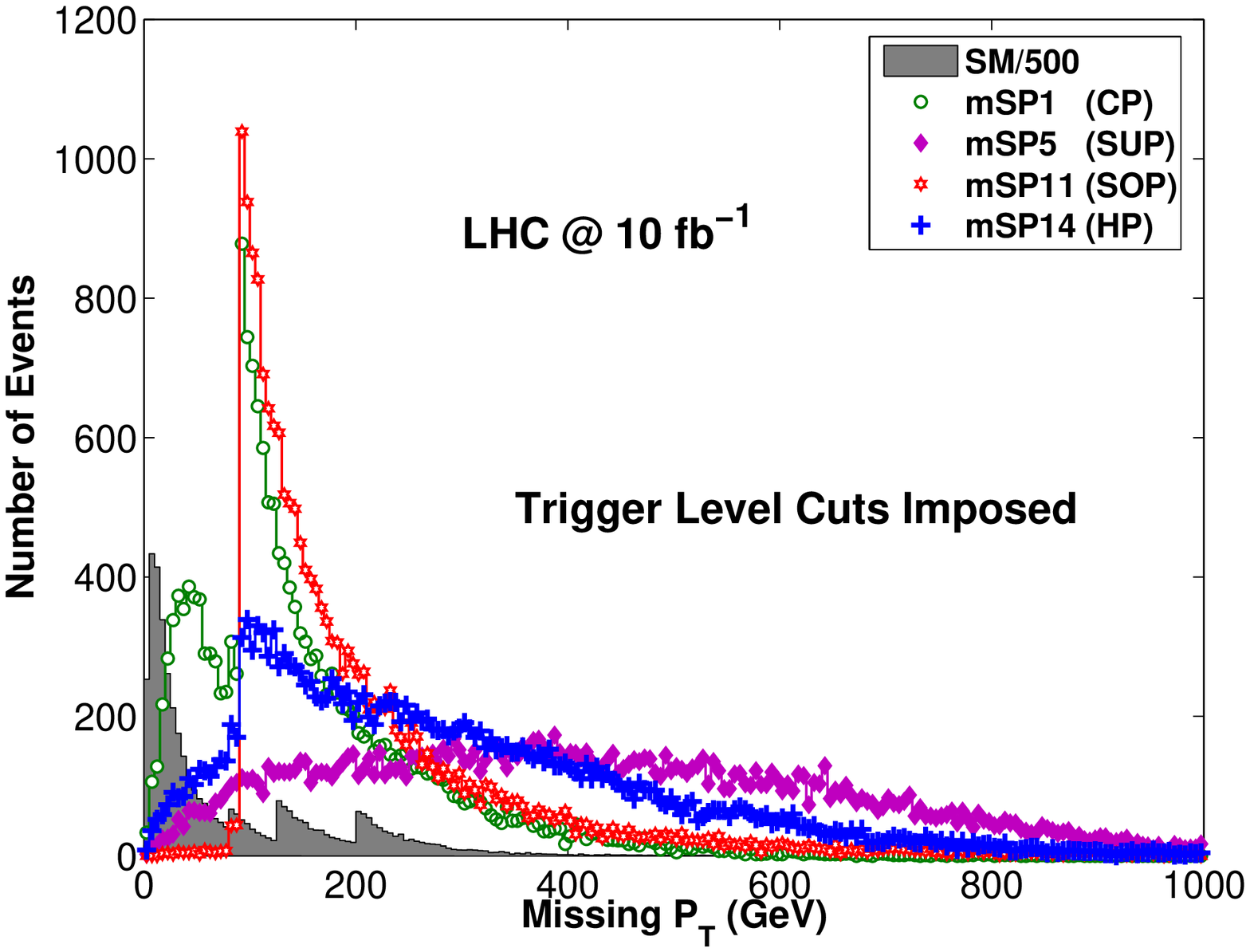}
\includegraphics[width=7.0cm,height=6.0cm]{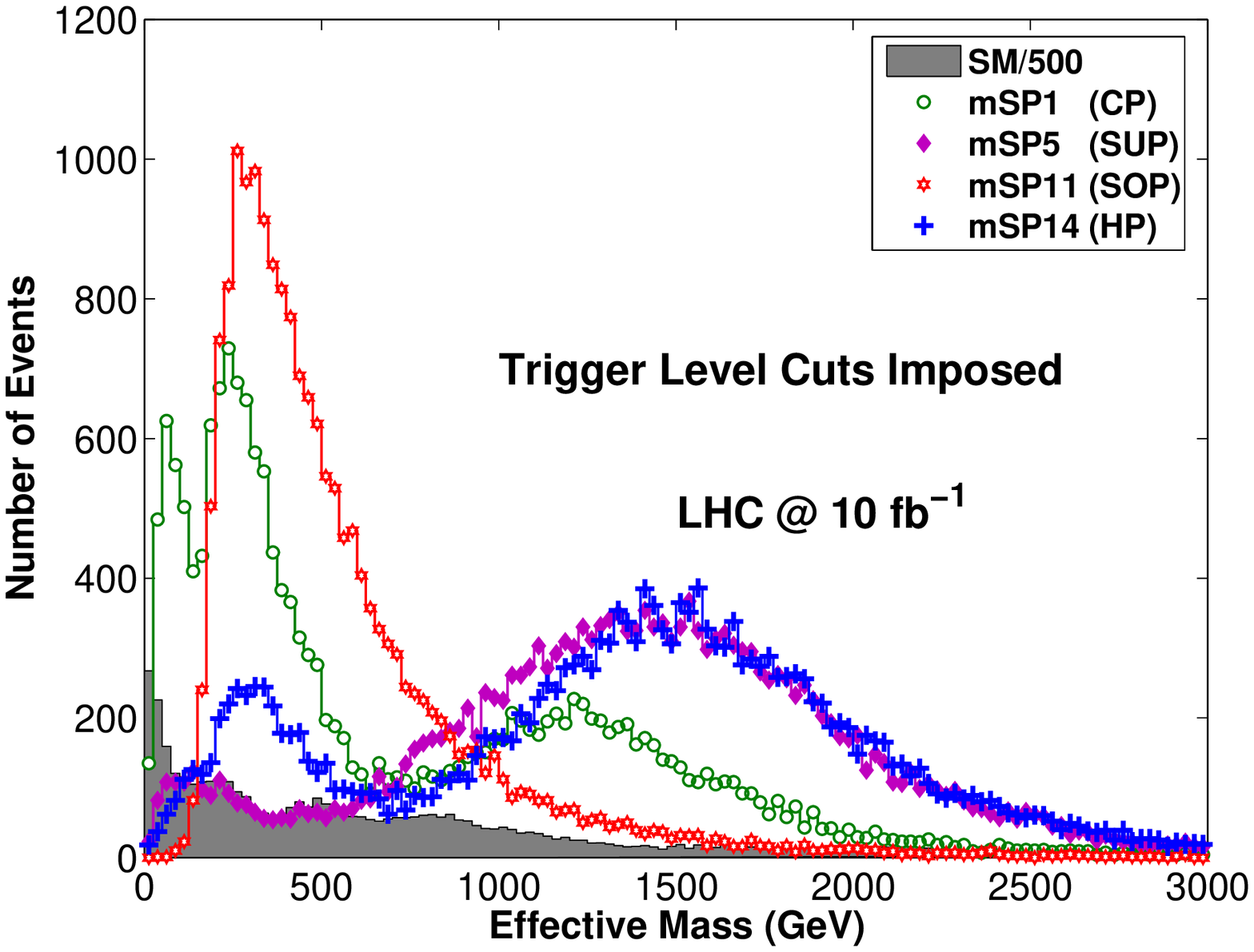}
\caption[]{
An exhibition of the missing $P_T$  and of the effective mass
 distributions for 4 different mSUGRA models
with each corresponding to one class of mSPs, and for the Standard Model.
In the missing $P_T$ distribution as well as in  the  effective mass distribution,
the Standard Model tends to produce events with a
 lower missing  $P_T$ and a lower effective mass relative to the  mSUGRA case which
 generates events at relatively higher missing $P_T$ and effective mass.
 Further,  there is  a large variation
between different mSUGRA models, as  can be seen above. Thus, for example
mSP5 (a stau pattern) and mSP14 (a Higgs pattern) have peaks at
larger values of missing $P_T$ and larger values of the effective mass relative to
mSP1 (a chargino pattern) and mSP11 (a stop pattern).
Additionally, the shapes of the distributions are also different.
Only trigger level cuts are employed here.
}
\label{kinpt}
\end{center}
\end{figure*}

In addition to the event counting signatures discussed above, the
kinematical signatures are an important tool for pattern
discrimination. We illustrate this using the kinematical variables
consisting of missing $P_T$ and the effective mass (see Table
(\ref{tab:counting}) for their definitions) and an illustration is
given in Fig.(\ref{kinpt}).
%%%%%%%%%%%%%%
%%%%%%%%%%%%%%
\begin{figure*}[htb]
  \begin{center}
\includegraphics[width=7.0cm,height=6.0cm]{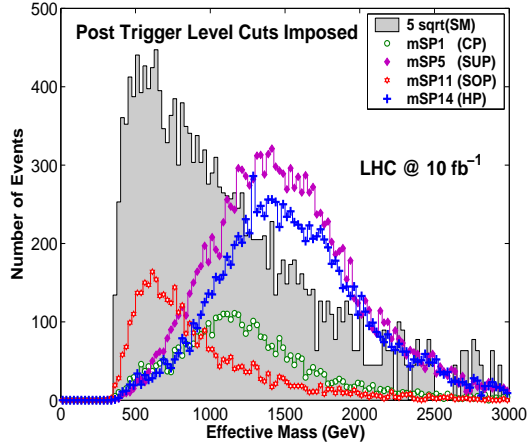}
\caption[]{
The effective mass distributions for 4 different mSUGRA models
with each corresponding to one class of mSPs, and for the Standard Model.
Post trigger level cuts are imposed here. The bin size used here is 25 GeV.
 We exhibit the mSUGRA points used here in
the order ($m_0$, $m_{1/2}$, $A_0$, $\tan\beta$, sign$\mu$):
CP (3206.9,  285.3,   -1319.8,  9.7,  +1),
SUP  (92.6,   462.1,   352.2,     4.5, +1),
SOP  (2296.9, 625.0,    -5254.9,  13.6, +1), and
HP   (756.8,     387.0,    1144.9,   56.5, +1).
}
\label{meff}
\end{center}
\end{figure*}
%%%%%%%%%%%%%
\begin{figure*}[htb]
  \begin{center}
\includegraphics[width=7.0cm,height=6.0cm]{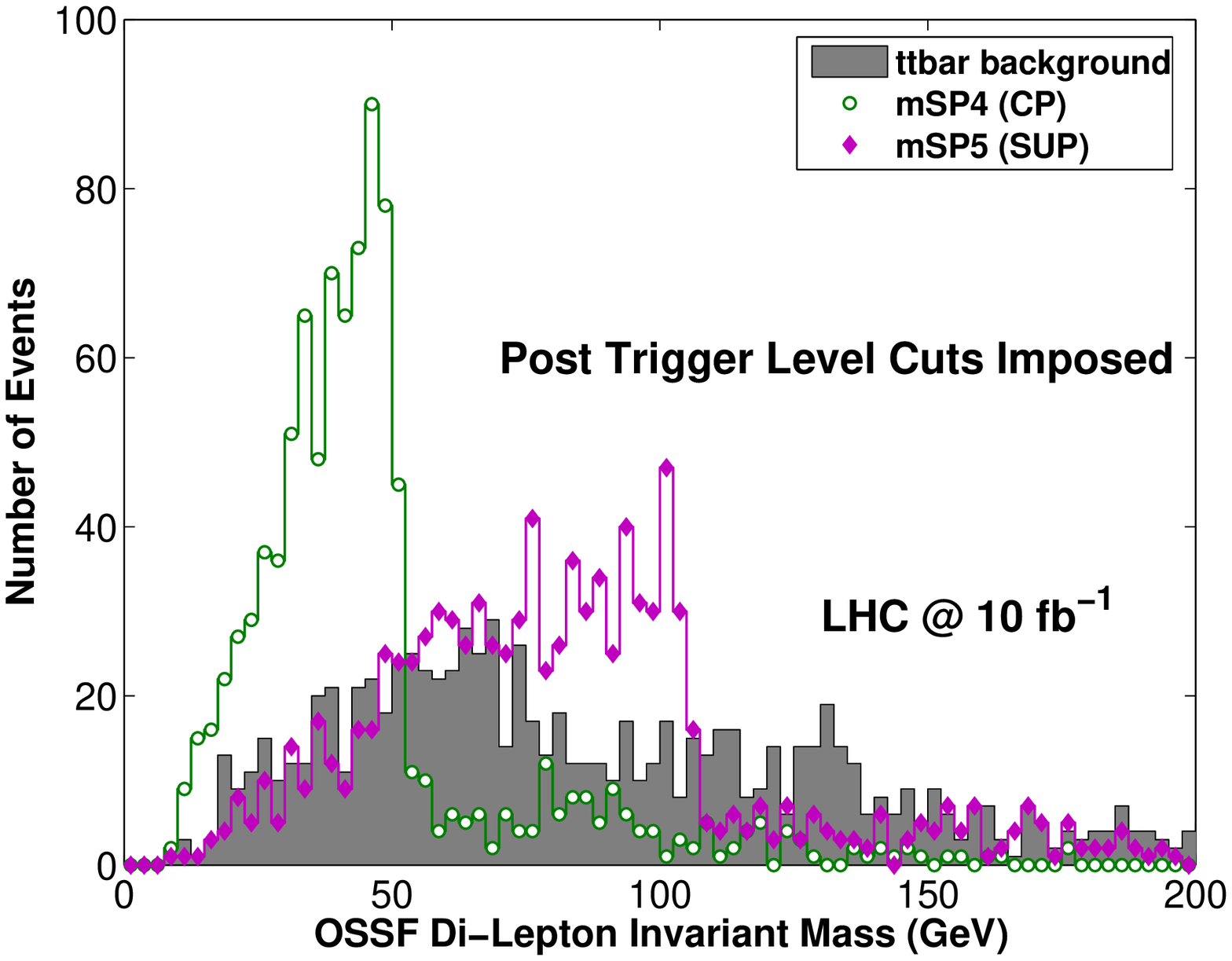}
\caption[]{
A plot of the opposite sign same flavor (OSSF) di-lepton invariant
mass distribution at LHC with 10 fb$^{-1}$ with the default post trigger cuts
imposed for two different mSP points on top of the SM $t\bar t$ background.
The two mSPs, mSP4 (1674.9, 137.6,  1986.5,  18.6,  +1) and mSP5 (84.4, 429.3, -263, 3.4, +1),
are clearly distinguishable from each other in the distribution. The mSP4 model point
shown here has recently been investigated  \cite{Gounaris:2008pa} in
the context of helicity  amplitudes as a discovery mechanism for supersymmetry.
}
\label{kinmass}
  \end{center}
\end{figure*}
Specifically the analysis of Fig.(\ref{kinpt}) uses  four mSUGRA points
 one each in the patterns CP, SUP, SOP and HP .
The analysis of Fig.(\ref{kinpt})  shows  that
the distributions for the CP, HP, SOP and SUP are substantially different.
It is  interesting  to note that
in the missing $P_T$ distribution, the HP and SUP model points have
a relatively flat distribution compared to the CP and SOP model points.
The missing $P_T$ distribution and the effective
mass distribution are useful when designing post trigger level
cuts to optimize the signal over the background.
For instance, one can take a 1 TeV effective mass cut to analyze the
SUP and HP signals shown in Fig.(\ref{kinpt}), but this method will
will not work well when it comes to the CP and SOP points since most of their events
have a rather small effective mass.  To illustrate that different models
have different effective mass distributions, and consequently
different effective mass cuts
are needed for different patterns, an analysis is given in Fig.(\ref{meff})
for the same set of points in Fig.(\ref{kinpt}) with post trigger level cuts imposed.

We also investigate the invariant mass distribution
for the opposite sign same flavor (OSSF) di-leptons ($e^+e^-, \mu^+\mu^-$)
in Fig.(\ref{kinmass}).
We applied the default post trigger cuts as in Sec.(\ref{D1}) to suppress
the SM background. As a comparison  the dominant Standard Model
$t\bar t$ background is also exhibited. We have cross checked our work with the CMS Note
\cite{CMSnote1}, and found good agreement regarding the SUSY signals and
the Standard Model background.
It is seen that the two mSP points plotted in Fig.(\ref{kinmass}) are easily distinguishable
from each other.

\subsection{A `global' analysis, fuzzy signature vectors, and  pattern discrimination \label{D7} }

In the above we have given specific examples of how  patterns can be
differentiated  from each other.
In the  previous sections we
used only  a few of the 40 signatures exhibited in  Table
(\ref{tab:counting}). However, in the analysis we have  carried out we have
examined all of them.
Thus for each parameter point we have analyzed 40 signatures. We
now define correlations among these  signatures.  Thus consider an
ordered set where the signatures are labeled $S_1, S_2, .., S_{40}$
and let the number of events in each  signature be
$N_1, N_2, .., N_{40}$. Define a signature vector for a given point
$x_{\alpha}$ ($\alpha =1, 2, .., p$) in the parameter space
\begin{equation}
 \xi^a=(\xi_1^a, \xi_2^a, .., \xi_{40}^a)
\end{equation}
where $\xi_i=N_i^a/N$ and $N$ is the total number of SUSY events.
As the parameter point $x_{\alpha}$ varies  over the allowed range
within a given pattern it generates a signature vector where the
elements trace out a given range. Thus  for a pattern X one generates
a  fuzzy pattern vector $\Delta\xi^X$ so that
\beqn
\Delta \xi^X=(\Delta\xi_1^X, \Delta\xi_2^X, .., \Delta\xi_{40}^X),
\eeqn
where $\Delta \xi_i^X$ is the range traced  out by the element
$\xi_i^X$ as the parameter point $x_{\alpha}$ moves in the allowed
parameter space of the pattern X.
What makes the vector $\Delta \xi^X$ fuzzy is that its elements are not single
numbers but a set which cover a range.
We define now the inner product of
two  such fuzzy pattern vectors  so that
 \beqn C_{XY}\equiv (\Delta\xi^X|\Delta\xi^Y)=0(1)
\label{inner}
\eeqn
where the inner product is 0 if the element
$\Delta\xi_i^X$ and $\Delta\xi_i^Y$ overlap for all $i$  $( i=1,.,40)$,
and 1 if at least one of the elements of pattern X, $\Delta \xi_j^X$
does not overlap with $\Delta\xi_j^Y$,  the element for pattern Y.
 Therefore,  if for two patterns
X and Y one finds there is no overlap at least for one signature component
$\Delta\xi_j$, then these two patterns can be distinguished in this
specific signature and one  obtains  $C_{XY}=1$. Otherwise $C_{XY}=0$
which means that all components of
$\Delta\xi^X$  and $\Delta\xi^Y$  have an overlap  and cannot be distinguished under
this critera.
%%%%%%%%%%%
\begin{figure*}[htb]
\begin{center}
\includegraphics[width=16.0cm,height=8.0cm]{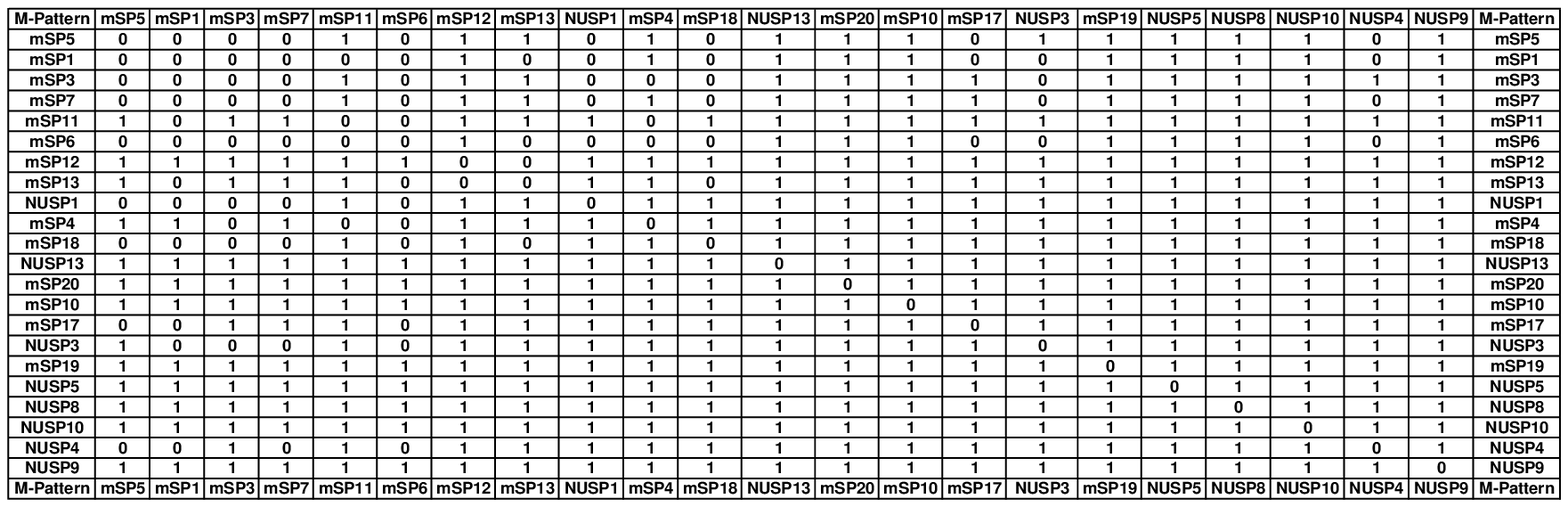}
\caption[]{
A table exhibiting the discrimination of patterns using the
criterion of  Eq.(\ref{inner})
where various signatures with both the default post trigger cuts and b jet cuts are utilized.
If the element of $i^{\rm th}$ row and $j^{\rm th}$ column is 1, i.e., $C_{i j} = 1$,
one can distinguish the $i^{\rm th}$ mass pattern from the $j^{\rm th}$ one.
}\label{grid}
\end{center}
\end{figure*}
%%%%%%%%%%%%
We can generalize the above procedure for the signatures
\begin{equation}
\zeta_{i,j}=\frac{N_i}{N_j}, ~~ (i,j=1,...,40). \label{tensor}
\end{equation}
Repeating the previous analysis, one can construct another fuzzy signature
vector for pattern X as
\begin{equation}
\Delta\zeta^X=(\Delta\zeta_{1,2}^X, ..,\Delta\zeta_{i,j}^X, ..,
\Delta\zeta_{39,40}^X)
\label{tensor1}
\end{equation}
where the elements have a range corresponding to the range spanned by
the soft parameters $x_{\alpha}$ as they move over the parameter space
specific to the pattern.
Further, the definition of the inner product Eq.~(\ref{inner}) still holds
for this new fuzzy signature vector.
We have carried out a full signature analysis of such comparisons, using 40
different signatures, and their combinations as defined in
Eq.~(\ref{tensor}) and Eq.~(\ref{tensor1}).
An illustration of the global analysis is given in Fig.(\ref{grid}). The analysis shows that
it is possible to often distinguish patterns using the criterion of Eq.(\ref{inner}).
We note that
the analyses exhibited in Fig.(\ref{fig:sugrafig}) are the special cases of the results in Fig.(\ref{grid}).
For instance, the clear separation between mSP7 and mSP11 in the signature
space shown in the top-left panel of Fig.(\ref{fig:sugrafig}) gives the
elements $C_{45}=C_{54}=1$ of Fig.(\ref{grid}).
As emphasized already
 the analysis of Fig.(\ref{grid})  is for illustrative purposes as
  we used a random sample of 22 patterns out of 37.
 Inclusion of each additional mass pattern brings in a significant set of model
 points which need to be simulated, and here
 one is limited by   computing power.
  The full analysis  including all the patterns can
 be implemented along similar lines with the necessary
 computing  power.
Finally we note that the analysis in Fig.(\ref{grid}) is done without statistical uncertainties.
Inclusion of uncertainties in pattern analysis would certainly be worthwhile in a future work.

%%% ----------------- Degeneracy & Resolution ------------------

\section{Signature Degeneracies and Resolution of Soft Parameters \label{E}}

\subsection{Lifting signature degeneracies  \label{E1}}
\begin{table}[htb]\scriptsize{
\begin{center}
\begin{tabular}{|l|l||r|r|r|r|r|r||r|r|r|r|r|r|}\hline\hline
$i$ &   $S_i$   &   $A$ &   $B$ &   $P_i$   &   $A'$    &   $B'$ &
$P_i$   &   $A$ &   $B$ &   $P_i$   &   $A'$    &   $B'$    & $P_i$
\\\hline 0   &   N   &   743 &   730 &   0.3 &   878 &   817 &   1.2
&   35947   &   35948   &   0.0 &   45479   &   41135   & 12.1
\\\hline 1   &   0L  &   430 &   414 &   0.4 &   484 &   437 &   1.3
&   20771   &   20592   &   0.7 &   25897   &   23427   & 9.1
\\\hline 2   &   1L  &   221 &   230 &   0.3 &   294 &   271 & 0.8 &
10641   &   10676   &   0.2 &   13669   &   12414   &   6.3
\\\hline 3   &   2L  &   78  &   71  &   0.5 &   83  &   96  &   0.8
&   3745    &   3933    &   1.8 &   4904    &   4369    &   4.5
\\\hline 4   &   3L  &   10  &   13  &   0.5 &   16  &   11  &   0.8
&   703 &   691 &   0.3 &   927 &   830 &   1.9 \\\hline 5   &   4L
&   4   &   2   &   0.6 &   1   &   2   &   0.4 &   87  &   56  &
2.1 &   82  &   95  &   0.8 \\\hline 6   &   0T  &   620 &   610 &
0.2 &   731 &   674 &   1.2 &   29533   &   30220   &   2.3 & 38213
&   34138   &   12.4    \\\hline 7   &   1T  &   112 &   104 &   0.4
&   137 &   129 &   0.4 &   5722    &   5140    &   4.6 & 6528    &
6296    &   1.7 \\\hline 8   &   2T  &   11  &   14  & 0.5 &   10  &
14  &   0.7 &   643 &   541 &   2.4 &   693 &   659 &   0.8 \\\hline
9   &   3T  &   0   &   2   &   1.1 &   0   &   0 &   0.0 &   44  &
45  &   0.1 &   43  &   40  &   0.3 \\\hline 10 &   4T  &   0   &
0   &   0.0 &   0   &   0   &   0.0 &   5   & 2   &   0.9 &   2   &
2   &   0.0 \\\hline 11  &   TL  &   38  & 26  &   1.2 &   50  &
45  &   0.4 &   1779    &   1597    &   2.6 &   2069    &   2029
&   0.5 \\\hline 12  &   OS  &   59  &   57 &   0.2 &   66  &   70
&   0.3 &   2755    &   2927    &   1.9 & 3665    &   3285    &
3.7 \\\hline 13  &   SS  &   19  &   14  & 0.7 &   17  &   26  &
1.1 &   990 &   1006    &   0.3 &   1239 &   1084    &   2.6
\\\hline 14  &   OSSF    &   40  &   46  &   0.5 &   49  &   52  &
0.2 &   2023    &   2112    &   1.1 &   2710 &   2389    &   3.7
\\\hline 15  &   SSSF    &   7   &   9   &   0.4 &   10  &   13  &
0.5 &   458 &   452 &   0.2 &   537 &   481 & 1.4 \\\hline 16  &
OST &   7   &   8   &   0.2 &   5   &   9   & 0.9 &   400 &   345 &
1.6 &   428 &   402 &   0.7 \\\hline 17  & SST &   4   &   6   &
0.5 &   5   &   5   &   0.0 &   243 &   196 &   1.8 &   265 &   257
&   0.3 \\\hline 18  &   0L1b    &   50  & 59  &   0.7 &   61  &
56  &   0.4 &   2586    &   2710    &   1.4 &   3527    &   3387
&   1.4 \\\hline 19  &   1L1b    &   45  & 39  &   0.5 &   48  &
53  &   0.4 &   1767    &   1799    &   0.4 &   2431    &   2268
&   1.9 \\\hline 20  &   2L1b    &   9   & 8   &   0.2 &   15  &
21  &   0.8 &   648 &   674 &   0.6 &   853 &   778 &   1.5 \\\hline
21  &   0T1b    &   86  &   88  &   0.1 & 100 &   110 &   0.6 &
4099    &   4357    &   2.3 &   5734    & 5353    &   3.0 \\\hline
22  &   1T1b    &   21  &   15  &   0.8 & 22  &   20  &   0.3 &
923 &   836 &   1.7 &   1150    &   1106 &   0.8 \\\hline 23  &
2T1b    &   3   &   3   &   0.0 &   4   & 2   &   0.6 &   115 &
109 &   0.3 &   111 &   129 &   0.9 \\\hline 24  &   0L2b    &   20
&   20  &   0.0 &   12  &   13  &   0.2 & 608 &   685 &   1.7 &
890 &   838 &   1.0 \\\hline 25  &   1L2b &   11  &   12  &   0.2 &
15  &   24  &   1.2 &   476 &   474 & 0.1 &   625 &   598 &   0.6
\\\hline 26  &   2L2b    &   3   &   5 &   0.6 &   1   &   2   &
0.4 &   175 &   190 &   0.6 &   251 & 227 &   0.9 \\\hline 27  &
0T2b    &   30  &   29  &   0.1 &   25 &   32  &   0.8 &   1027    &
1115    &   1.6 &   1481    &   1379 &   1.6 \\\hline 28  &   1T2b
&   4   &   6   &   0.5 &   4   & 6   &   0.5 &   242 &   234 &
0.3 &   300 &   297 &   0.1 \\\hline 29  &   2T2b    &   0   &   2
&   1.1 &   0   &   1   &   0.7 & 30  &   40  &   1.0 &   28  &   27
&   0.1 \\\hline 30  &   ep  & 71  &   71  &   0.0 &   93  &   83  &
0.6 &   3240    &   3219 &   0.2 &   4251    &   3957    &   2.6
\\\hline 31  &   em  &   47 &   44  &   0.3 &   52  &   51  &   0.1
&   2055    &   1994    & 0.8 &   2618    &   2358    &   3.0
\\\hline 32  &   mp  &   60  & 70  &   0.7 &   103 &   78  &   1.5 &
3338    &   3442    &   1.0 &   4236    &   3821    &   3.8 \\\hline
33  &   mm  &   43  &   45 &   0.2 &   46  &   59  &   1.0 &   2008
&   2021    &   0.2 & 2564    &   2278    &   3.4 \\\hline 34  &
tp  &   60  &   53  & 0.5 &   69  &   80  &   0.7 &   3203    &
2803    &   4.2 &   3564 &   3504    &   0.6 \\\hline 35  &   tm  &
52  &   51  &   0.1 & 68  &   49  &   1.4 &   2519    &   2337    &
2.1 &   2964    & 2792    &   1.9 \\\hline 36  &   0b  &   597 &
585 &   0.3 &   717 &   642 &   1.7 &   29276   &   29072   &   0.7
&   36432   & 32602   &   11.9    \\\hline 37  &   1b  &   110 &
107 &   0.2 & 126 &   132 &   0.3 &   5150    &   5314    &   1.3 &
7003    & 6593    &   2.9 \\\hline 38  &   2b  &   34  &   37  &
0.3 &   29 &   39  &   1.0 &   1302    &   1389    &   1.4 &   1810
&   1706 &   1.4 \\\hline 39  &   3b  &   1   &   1   &   0.0 &   6
&   2 &   1.1 &   192 &   153 &   1.7 &   215 &   205 &   0.4
\\\hline 40 &   4b  &   1   &   0   &   0.7 &   0   &   2   &   1.1
&   27  & 20  &   0.8 &   19  &   29  &   1.2 \\\hline \hline
\end{tabular}
\caption{ An exhibition of lifting the degeneracy of two points in
the mSUGRA parameter space using luminosity. Two pairs of points
($A$, $B$) and ($A'$, $B'$) are indistinguishable under the 2 sigma
criteria at 10 fb$^{-1}$ luminosity (column 3-8), but can be clearly
separated when the luminosity increases to 500 fb$^{-1}$ (column
9-14).  The Standard Model uncertainty is estimated as $\delta
n_i^{SM}= (\delta n_i^{A}+\delta n_i^{B})/2$. } \label{tbl:degelum}
\end{center}}
 \end{table}

It may happen that two distinct points in the soft parameter space
may lead to the same  set of signatures for a given integrated
luminosity within some predefined notion of indistinguishability.
Thus consider two parameter points $A$ and $B$ and define
the `pulls' in each of their signatures by
\begin{eqnarray}
P_i & = & \frac{|n^A_i-n^B_i|}{\sigma_{AB}},\nonumber\\
\sigma_{AB} & = & \sqrt{(\delta n^A_i)^2+ (\delta n^B_i)^2+ (\delta
n_i^{SM})^2}. \label{pulls1}
 \end{eqnarray}\\
Here $\delta n_i^A\sim\sqrt{n_i^A}$ is the uncertainty in the signature
events $n_i^A$, and we estimate the SM uncertainty as $\delta
n_i^{SM}\sim\sqrt{y}(\delta n_i^A+\delta n_i^B)/2$. Here the
parameter $y$ parameterizes the effect of the SM events, and for the analysis
in this section, we take $y=1$. In other words, if the pulls in each of the
signatures is less than 5, then the two SUGRA parameter space points are
essentially indistinguishable in the signature space. In such a situation
one could still distinguish model points  either by including more signatures,
or by an increase in luminosity. Thus, for example,
inclusion of the Higgs production cross sections, $B_s\to \mu^+\mu^-$ constraints,
as well as the inclusion of neutralino proton scattering cross sections constraints
tend to discriminate among the model parameter points as shown in
Ref.~\cite{Feldman:2007fq}. Here we point out that in some cases
increasing the luminosity can allow one to lift the degeneracies
enhancing a subset of signatures in one case relative to the other.
For illustration we consider the following two sets of
points in the pattern mSP5 in the mSUGRA parameter space in the
following order ($m_0$, $m_{1/2}$, $A_0$, $\tan\beta$, sign$\mu$).
\begin{equation}
\begin{array}{rl}
{\rm Point ~A} & (192.6, 771.3, ~1791.1, 8.8, +1),\\
{\rm Point ~B} & (163.0, 761.3, -775.8, 4.7, +1);
\end{array}
\end{equation}
\begin{equation}
\begin{array}{rl}
{\rm Point ~A'} & (159.3, 732.3, -783.1, 5.6, +1),\\
{\rm Point ~B'} & (163.5, 753.3, -918.2, 3.3, +1).
\end{array}
\end{equation}
In  Table(\ref{tbl:degelum}) we compare the pulls for the pairs of
points ($A$, $B$) and ($A'$, $B'$) at an integrated luminosity of 10
fb$^{-1}$ and 500 fb$^{-1}$. For points $A$ and $B$, one finds that
the pulls are all less than 2 for an integrated  luminosity of 10
fb$^{-1}$. However,  for an integrated luminosity of 500 fb$^{-1}$,
the  pulls for signatures $(6,7,8,11,21,34)$ increase significantly
and the pull for signature number 7 is in excess of 4.5 allowing one
to discriminate between the two parameter points $A$ and $B$. A very
similar analysis is carried out for parameter points $A'$ and $B'$.
Here one finds that the signature $(0, 1, 2, 3, 6, 12, 14, 32, 33,
36)$ receive a big boost as we go from 10 fb$^{-1}$ to 500
fb$^{-1}$, and the signatures $(0, 1, 2, 6, 36)$ give pulls greater
than 5, with the largest pulls being in excess of 12, allowing one
to discriminate between the parameter points $A'$ and $B'$.
 We note the analysis ignores
systematic errors and also does not consider an ensemble of simulations.
Nonetheless it  does  illustrate the effects of moving from
a  low to a high LHC luminosity
allowing one to discriminate some model pairs, which appear degenerate in
the signature space at one luminosity, but can become distinct from each other
at a larger luminosity.

\subsection{Resolving soft parameters using LHC data\label{E2}}

%%% ---  resolution figures ---
\begin{figure*}[htb]
\centering
\includegraphics[width=7.0cm,height=6.0cm]{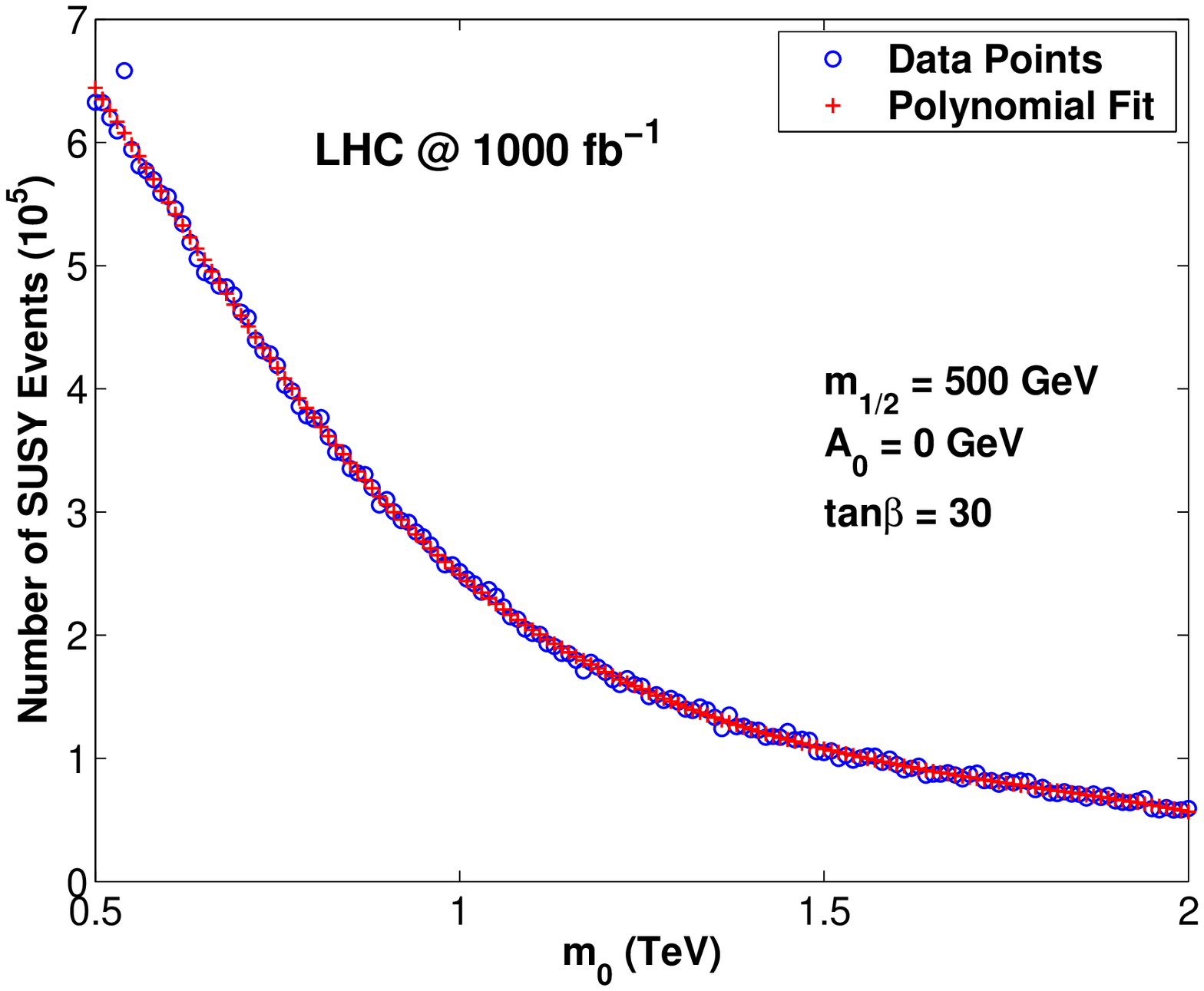}
\includegraphics[width=7.0cm,height=5.8cm]{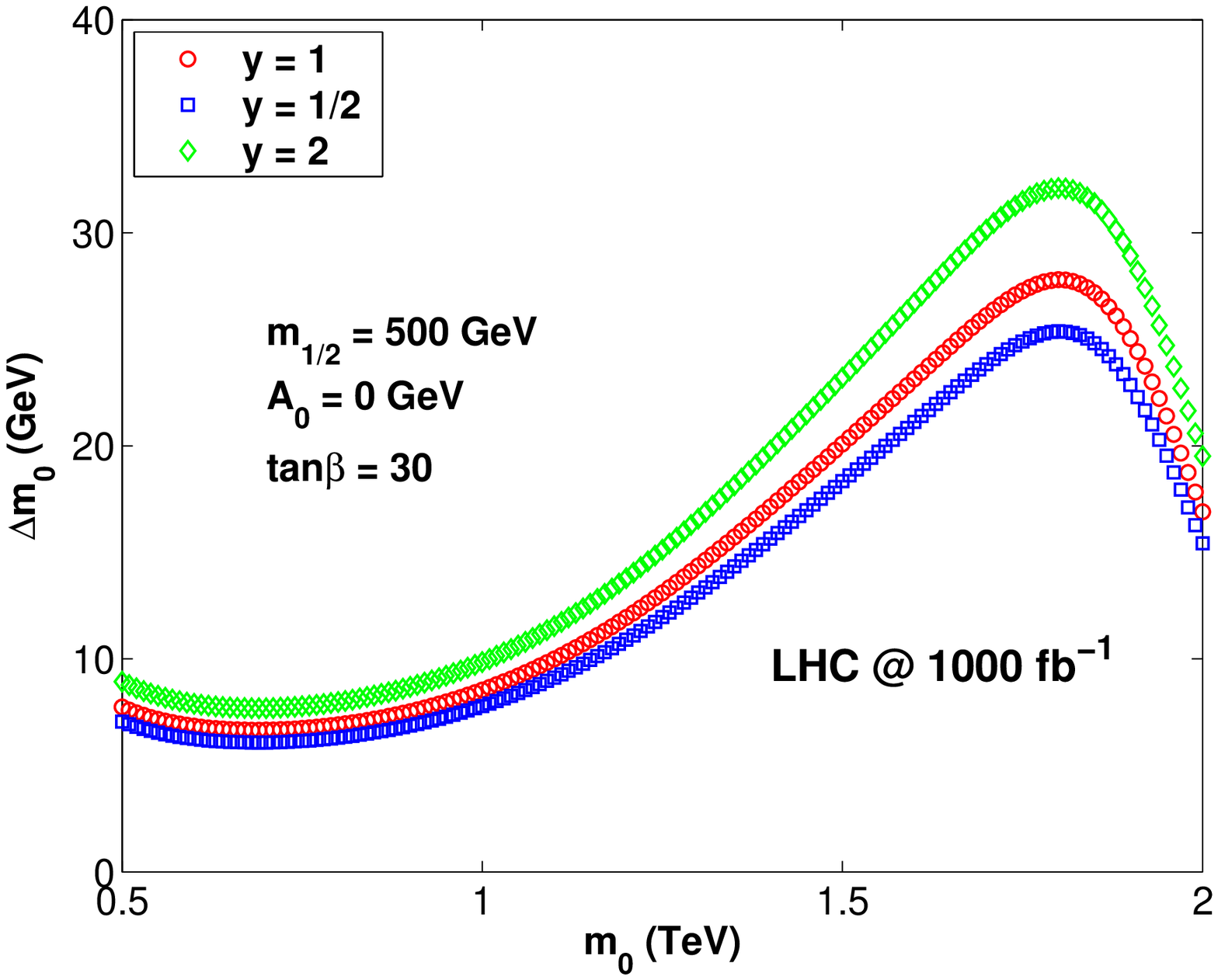}
\includegraphics[width=7.0cm,height=6.0cm]{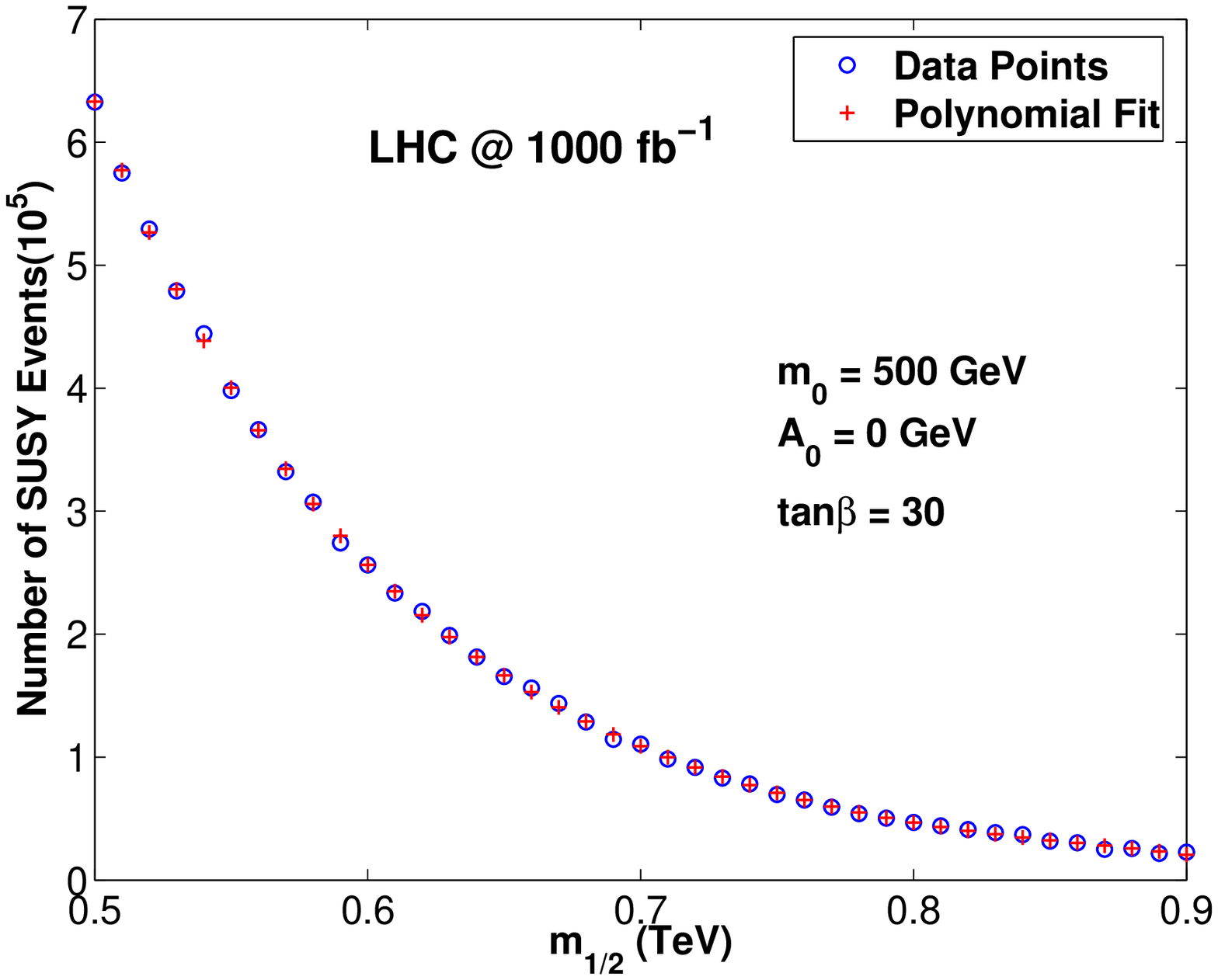}
\includegraphics[width=7.0cm,height=5.8cm]{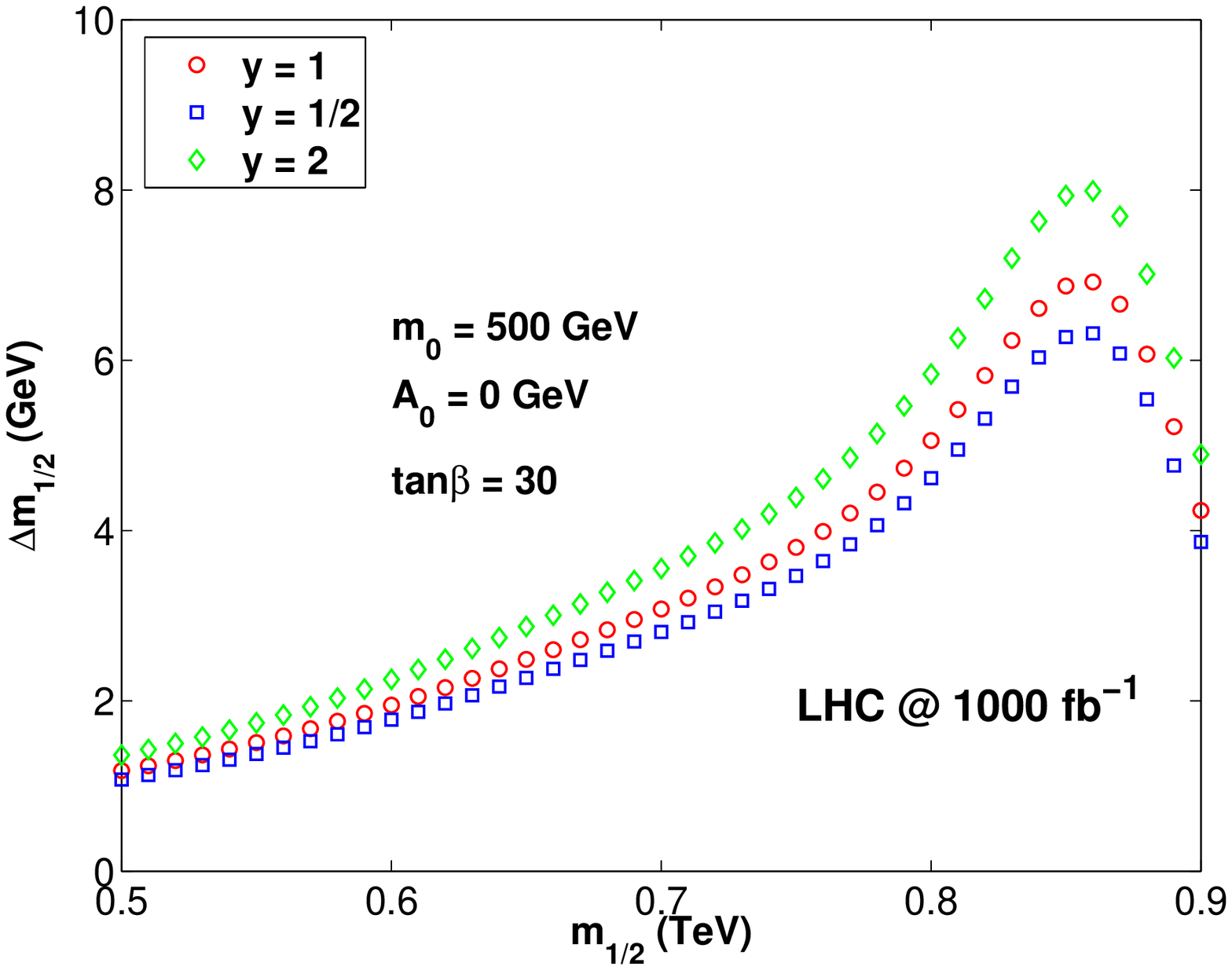}
\caption{An analysis showing the resolutions in $m_0$ and $m_{1/2}$
that can be reached with 1000 fb$^{-1}$ of integrated luminosity
under the REWSB constraints.
The two left panels give the number of SUSY events vs
$m_0$ (top left panel) and vs $m_{1/2}$ (lower left panel) for 1000
fb$^{-1}$ of integrated luminosity. The right panels give the
resolutions in $m_0$ (top right panel) and in $m_{1/2}$ (lower right
panel) using the left panels.
}
\label{fig:resolution}
\end{figure*}

We discuss now the issue of how well  we can resolve the points in
the parameter space $x_{\alpha}$ ($\alpha =1,..,p$) for a given
luminosity.
Consider Eq.(\ref{pulls1}) and set  $\delta N=\sqrt N$, and parameterize
the standard  model uncertainty  by $\delta N^{SM}=\sqrt y \delta N$.
Next we set the criterion for the resolution of two
adjacent points in the SUGRA parameter space separated by $\Delta x_{\alpha} $ so that
the separation in the signature space satisfies
\begin{equation}
\frac{\Delta N}{\sqrt{2N+yN}}= 5.
\end{equation}
Since $N=\sigma_{\rm susy}(x_{\alpha}){\cal{L}}_{\rm LHC}$, where $\sigma_{\rm susy}$ is the cross section
for the production of sparticles, and
${\cal{L}}_{\rm LHC}$
is the LHC integrated luminosity,
  the resolution achievable
   in the vicinity of SUGRA parameter point $x_{\alpha}$  at that luminosity
   is given by
\begin{equation}
\Delta x_{\alpha} = \frac{5}{2}{(2+y)}^{1/2}
{\cal{L}}_{\rm LHC}^{-1/2}
\left(\frac{\partial \sigma^{1/2}_{\rm susy}(x)}{\partial
x_{\alpha}}\right)^{-1}. \label{resolve}
\end{equation}
In Fig.(\ref{fig:resolution}) we give an illustration of the above
 when $m_0$ varies between
500 GeV and 2000 GeV while $m_{1/2}=500$ GeV, $A_0=0, \tan\beta=30$,
and $\mu>0$. From Fig.(\ref{fig:resolution}) one finds that the
resolution in $m_0$ strongly depends on the point in the parameter
space and on the luminosity. Quite interestingly a resolution as
small as  a few GeV can be achieved for $m_0$ in the  range 500-1000
GeV with 1000 fb$^{-1}$ of integrated luminosity. A similar analysis
varying  $m_{1/2}$ in the range 500-900 GeV for the case when
$m_{0}=500$ GeV, $A_0=0, \tan\beta=30$ and $\mu>0$, shows that  a
resolution in $m_{1/2}$ as low as 1 GeV can be achieved with 1000
fb$^{-1}$ of integrated luminosity.

%%% ----------- Conclusion -------------------

\section{Conclusion   \label{F}}

The minimal supersymmetric Standard Model has 32 sparticle masses.
Since the soft breaking sector MSSM is arbitrary, one is led to a
landscape of as many as $10^{28}$ or more possibilities for the
sparticle mass hierarchies. The number of  possibilities is
drastically reduced in well motivated models such as supergravity
models, and one expects similar reductions to occur also in
 gauge and  anomaly mediated models, and in string and brane
models.  In this work we have analyzed the  mass hierarchies for the
first four lightest sparticle (aside from the lightest Higgs boson)
for supergravity models.
  Specifically,  in  Sec.(\ref{A}) we analyzed the
mass hierarchies for the
mSUGRA model and for supergravity models with \non in
the soft breaking in the Higgs sector, \non in the soft breaking in
the third generation sector, and \non in the soft breaking in the
gaugino sector.
 It is found that in each case only a small number of mass  hierarchies or patterns survive the rigorous
 constraints of radiative breaking of the electroweak symmetry, relic density constraints on cold dark matter
 from the WMAP data, and other experimental constraints from colliders.  These mass hierarchies can be
 conveniently put into different classes labeled by the sparticle which is next heavier
 after the LSP. For
 the SUGRA models we find six different classes: chargino patterns, stau patterns, stop patterns, Higgs
 patterns, neutralino patterns, and gluino patterns.  Benchmarks for each of these  patterns were given
 in  Sec.(\ref{B}).
  In Sec.(\ref{D})  we discussed the techniques for the analysis of the  signatures and the
 technical details on simulations of sparticle events.
 In this section we also discuss the backgrounds to
 the SUSY phenomena arising from the Standard Model processes.
 Additionally we discussed here the  identification of patterns based on 40 event
 identification criteria listed in Fig.(\ref{grid}). It is found that these  criteria allow one to discriminate
 among most of the patterns.
An analysis of how one may lift degeneracies in the signature space, and how accurately
one can determine the soft parameters using the LHC luminosities is given in Sec.(\ref{E}).
It is hoped that the analysis of the type discussed here would help not only in the search
for supersymmetry but also allow one to use the signatures to extrapolate back
to the underlying supersymmetric model using the experimental data when such data
 from the LHC comes in.
  In the above our analysis was focused on supergravity unified models. However, the
 techniques discussed here have a much wider applicability to other models, including models
 based on  gauge and anomaly mediated breaking, as well as  string and brane based models.

\section{Acknowledgements}
This work is supported in part by NSF grant  PHY-0456568.
Some of the results given here were presented  at the workshop ``LHC New Physics Signatures
Workshop'', at the Michigan Center for Theoretical Physics, January 5-11, 2008, and at the
conference ``From Strings to LHC-II'', Bangalore,
December 19-23, 2007.
The authors have benefited from interactions at these workshops.

\clearpage \section{Appendix: Benchmarks for sparticle mass
hierarchies  \label{AAA}}

%%-------------- TABLEs Bench --------------
%%

\begin{table}[htbp]
    \begin{center}
   {Chargino  Patterns (CPs)}
       \scriptsize{
\begin{tabular}{|c|c|c|c|c|c|c|c|c|c}                                                                  \hline  \hline \hline
 {\bf  \rm SUGRA}    &   $m_0$   &   $m_{1/2}$   &   $A_{0}$ &   $\tan {\beta}$  &   $\mu$   &   NUH &   NU3 &   NUG \\
{\bf  \rm Pattern}   &   (GeV)   &   (GeV)   &   (GeV)       &   ($v_u/v_d$)        & (sign)   & $(\delta_{H_u},\delta_{H_d})$ & $(\delta_{q3},\delta_{tbR})$ & $(\delta_{M_2},\delta_{M_3})$   \\
\hline  \hline
 ${\bf mSP1 }$  &   2001    &   411 &   0   &   30.0    &   +   &   (0,0)   &   (0,0)   &   (0,0)   \\ \hline
 ${\bf mSP1 }$  &   2366    &   338 &   -159    &   9.8 &   -   &   (0,0)   &   (0,0)   &   (0,0)   \\ \hline
 ${\bf mSP1 }$  &   1872    &   327 &   -1893   &   14.9    &   +   &   (0.107,0.643)   &   (0,0)   &   (0,0)   \\ \hline
 ${\bf mSP1 }$  &   1041    &   703 &   1022    &   11.6    &   +   &   (0,0)   &   (-0.524,-0.198) &   (0,0)   \\ \hline
 ${\bf mSP1 }$  &   1361    &   109 &   1058    &   14.4    &   +   &   (0,0)   &   (0,0)   &   (0.929,0.850)   \\ \hline \hline
 ${\bf mSP2 }$  &   1125    &   614 &   2000    &   50.0    &   +   &   (0,0)   &   (0,0)   &   (0,0)   \\ \hline
 ${\bf mSP2 }$  &   2365    &   1395    &   3663    &   42.2    &   -   &   (0,0)   &   (0,0)   &   (0,0)   \\ \hline
 ${\bf mSP2 }$  &   1365    &   595 &   3012    &   35.1    &   +   &   (0.116,-0.338)  &   (0,0)   &   (0,0)   \\ \hline
 ${\bf mSP2 }$  &   1166    &   507 &   -954    &   59.6    &   +   &   (0,0)   &   (0.325,0.458) &   (0,0)   \\ \hline
 ${\bf mSP2 }$  &   1414    &   221 &   -551    &   54.3    &   +   &   (0,0)   &   (0,0)   &   (0.156,0.968)   \\ \hline \hline
 ${\bf mSP3 }$  &   741 &   551 &   0   &   50.0    &   +   &   (0,0)   &   (0,0)   &   (0,0)   \\ \hline
 ${\bf mSP3 }$  &   1585    &   1470    &   3133    &   39.1    &   -   &   (0,0)   &   (0,0)   &   (0,0)   \\ \hline
 ${\bf mSP3 }$  &   694 &   674 &   -1564   &   27.0    &   +   &   (0.922,-0.293)  &   (0,0)   &   (0,0)   \\ \hline
 ${\bf mSP3}$   &   570 &   559 &   1042    &   41.3    &   +   &   (0,0)   &   (-0.482,-0.202) &   (0,0)   \\ \hline
 ${\bf mSP3 }$  &   392 &   312 &   320 &   41.3    &   +   &   (0,0)   &   (0,0)   &   (-0.404,0.908)  \\ \hline \hline
 ${\bf mSP4 }$  &   1674    &   137 &   1985    &   18.6    &   +   &   (0,0)   &   (0,0)   &   (0,0)   \\\hline
 ${\bf mSP4 }$  &   1824    &   127 &   -1828   &   6.4 &   -   &   (0,0)   &   (0,0)   &   (0,0)   \\ \hline
 ${\bf mSP4 }$  &   1021    &   132 &   -638    &   6.6 &   +   &   (0,0)   &   (-0.020,0.963)  &   (0,0)   \\ \hline
 ${\bf mSP4 }$  &   2181    &   127 &   -3859   &   3.9 &   +   &   (0,0)   &   (0,0)   &   (0.836,-0.248)  \\ \hline \hline
${\bf NUSP1}$   &   2738    &   1689    &   -4243   &   42.4    & + & (0,0)   &   (-0.828,-0.899) &   (0,0)   \\ \hline
${\bf NUSP1}$   &   540 &   1190    &   2516    &   13.9    &   +   & (0,0)   & (0,0) & (-0.408,-0.660) \\ \hline
${\bf NUSP2}$   &   845 &   726 &   -75 & 48.4    &   +   &   (0,0)   & (-0.694,-0.400) &   (0,0) \\ \hline
${\bf NUSP3}$  &   396 &   1018    &   -179    & 18.3    &   +   &   (0,0)   &   (0,0) &   (0.250,-0.452)  \\ \hline
${\bf NUSP4}$   &   400 &   1558    & 2511    &   5.9 &   +   & (0,0)   &   (0,0)   &   (-0.401,-0.607)
\\ \hline
                                                                    \hline \hline
\end{tabular}
}
 \caption[]{ Benchmarks for the class CP where the chargino $\cha$ is the NLSP
in mSUGRA and  in NUSUGRA  models. Benchmarks are computed with ${m_b}^{\overline{\rm MS}}(m_b) = 4.23$ {\rm GeV},
${\alpha_s}^{\overline{\rm MS}}(M_Z)=.1172$, and $m_t({\rm pole}) =
170.9$ ${\rm GeV}$ with SuSpect
2.34 interfaced to micrOMEGAS 2.07. } \label{b1}
    \end{center}
 \end{table}

\begin{table}[h]
\begin{center}
  {Gluino  Patterns (GPs)}
    \small{
\begin{tabular}{|c|c|c|c|c|c|c|c|c|}\hline \hline\hline
   {\bf  \rm SUGRA}     &   $m_0$   &   $m_{1/2}$   &   $A_{0}$ &   $\tan {\beta}$  &   $\mu$   &   NUH &   NU3 &   NUG \\
   {\bf  \rm Pattern}    &   (GeV)   &   (GeV)   &   (GeV)   &   ($v_u/v_d$)   &   (sign)  &   $(\delta_{H_u},\delta_{H_d})$   &   $(\delta_{q3},\delta_{tbR})$   &   $(\delta_{M_2},\delta_{M_3})$   \\ \hline  \hline
 {\bf NUSP13}    &  2006    &   1081 &   -2027    &   21.1   &   +   &   (0,0)   &   (0,0)   &   (0.207,-0.844)  \\ \hline
  {\bf NUSP14}    &   3969    &   1449    &   -6806   &   29.3   &   +   &   (0,0)   &   (0,0)   &   (0.611,-0.834)   \\ \hline
 {\bf NUSP15}    &   1387    &   695 &   2781    &   50.5   &   +   &   (0,0)   &   (0,0)   &   (0.136,-0.827)  \\ \hline
                                                                        \hline \hline
\end{tabular}
} \caption[]{Benchmarks for the class GP where the  gluino $~\g  $ is
the NLSP.  Such a pattern was only seen to appear in NUSUGRA models with non universal
gaugino masses.
} \label{b4}
    \end{center}
 \end{table}

% ---------------------------- TABLE of benchmarks  ----------------------------

\begin{table}[htbp]
\begin{center}
  {Stau  Patterns (SUPs)}
%\end{center}    \begin{center}
   \scriptsize{
\begin{tabular}{|c|c|c|c|c|c|c|c|c|}                                                                  \hline  \hline \hline
{\bf  \rm SUGRA}   &   $m_0$   &   $m_{1/2}$   &   $A_{0}$ &   $\tan {\beta}$  &   $\mu$   &   NUH &   NU3 &   NUG \\
{\bf  \rm Pattern}  &   (GeV)   &   (GeV)   &   (GeV)   &   ($v_u/v_d$) & (sign)  &   $(\delta_{H_u},\delta_{H_d})$   & $(\delta_{q3},\delta_{tbR})$ & $(\delta_{M_2},\delta_{M_3})$   \\
\hline  \hline
${\bf mSP5}$    & 111 & 531 &   0   &   5.0 &   +   &   (0,0)   &   (0,0)   & (0,0)   \\ \hline
${\bf mSP5}$    &   162 &   569 &   1012    & 15.8    &   - & (0,0)   &   (0,0)   &   (0,0)   \\ \hline
${\bf mSP5}$    & 191 & 545 &   -722    &   17.2    &   +   & (-0.340,-0.332) & (0,0)   & (0,0)   \\ \hline
${\bf mSP5}$    & 114 &   440 &   -50 &   15.2    &   +   &   (0,0)   & (-0.204,-0.846) &   (0,0)   \\ \hline
${\bf mSP5}$    &   75  & 348 &   301 &   12.0    &   +   & (0,0)   &   (0,0)   & (0.234,-0.059)  \\ \hline\hline
${\bf mSP6}$ & 245 &   370 & 945 &   31.0    &   +   &   (0,0)   &   (0,0)   &   (0,0)   \\ \hline
${\bf mSP6}$    &   1452    &   1651    &   2821    &   38.5 &   -   &   (0,0)   &   (0,0)   &   (0,0)   \\ \hline
${\bf mSP6}$ & 356 &   545 &   927 &   31.7    &   +   &   (0.667,0.055)   & (0,0) &   (0,0)   \\ \hline
${\bf mSP6}$    &   442 &   463 & 1150    & 41.0    &   +   &   (0,0)   &   (-0.187,-0.546) & (0,0)   \\ \hline
${\bf mSP6}$    &   308 &   307 &   965 &   35.6 &   +   &   (0,0) & (0,0)   &   (-0.383,0.405)  \\ \hline\hline
${\bf mSP7}$    & 75  & 201 &   230 &   14.0    &   +   & (0,0)   &   (0,0)   & (0,0)   \\ \hline
${\bf mSP7}$    &   781 & 1423    &   983 &   36.8 &   -   & (0,0)   &   (0,0)   & (0,0)   \\\hline
${\bf mSP7}$ &   428 &   671 &   484 &   43.8 &   +   &   (-0.392,-0.808) & (0,0)   &   (0,0) \\ \hline
${\bf mSP7}$    &   226 &   426 & 944 &   27.1    &   +   &   (0,0)   & (0.176,-0.430)  &   (0,0)   \\ \hline
${\bf mSP7}$    &   143 & 425 &   266 &   23.4    &   +   & (0,0)   &   (0,0)   & (0.718,0.100)   \\ \hline\hline
${\bf mSP8}$ & 1880    &   877 &   4075    &   54.8    &   +   &   (0,0)   & (0,0) &   (0,0) \\ \hline
${\bf mSP8}$    &   994 &   1073    &   3761    &   38.1 &   -   &   (0,0)   &   (0,0)   &   (0,0)   \\ \hline
${\bf mSP8}$ & 602 &   684 &   805 &   49.6    &   +   &   (0.490,0.326)   & (0,0) &   (0,0)   \\ \hline
${\bf mSP8}$    &   470 &   624 & -88 &   55.4 &   +   &   (0,0)   &   (-0.531,-0.075) &   (0,0)  \\ \hline
${\bf mSP8}$    &   525 &   450 &   642 &   56.4    &   + &   (0,0)   &   (0,0)   &   (0.623,0.246)   \\ \hline\hline
${\bf mSP9}$    &   667 &   1154    &   -125    &   51.0    &   +   & (0,0)   &   (0,0)   &   (0,0)   \\ \hline
${\bf mSP9}$    &   560 & 1156    &   -1092   &   39.5    &   -   &   (0,0)   &   (0,0)   & (0,0)   \\ \hline
${\bf mSP9}$    &   362 &   602 &   268 &   37.0 & +   &   (0.969,-0.232)  &   (0,0)   &   (0,0)   \\ \hline
${\bf mSP9}$    &   496 &   731 &   679 &   49.3    &   +   &   (0,0)   & (-0.241,-0.452) &   (0,0)   \\ \hline
${\bf mSP9}$    &   485 & 478 &   -128    &   52.8    &   +   &   (0,0)   &   (0,0)   & (0.971,0.653)   \\ \hline\hline
${\bf mSP10}$   &   336 &   772 & -3074   &   10.8    &   +   &   (0,0)   &   (0,0)   &   (0,0)   \\ \hline
${\bf mSP10}$   &   738 &   1150    &   -4893   &   15.5    & +   &   (0,0)   &   (0.802,0.343)   &   (0,0)   \\ \hline\hline
${\bf mSP17}$   &   908 &   754 &   5123    &   25.4    &   -   & (0,0)   &   (0,0)   &   (0,0)   \\ \hline\hline
${\bf mSP18}$   & 344 &   686 &   -2718   &   13.8    &   -   &   (0,0)   &   (0,0) & (0,0)   \\ \hline
${\bf mSP18}$   &   322 &   806 &   -3069   & 9.3 &   +   &   (0.526,-0.707)  &   (0,0)   &   (0,0)   \\ \hline
${\bf mSP18}$   &   60  &   290 &   -339    &   5.2 &   +   & (0,0)   & (0,0)   &   (0.967,-0.074)  \\ \hline\hline
${\bf mSP19}$   &   1530 &   1875    &   13081   &   16.3    &   -   & (0,0)   &   (0,0)   & (0,0)   \\ \hline
${\bf mSP19}$   &   1828 &   1326    &   -5102   & 32.3    &   +   &   (0.592,-0.213)  & (0,0)   &   (0,0)   \\ \hline
${\bf mSP19}$   &   782 &   637 & 2688    &   37.9    &   +   & (0,0)   &   (0,0)   & (0.451,-0.551)  \\ \hline\hline
${\bf NUSP5}$ & 649 &   955 & -1984   &   33.5    &   +   &   (0,0)   & (-0.763,0.701)  & (0,0)   \\ \hline
${\bf NUSP6}$ &   1360    &   1736 &   -2871   & 46.1    &   +   &   (0,0)   &   (-0.466,0.694)  &   (0,0)   \\\hline
${\bf NUSP7}$ &   1481    &   1531    &   -3169   &   42.2    & +   &   (0,0)   &   (0,0)   &   (0.117,-0.463)  \\ \hline
${\bf NUSP8}$ &   670 &   1788    &   371 &   57.9    &   +   &   (0,0)   & (0,0)   &   (-0.223,0.931)  \\ \hline
${\bf NUSP9}$ &   46  &   1938 & -48  &   13.0    &   +   &   (0,0)   &   (0,0)   & (-0.412,-0.650) \\ \hline
                                                                    \hline \hline
\end{tabular}
} \caption[]{ Benchmarks for the class SUP where the stau $\sta$ is
the NLSP  in mSUGRA and in NUSUGRA. } \label{b2}
    \end{center}
 \end{table}

% ---------------------------- TABLE of benchmarks  ----------------------------

\begin{table}[htbp]
\begin{center}
  {Stop Patterns (SOPs)}
%\end{center}    \begin{center}
       \scriptsize{
\begin{tabular}{|c|c|c|c|c|c|c|c|c|}\hline  \hline \hline
 {\bf  \rm SUGRA}   &   $m_0$   &   $m_{1/2}$   &   $A_{0}$ &   $\tan {\beta}$  &   $\mu$   &   NUH &   NU3 &   NUG \\
{\bf  \rm Pattern}   &   (GeV)   &   (GeV)   &   (GeV)   &   ($v_u/v_d$) &
(sign)  &   $(\delta_{H_u},\delta_{H_d})$   &
$(\delta_{q3},\delta_{tbR})$ & $(\delta_{M_2},\delta_{M_3})$   \\
\hline  \hline
 ${\bf mSP11 }$ &   871 &   1031    &   -4355   &   10.0    &   +   &   (0,0)   &   (0,0)   &   (0,0)   \\ \hline
 ${\bf mSP11 }$ &   1653    &   909 &   7574    &   5.9 &   -   &   (0,0)   &   (0,0)   &   (0,0)   \\ \hline
 ${\bf mSP11 }$ &   1391    &   1089    &   8192    &   14.9    &   +   &   (0.470,0.632)   &   (0,0)   &   (0,0)   \\ \hline
 ${\bf mSP11 }$ &   2204    &   933 &   -1144   &   35.6    &   +   &   (0,0)   &   (0.642,-0.400)  &   (0,0)   \\ \hline
 ${\bf mSP11 }$ &   1406    &   1471    &   -2078   &   8.3 &   +   &   (0,0)   &   (0,0)   &   (-0.130,-0.690) \\ \hline \hline
 ${\bf mSP12 }$ &   1371    &   1671    &   -6855   &   10.0    &   +   &   (0,0)   &   (0,0)   &   (0,0)   \\ \hline
 ${\bf mSP12 }$ &   1054    &   1372    &   -5754   &   13.7    &   -   &   (0,0)   &   (0,0)   &   (0,0)   \\ \hline
 ${\bf mSP12 }$ &   915 &   927 &   -3993   &   20.7    &   +   &   (0.078,0.833)   &   (0,0)   &   (0,0)   \\ \hline
 ${\bf mSP12 }$ &   826 &   1016    &   -3926   &   12.8    &   +   &   (0,0)   &   (-0.630,-0.490) &   (0,0)   \\ \hline
 ${\bf mSP12 }$ &   1706    &   1287    &   -4436   &   29.7    &   +   &   (0,0)   &   (0,0)   &   (0.416,-0.260)  \\ \hline \hline
 ${\bf mSP13 }$ &   524 &   800 &   -3315   &   15.0    &   +   &   (0,0)   &   (0,0)   &   (0,0)   \\ \hline
 ${\bf mSP13 }$ &   765 &   1192    &   -4924   &   12.0    &   -   &   (0,0)   &   (0,0)   &   (0,0)   \\ \hline
 ${\bf mSP13 }$ &   1055    &   1601    &   -6365   &   13.6    &   +   &   (0.277,-0.820)  &   (0,0)   &   (0,0)   \\ \hline
 ${\bf mSP13}$  &   1073    &   1664    &   -6528   &   11.6    &   +   &   (0,0)   &   (0.728,0.060)   &   (0,0)   \\ \hline
 ${\bf mSP13 }$ &    540 &   774 &   -2432   &   5.3 &   +   &   (0,0)   &   (0,0)   &   (0.705,-0.201)  \\ \hline \hline
 ${\bf mSP20 }$ &   1754    &   840 &   7385    &   13.3    &   -   &   (0,0)   &   (0,0)   &   (0,0)   \\\hline
 ${\bf mSP21 }$ &   792 &   845 &   6404    &   12.6    &   -   &   (0,0)   &   (0,0)   &   (0,0)   \\ \hline
${\bf NUSP10}$  &   718 &   467 &   1657    &   19.0    &   +   & (0,0)   &   (0,0)   &   (0.023,-0.810)  \\ \hline
                                                                    \hline \hline
\end{tabular}
} \caption[]{Benchmarks for the class SOP where the stop  $\ta$
is the NLSP  in mSUGRA and in NUSUGRA models.  } \label{b3}
    \end{center}
 \end{table}

% ---------------------------- TABLE of benchmarks  ----------------------------

\begin{table}[h]
    \begin{center}
    Higgs Patterns (HPs)
    \scriptsize{
\begin{tabular}{|c|c|c|c|c|c|c|c|c|}
                                                                    \hline  \hline \hline
 {\bf  \rm SUGRA}   &   $m_0$   &   $m_{1/2}$   &   $A_{0}$ &   $\tan {\beta}$  &   $\mu$   &   NUH &   NU3 &   NUG \\
{\bf  \rm Pattern}   &   (GeV)   &   (GeV)   &   (GeV)   &   ($v_u/v_d$) &
(sign)  &   $(\delta_{H_u},\delta_{H_d})$   &
$(\delta_{q3},\delta_{tbR})$ & $(\delta_{M_2},\delta_{M_3})$   \\
\hline  \hline
 ${\bf mSP14 }$ & 1040 & 560 & 450 & 53.5 & + & (0,0)          & (0,0) & (0,0) \\ \hline
 ${\bf mSP14 }$ & 760 & 515 & 2250 & 31.0 & + & (0.255,-0.500) & (0,0) & (0,0) \\ \hline
 ${\bf mSP14 }$ & 740 & 620 & 840   & 53.1 & + & (0,0)     & (-0.530,-0.249) & (0,0)\\ \hline
 ${\bf mSP14 }$ & 1205 & 331 & -710 & 55.0 & + & (0,0) &(0,0) & (0.380,0.250)\\ \hline\hline
 ${\bf mSP15 }$ & 1110 & 760 & 1097 & 51.6 & + & (0,0)          & (0,0) & (0,0) \\ \hline
 ${\bf mSP15 }$ & 1395 & 554 &-175 & 59.2  & + &  (0,0)& (-0.040,0.918) & (0,0)\\ \hline
 ${\bf mSP15 }$ & 905  & 500 & 1460 & 54.8 & + & (0,0) & (0,0)& (-0.350,-0.260)\\ \hline\hline
${\bf mSP16 }$ & 520 & 455 &620 &55.5 & + & (0,0) & (0,0)& (0,0)\\ \hline
 ${\bf mSP16 }$ &   282 &   464 &   67  &   43.2    &   + &   (0.912,-0.529)  &   (0,0)   &   (0,0)   \\ \hline
${\bf NUSP12}$  &   2413    &   454 &   -2490   &   48.0    & + & (0,0)   &   (0,0)&   (-0.285,-0.848) \\ \hline
                                                                    \hline \hline
                                                                     \end{tabular}
} \caption[]{ Benchmarks for  the class HP where the Higgs boson
$(A,H)$ is the next nearest  heavy particle after the LSP
 in mSUGRA and in NUSUGRA. The LSP and $(A,H)$ sometimes are seen to switch.} \label{b5}
    \end{center}
 \end{table}

\clearpage

\end{document}